\documentclass[12pt]{article}
\pdfoutput=1
\usepackage{makecell}
\usepackage{booktabs}
\usepackage{physics}
\usepackage{tabulary}
\usepackage{multirow}
\usepackage{rotating}
\usepackage{epsfig}
\usepackage{psfrag}
\usepackage{latexsym}
\usepackage{indentfirst}
\usepackage{fancyhdr}
\usepackage{dsfont}
\usepackage{amsmath}
\usepackage{amssymb}
\usepackage{amsfonts}
\usepackage{mathrsfs}
\usepackage{amsthm}
\usepackage{pifont}
\usepackage{dsfont}
\usepackage{multirow}
\usepackage{array}
\usepackage{chngpage}
\usepackage{longtable}
\usepackage{cite}
\usepackage{bbold}
\usepackage{color}
\usepackage{braket}
\usepackage{colordvi}
\usepackage{color}
\usepackage{fancybox}
\usepackage[footnotesize]{caption2}
\usepackage{graphicx}
\usepackage[center,footnotesize,hang]{subfigure}
\usepackage{bbm}
\usepackage{bm}
\usepackage{url}
\usepackage[colorlinks, linkcolor=red, anchorcolor=black, citecolor=green]{hyperref}
\usepackage{multirow}
\usepackage{fancybox}
\usepackage{tabularx}
\usepackage{harpoon}
\usepackage{textcomp}
\usepackage{enumitem}
\usepackage{mathrsfs}
\usepackage{multicol}
\usepackage{fancybox}
\usepackage{supertabular}
\usepackage{tabularx}
\usepackage[table]{xcolor}
\usepackage{longtable}
\newcommand{\PreserveBackslash}[1]{\let\temp=\\#1\let\\=\temp}
\newcolumntype{C}[1]{>{\PreserveBackslash\centering}p{#1}}
\newcolumntype{R}[1]{>{\PreserveBackslash\raggedleft}p{#1}}
\newcolumntype{L}[1]{>{\PreserveBackslash\raggedright}p{#1}}
\allowdisplaybreaks

\newcommand{\bq}{\begin{eqnarray}}
\newcommand{\nq}{\end{eqnarray}}

\newcommand{\ignore}[1]{}

\makeatletter
\@addtoreset{equation}{section}
\makeatother

\begin{document}

\title{
{\Large \bf
New modular fixed point neutrino models and their phenomenological implications for JUNO, T2HK and DUNE
\\[2mm]}
\date{}
\author{Er-Hao Shang$^{a}$\footnote{E-mail: {\tt
shangeth@mail.ustc.edu.cn}}\,, \
Jun-Nan Lu$^{a}$\footnote{E-mail: {\tt
hitman@mail.ustc.edu.cn}}\,,  \
Gui-Jun Ding$^{a,b}$\footnote{E-mail: {\tt
dinggj@ustc.edu.cn}}\,, \
Stephen F. King$^{c}$\footnote{E-mail: {\tt king@soton.ac.uk}} \,
\\*[20pt]
\centerline{
\begin{minipage}{\linewidth}
\begin{center}
$^a${\it \small Department of Modern Physics,  and Anhui Center for fundamental sciences in theoretical physics,\\
University of Science and Technology of China, Hefei, Anhui 230026, China}\\[2mm]
$^b${\it \small  College of Physics, Guizhou University, Guiyang 550025, China}\\[2mm]
$^c${\it \small Physics and Astronomy, University of Southampton, Southampton, SO17 1BJ, U.K.}
\end{center}
\end{minipage}}
\\[10mm]}}

\maketitle

\thispagestyle{empty}

\begin{abstract}
We perform a general analysis of minimal modular fixed point models based on two right-handed neutrinos (2RHNs) and three modular fixed points, and find that the only viable possibilities are based on modular $S_4'$ and $A_5$ symmetry. Such models are highly predictive, with neutrino masses and the lepton mixing mixing matrix being fixed by three real parameters, as in the Littlest Seesaw Models. We perform an exhaustive scan over all possible models in this class and find many viable fixed points and
modular form alignments, after confronting them  with the latest neutrino oscillation global fits. The resulting models have the new feature that the two Dirac columns take more general forms than traditional Littlest Seesaw models, resulting in new sum rule relations between the solar and reactor angles, beyond those associated with TM1 (where the first column of the tri-bimaximal mixing matrix is preserved), which are compared to present and future projected JUNO results. We also compare the predictions of these models for the atmospheric angle and CP violating phase to current global fits and future T2HK and DUNE sensitivities.
\end{abstract}

\clearpage

\section{Introduction}

Since the discovery of neutrino oscillations~\cite{McDonald:2016ixn,Kajita:2016cak}, there has been remarkable progress in the experimental understanding of
neutrino mass and lepton mixing. The three neutrino mixing  paradigm~\cite{ParticleDataGroup:2024cfk} has emerged as the leading explanation of all oscillation data. The PMNS lepton mixing matrix involving the solar, reactor and atmospheric mixing angles
$\theta_{12}$,  $\theta_{13}$ and $\theta_{23}$, together with the two neutrino mass squared differences  $\Delta m^2_{31}$ (whose sign is undetermined) and the positive $\Delta m^2_{21}$ provide a good fit to all the experimental information~\cite{Capozzi:2025wyn,Capozzi:2025ovi,Esteban:2024eli,deSalas:2020pgw}.
The Dirac CP violating phase $\delta$ appearing in the PMNS matrix is currently only weakly constrained by oscillation data. The Dirac or Majorana nature of neutrino masses is currently undetermined, while the absolute scale of neutrino mass is only constrained by cosmology, though (neutrinoless double) beta decay experiments are underway to determine these unknowns.

The latest oscillation experiment to present new results is the
 Jiangmen Underground Neutrino Observatory (JUNO) experiment,
 which provides a world leading precision measurement of the solar oscillation parameters $\sin^2\theta_{12}$ and
 $\Delta m^2_{21}$ using the first 59.1 days of data~\cite{JUNO:2025gmd},
 \begin{equation}
\label{JUNO_59day}
\sin^2\theta_{12}=0.3092\pm0.0087,\ \ \ \  \Delta m^2_{21}=(7.50\pm0.12)\times 10^{-5} \text{eV}^2\,.
\end{equation}
Some implications of these results for theoretical models of lepton mixing have already been considered in~\cite{Xing:2025xte,Ge:2025cky,Ge:2025csr,Chao:2025sao,Huang:2025znh,Zhang:2025jnn,Li:2025hye,He:2025idv,Chen:2025afg,Jiang:2025hvq,Luo:2025pqy,Petcov:2025aci,Capozzi:2025ovi,Ding:2025dzc,Goswami:2025wla,Ding:2025dqd}.

Here we consider models of neutrino mass and lepton mixing based on modular symmetry~\cite{Feruglio:2017spp} (for reviews see e.g.~\cite{Kobayashi:2023zzc,Ding:2023htn}). In the modular symmetry approach to neutrino models, the flavour symmetry emerges as a finite quotient group $\Gamma'_N$ (or $\Gamma_N$) of the modular symmetry, broken by the vacuum expectation value (VEV) of a modulus field $\tau$. In particular we focus on a highly predictive sub-class of models in which the neutrino parameters are governed by special values of $\tau$ where residual subgroups of $\Gamma_N$ would be preserved,
known as fixed points. In the fundamental domain such fixed point values occur at $\tau_S=i$, $\tau_{ST}=(-1+i\sqrt{3})/2$, $\tau_{TS}=(1+i\sqrt{3})/2$, $\tau_T=i\infty$, leading to special modular form alignments for example in
modular $A_4$~\cite{Novichkov:2018yse}, $S_4$~\cite{Novichkov:2018ovf} and $A_5$~\cite{Novichkov:2018nkm,deMedeirosVarzielas:2022ihu} symmetries. If the modular symmetry is broken down to a residual $Z_3$ (or $Z_5$) subgroup in charged lepton sector and to a $Z_2$ subgroup in the neutrino sector, the trimaximal TM1 and TM2 mixing patterns can be obtained~\cite{Novichkov:2018yse,Novichkov:2018ovf}, although this requires the use of multiple moduli~\cite{deMedeirosVarzielas:2019cyj,King:2019vhv}.

With multiple moduli, the choice of fixed points is no longer restricted to the fundamental domain, and many more fixed points become relevant outside of it, namely
$\tau_f=\gamma\tau_S$, $\gamma\tau_{ST}$, $\gamma\tau_{TS}$ and $\gamma\tau_{T}$ in the upper half complex plane, where it is sufficient to consider $\gamma\in\Gamma_{N}$~\cite{Ding:2019gof}.
Such an approach yields a correspondingly larger range of modular form alignments, opening up the choices for model building
considerably, as was studied for $S_4$ and $A_4$ modular symmetries~\cite{Ding:2019gof}.
For example, in the case of $S_4$, one may apply the results to lepton mixing, with different residual subgroups associated fixed points, one for the charged lepton sector and one for each of the right-handed neutrinos sectors. In the minimal case of two right-handed neutrinos,
referred to as the Littlest Modular Seesaw,
there are thus three different modular fixed points involved in the lepton sector and
the light neutrino mass matrix only depends on three free parameters. In the Littlest Modular Seesaw, there are distinct predictions
for $\theta_{23}$ and $\delta_{CP}$~\cite{Ding:2019gof} which differ from those of the usual
non-modular Littlest Seesaw cases~\cite{King:2013iva,King:2013xba,King:2015dvf,Chen:2019oey},
based on non-abelian discrete flavour symmetry with flavons~\cite{Altarelli:2010gt,Ishimori:2010au,King:2013eh,King:2014nza,King:2015aea,King:2017guk,Feruglio:2019ybq,Ding:2024ozt}.

The Littlest Seesaw approach arose from the idea of sequential dominance (SD)~\cite{King:1998jw,King:1999cm,King:1999mb,King:2002nf} of right-handed neutrinos. The basic idea of SD is that, in the flavour basis (diagonal RHNs and charged lepton masses), assuming a normal and hierarchical mass ordering, one of the (CP conjugated) RHNs $N^c_{\mathrm{atm}}$ with mass $M_{\mathrm{atm}}$ is dominantly responsible for the heaviest physical neutrino mass $m_3$, while a second subdominant RHN $N^c_{\mathrm{sol}}$ with mass $M_{\mathrm{sol}}$ is mainly responsible for the second heaviest physical mass $m_2$, and a third essentially decoupled RHN $N^c_{\mathrm{dec}}$ of mass $M_{\mathrm{dec}}$ gives a very suppressed lightest neutrino mass $m_1$. In the limit that the third right-handed neutrino responsible for the lightest light neutrino mass is decoupled from the seesaw mechanism,
this leads to an effective 2RHN model with a neutrino mass hierarchy,
with $m_1 \approx 0$, where the large neutrino mixing angles arise in a natural way from ratios of couplings to the same right-handed neutrino~\cite{King:1998jw,King:1999cm,King:1999mb,King:2002nf}.
The predictivity of SD may be enhanced if certain relations between the
couplings can be enforced by some symmetry. In constrained sequential dominance (CSD), the dominant column of the Dirac mass matrix is proportional to
$(0,1,1)$ and the subdominant column to $(1,n,n-2)$ or $(1,n-2,n)$
for some real number $n$~~\cite{King:2005bj,Antusch:2011ic,King:2013iva,King:2015dvf,King:2016yvg,Ballett:2016yod,King:2018fqh,King:2013xba,King:2013hoa,Bjorkeroth:2014vha}. The choice $n\approx 3$ provides a particularly good fit to neutrino oscillation data and is called the Littlest Seesaw (LS) \cite{King:2015dvf}. For example models based on CSD($3$)~\cite{King:2013iva,King:2015dvf,King:2016yvg,Ballett:2016yod,King:2018fqh}, CSD($2.5$)~\cite{Chen:2019oey} may arise from vacuum alignment. For a given value of $n$, predictions for the PMNS matrix and the three neutrino masses can be analytically derived from the three real input parameters~\footnote{In the tri-direct CP approach~\cite{Ding:2018fyz,Ding:2018tuj,Chen:2019oey,Yan:2025itm}, other variants of the Littlest Seesaw can emerge.}. For the viable $S_4$ Littlest Modular Seesaw cases discussed in the literature so far, $n=1+\sqrt{6}$ and TM1 mixing is predicted, leading to
CSD($1+\sqrt{6}$) $\approx$ CSD($3.45$)~\cite{Ding:2019gof,Ding:2021zbg,deMedeirosVarzielas:2022fbw,deMedeirosVarzielas:2023ujt,deAnda:2023udh} from modular symmetry.

In this paper we extend the previous modular Littlest seesaw analysis, based on two right-handed neutrinos and three modular fixed points, from $S_4$ to the modular double cover group $S_4'$, resulting in new viable examples. We also include for the first time modular Littlest seesaw models based on $A_5$~\footnote{It turns out that the double cover group $A_5'$ does not yield any new interesting examples. We also studied the modular groups $A_4\times Z_2$, $GL(2,3)$ and $2O$, yet no new phenomenologically viable case beyond $S_4$ is found.}, leading to many more viable fixed points and modular form alignments than before. We perform an exhaustive scan over all possible Littlest Modular Seesaw models, where the resulting plethora of models are confronted with the latest neutrino oscillation global fits. The resulting models have the new feature that the two Dirac columns take more general forms than in CSD$(n)$, resulting in new sum rules beyond those associated with TM1, which are then confronted by the JUNO first results and the projected precision of 6 years data taking. We also compare the predictions of these models for the atmospheric angle $\theta_{23}$ and CP violating phase $\delta_{CP}$ to current global fits and future T2HK and DUNE sensitivities.

The layout of the remainder of the paper is as follows. We review the modular symmetry and modular fixed points in section~\ref{sec:modular-sym-fixed-points}. The framework of modular Littlest seesaw is presented in section~\ref{sec:modular-LSS}.
An exhaustive scan over all possible modular Littlest Seesaw constructions is performed, the phenomenologically viable models based on $S'_4$ and $A_5$ are presented in section~\ref{sec:MLS-N4} and section~\ref{sec:MLS-N5} respectively. The phenomenological implications of the these modular Littlest seesaw models at JUNO and long baseline neutrino oscillation experiments such as DUNE and T2HK are investigated in section~\ref{sec:pheno}. We conclude the paper in section~\ref{sec:conclusion}. The finite modular group $S'_4$ and modular forms of level $N=4$ are presented in the Appendix~\ref{app:N4-group-MF}, the alignments of modular form triplets at fixed points are tabulated up to weight 6. The $A'_5$ modular group and modular forms of level $N=5$ are given in the Appendix~\ref{app:N5}. The origin of lepton mixing is more clear in the charged lepton diagonal basis, although the physics results are independent of basis. The procedure of performing basis transformation is sketched in the Appendix~\ref{app:basis-transformation}.

\section{\label{sec:modular-sym-fixed-points}Modular symmetry and modular fixed points  }

The modular symmetry group $SL(2,\,\mathbb{Z})\equiv \Gamma$ is the special linear group of two-by-two integer matrices with unit determinant~\cite{Feruglio:2017spp,Ding:2023htn},
\begin{eqnarray}
SL(2,\,\mathbb{Z})=\left\{\begin{pmatrix}
a ~&~ b\\
c ~&~ d
\end{pmatrix}\Bigg| a, b,c,d\in \mathbb{Z}, ad-bc=1\right\}\equiv\Gamma\,.
\end{eqnarray}
The modular group has infinite group elements, nevertheless it can be generated by two generators $S$ and $T$,
\begin{eqnarray}
S=\begin{pmatrix}
0 ~&~ 1 \\
-1  ~&~ 0
\end{pmatrix}\,,~~~T=\begin{pmatrix}
1 ~&~ 1 \\
0  ~&~ 1
\end{pmatrix}\,,
\end{eqnarray}
which satisfy the following multiplication rules
\begin{equation}
S^4=(ST)^3=1,\quad  S^2T=TS^2\,.\label{eq-ST-rules}
\end{equation}
The modular group acts on the complex modulus $\tau$ via the linear fraction transformation,
\begin{equation}
\tau\mapsto\gamma\tau=\dfrac{a\tau+b}{c\tau+d}\,,
\end{equation}
where the imaginary part of the modulus $\tau$ is positive with $\text{Im}(\tau)>0$. Then one can read out the action of the modular generators $S$ and $T$ on $\tau$ as
\begin{equation}
S:\tau\mapsto-\dfrac{1}{\tau}\,,\qquad T:\tau\mapsto\tau+1\,.
\end{equation}
Each point of $\tau$ in the upper half complex plane can be
mapped into the fundamental domain $\mathcal{D}$ defined as
\begin{equation}
\mathcal{D}=\left\{\tau\Big|\text{Im}(\tau)>0,~ -\frac{1}{2}\leq\text{Re}(\tau)\leq \frac{1}{2}, ~|\tau|\geq 1 \right\}\,.
\end{equation}
The left half of the boundary of $\mathcal{D}$ is related to the left half boundary by $S$ and $T$ modular transformations. Nevertheless no two points in the interior of $\mathcal{D}$ are related under the modular transformation. In modular symmetry, the modular transformation of the matter field $\Phi_I$ is characterized by the modular weight $k_I$ and its transformation under the finite modular group $\Gamma'_N$  or $\Gamma_N$,
\begin{equation}
\gamma: \Phi_I\mapsto (c\tau+d)^{-k_I}\rho_I(\gamma)\Phi_I\,,
\end{equation}
where $\rho_I$ is an irreducible representation of $\Gamma'_N$ or $\Gamma_N$. Notice that $\Gamma'_N$ and $\Gamma_N$ are the quotient groups $\Gamma'_N=\Gamma/\Gamma(N)$ and $\Gamma_N=\Gamma/\pm\Gamma(N)$ respectively, where $\Gamma(N)$ for any positive integer $N$ is the principal congruence subgroup of level $N$ defined as
\begin{equation}
\Gamma(N) = \left\{ \begin{pmatrix}
a ~&~ b \\
c ~&~ d
\end{pmatrix} \in SL(2, \mathbb{Z}) \Bigg| \begin{pmatrix}
a ~&~ b \\
c ~&~ d
\end{pmatrix} = \begin{pmatrix}
1 ~&~ 0 \\
0 ~&~ 1
\end{pmatrix}~~({\rm mod}\, N) \right\} \,,
\end{equation}
which implies $T^N\in \Gamma(N)$. As a consequence, the homogeneous finite modular group $\Gamma'_N$ can be generated by the modular generator $S$ and $T$ obeying the rules $S^4=(ST)^3=T^N=1, S^2T=TS^2$, and additional relations are required to render the group $\Gamma'_N$ finite~\cite{deAdelhartToorop:2011re,Li:2021buv,Ding:2020msi}. Analogously the multiplication rules of the inhomogeneous finite modular group $\Gamma_N$ is $S^2=(ST)^3=T^N=1$ for $N\leq5$. The group $\Gamma'_N$ has twice as many elements as the group $\Gamma_N$, and it is the double cover of $\Gamma_N$~\cite{Liu:2019khw}.

In the framework of modular flavor symmetry, modular symmetry together with supersymmetry constrains the Yukawa couplings to be modular forms of level $N$. The modular forms of a given level $N$ at any non-negative integer weight $k$ span a linear space of finite dimension. For small values of $N$ and $k$, the dimension of the linear space of modular forms is smaller. Consequently only a small number of independent Yukawa couplings compatible with modular symmetry can be constructed, resulting in predictive flavor models. In fact, the modular forms of level $N$ and weight $k$, can be arranged into different irreducible multiplets $Y^{(k)}_{\bm{r}}(\tau)$ of $\Gamma'_N$ up to the automorphic factor~\cite{Feruglio:2017spp,Liu:2019khw},
\begin{eqnarray}
\label{eq:MF-decom}
Y^{(k)}_{\bm{r}}(\gamma\tau)=(c\tau+d)^k\rho_{\bm{r}}(\gamma)Y^{(k)}_{\bm{r}}(\tau)\,,
\end{eqnarray}
where $\gamma $ is the representative element of the coset $\gamma\Gamma(N)$ in $\Gamma'_N$, and $\rho_{\bm{r}}(\gamma)$ is the representation matrix of the element $\gamma$ in the irreducible representation $\bm{r}$ of $\Gamma_N$. When the modular weight $k$ is an even integer,  $\rho_{\bm{r}}(\gamma)$ reduces to irreducible representation of $\Gamma_N$ so that the even weight modular forms of level $N$ can be organized into irreducible multiplets of $\Gamma_N$.

In the context of $\mathcal{N}=1$ global supersymmetry, the
superpotential can be expanded in powers of matter superfields $\Phi_I$ as follow,
\begin{eqnarray}
\mathcal{W}=\sum\left(Y_{I_1I_2\ldots I_n}(\tau)\Phi_{I_1}\Phi_{I_2}\ldots\Phi_{I_n}\right)_{\bm{1}}\,,
\end{eqnarray}
where one should sum over all possible field combinations and independent singlet contractions of $\Gamma'_N$. Holomorphicity of the superpotential $\mathcal{W}$ implies that $Y_{I_1I_2\ldots I_n}(\tau)$ is a holomorphic function of the complex modulus $\tau$. Furthermore, invariance of the superpotential $\mathcal{W}$ under the action of modular symmetry constrains the Yukawa couplings $Y_{I_1I_2\ldots I_n}(\tau)$ to be modular forms of level $N$ and they should transform under $SL(2,\,\mathbb{Z})$ as follow,
\begin{eqnarray}
Y_{I_1I_2\ldots I_n}(\tau) \rightarrow Y_{I_1I_2\ldots I_n}(\gamma\tau)=(c\tau+d)^{k_Y}\rho_Y(\gamma)Y_{I_1I_2\ldots I_n}(\tau)\,,
\end{eqnarray}
where $k_Y=k_{I_1}+k_{I_2}+\ldots+k_{I_n}$ is the modular weight of the modular form $Y_{I_1I_2\ldots I_n}(\tau)$ which is in the irreducible representation $\rho_Y$ of $\Gamma'_N$ such that $\rho_Y\otimes \rho_{I_1}\otimes \rho_{I_2}\otimes\ldots\otimes\rho_{I_n}\supset \bm{1}$.

Modular invariance of the K\"ahler potential can be easily achieved, and the minimal form of the K\"ahler potential is adopted in this work,
\begin{eqnarray}
\mathcal{K}\left(\tau, \bar{\tau}; \Phi_I, \bar{\Phi}_I\right)=-\Lambda^2_K\log(-i\tau+i\bar{\tau})+\sum \frac{\bar{\Phi}_I\Phi_I}{(-i\tau+i\bar{\tau})^{k_I}}\,,
\end{eqnarray}
where $\Lambda_K$ is some mass parameter. Therefore we have to redefine the matter fields $\Phi_I\rightarrow (-i\tau+i\bar{\tau})^{k_I/2}\Phi_I$ to yield canonical kinetic term. This effect of field redefinition can be absorbed into the unknown couplings of the superpotential. Note that non-minimal Kahler potentials are allowed by modular symmetry, but they would in general introduce new input parameters affecting our
predictions and reducing predictability~\cite{Chen:2019ewa,Lu:2019vgm}.

\subsection{\label{subsec:modular-fixed-points}Modular fixed points and residual modular symmetry }

There is no value of $\tau$ which leave the whole modular group $SL(2, \mathbb{Z})$ unbroken. However, certain value of $\tau$ breaks the modular group partially, and it is called modular fixed point. In the fundamental domain, there are only four fixed points $\tau_0=i$, $e^{i2\pi/3}$, $e^{i\pi/3}$ and $i\infty$, which are invariant under the action of modular transformation $\gamma_0=S$, $ST$, $TS$ and $T$ respectively, i.e. $\gamma_0\tau_0=\tau_0$~\cite{cohen2017modular}. For clarity we denote $\tau_S=i$, $\tau_{ST}=e^{i2\pi/3}\equiv\omega$, $\tau_{TS}=e^{i\pi/3}$ and $\tau_{T}=i\infty$ in the following. Notice that $\tau_{ST}$ and $\tau_{TS}$ are related by the modular generator $T$ via $T\tau_{ST}=\tau_{TS}$ and $TS=T(ST)T^{-1}$. From Eq.~\eqref{eq:MF-decom}, we know that the modular form multiplet $Y^{(k)}_{\bm{r}}(\tau_0)$ at the fixed point $\tau_0$ takes a specific form satisfying,
\begin{eqnarray}
Y^{(k)}_{\bm{r}}(\tau_0)=Y^{(k)}_{\bm{r}}(\gamma_0\tau_0)=J^{k}(\gamma_0,\,\tau_0)\rho_{\bm{r}}(\gamma_0)Y^{(k)}_{\bm{r}}(\tau_0)\,,
\label{eq-modular-multiplet-0}
\end{eqnarray}
where $J(\gamma_0, \tau_0)$ denotes the automorphy factor,
\begin{equation}
J(\gamma_0, \tau_0)=c_0\tau_0+d_0\,,
\end{equation}
with
\begin{eqnarray}
J(S, \tau_S)=-i\,,~~~ J(ST, \tau_{ST})=\omega^2\,,~~~ J(TS, \tau_{TS})=\omega^2\,,~~~ J(T, \tau_{T})=1\,.
\end{eqnarray}
From Eq.~\eqref{eq-modular-multiplet-0}, we can see that $Y^{(k)}_{\bm{r}}(\tau_0)$ is an eigenvector of $\rho_{\bm{r}}(\gamma_0)$ corresponding to the eigenvalue $J^{-k}(\gamma_0,\,\tau_0)$. In fact, there are many other modular fixed points in the upper half complex plane, and they are related to $\tau_0$ by certain modular transformation~\cite{Ding:2019gof}, i.e.
\begin{equation}
\label{eq:tauf-Hplane}\tau_f=\gamma\tau_0,~~~\gamma_f=\gamma\gamma_0\gamma^{-1}\,,~~\gamma\in\Gamma\,,
\end{equation}
which fulfills $\gamma_f\tau_f=\tau_f$, and $\gamma$ is an arbitrary element of the modular group $SL(2,\,\mathbb{Z})$. The modular multiplet $Y^{(k)}_{\bm{r}}(\tau_f)$ at the fixed point $\tau_f$ is proportional to $\rho_{\bm{r}}(\gamma)Y^{(k)}_{\bm{r}}(\tau_0)$,
\begin{equation}
\label{eq:Ytauf}Y^{(k)}_{\bm{r}}(\tau_f)=
J^{k}(\gamma,\,\tau_0)\rho_{\bm{r}}(\gamma)Y^{(k)}_{\bm{r}}(\tau_0)\,,
\end{equation}
Consequently the modular multiplet $Y^{(k)}_{\bm{r}}(\tau_f)$ is an eigenvector of $\rho_{\bm{r}}(\gamma_f)$ corresponding to the eigenvalue $J^{-k}(\gamma_0,\,\tau_0)$ and the identity $\rho_{\bm{r}}(\gamma_f)Y^{(k)}_{\bm{r}}(\tau_f)=J^{-k}(\gamma_0,\,\tau_0)Y^{(k)}_{\bm{r}}(\tau_f)$ is fulfilled. Since $\rho_{\bm{r}}(\gamma)$ is a unit matrix for $\gamma\in\Gamma(N)$, there are a finite number of independent alignments of modular multiplet $Y^{(k)}_{\bm{r}}(\tau_f)$ at the fixed points, although there are infinite number of fixed points in the upper half plane. It is sufficient to only consider the representative elements $\gamma\in\Gamma'_N=\Gamma/\Gamma(N)$ and the fixed points $\tau_f=\Gamma'_N\tau_0$.

\section{\label{sec:modular-LSS}Framework of modular Littlest seesaw}

The modular Littlest seesaw approach is based on the minimal seesaw model with two right-handed neutrinos which are denoted as $N^c_{\mathrm{atm}}$ and $N^c_{\mathrm{sol}}$. The three generations of left-handed lepton doublets $L$ are assumed to transform as an irreducible triplet $\bm{3}$ of the finite modular group $\Gamma'_N$, the right-handed lepton fields $E^c$ can be either triplet or singlets of $\Gamma'_N$. Both right-handed neutrinos $N^c_{\mathrm{atm}}$ and $N^c_{\mathrm{sol}}$ are assigned to be singlets of $\Gamma'_N$, the Higgs fields $H_u$ and $H_d$ are invariant under $\Gamma'_N$. The modular invariant superpotential for the charged lepton and neutrino masses can be written as follow,
\begin{eqnarray}
\nonumber \mathcal{W}&=&-E^{c}_i Y^{\ell}_{ij}(\tau)L_jH_d +y_{\mathrm{atm}}LY_{\mathrm{atm}}(\tau)N^c_{\mathrm{atm}}H_u+y_{\mathrm{sol}}LY_{\mathrm{sol}}(\tau)N^c_{\mathrm{sol}}H_u \\
\label{eq:spp-MLSS}&&+\frac{1}{2}M_{\mathrm{atm}}N^c_{\mathrm{atm}}N^c_{\mathrm{atm}}
+\frac{1}{2}M_{\mathrm{sol}}N^{c}_{\mathrm{sol}}N^c_{\mathrm{sol}}\,,
\end{eqnarray}
where each term is modular invariant, and all possible contractions into singlet should be considered. The charged lepton Yukawa couplings $Y^{e}_{ij}(\tau)$ are modular forms. Obviously both $Y_{\mathrm{atm}}(\tau)$ and $Y_{\mathrm{sol}}(\tau)$ should transform as a triplet under the finite modular group $\Gamma'_N$ in order to fulfill the requirement of modular invariance, and the cross term $N^{c}_{\mathrm{atm}}N^c_{\mathrm{sol}}$ is forbidden by proper weight and representation assignments of $N^c_{\mathrm{atm}}$ and $N^c_{\mathrm{sol}}$ in the following. In the framework of modular Littlest seesaw, it is assumed that three complex moduli $\tau_{\ell}$, $\tau_{\text{atm}}$ and $\tau_{\text{sol}}$ are involved~\cite{Ding:2019gof}, $\tau_{\ell}$ is responsible for the breaking of modular symmetry in the charged lepton sector via its VEV, $\tau_{\text{atm}}$ and $\tau_{\text{sol}}$ break the modular symmetry in the atmospheric and solar neutrino sectors respectively. In order to increase the predictivity of the model, the VEV of the moduli $\langle\tau_{\ell}\rangle$, $\langle\tau_{\text{atm}}\rangle$ and $\langle\tau_{\text{sol}}\rangle$ are restricted to fixed points~\cite{Ding:2019gof}, the corresponding stabilizers are denoted as $\gamma_{\ell}$, $\gamma_{\text{atm}}$ and $\gamma_{\text{sol}}$ respectively satisfying $\gamma_{\ell}\langle\tau_{\ell}\rangle=\langle\tau_{\ell}\rangle$, $\gamma_{\text{atm}}\langle\tau_{\text{atm}}\rangle=\langle\tau_{\text{atm}}\rangle$ and $\gamma_{\text{sol}}\langle\tau_{\text{sol}}\rangle=\langle\tau_{\text{sol}}\rangle$. As a consequence, a residual subgroup generated by the stabilizer is preserved in each of the charged lepton, atmospheric neutrino and solar neutrino sectors. The above scenario of modular Littlest seesaw can be realized in the context of multiple modular symmetries~\cite{deMedeirosVarzielas:2022fbw,deMedeirosVarzielas:2023ujt} or in the orbifold construction with three factorizable tori~\cite{deAnda:2023udh}.

We first consider the charged lepton sector, the charged lepton Yukawa interactions are invariant under the following modular transformations of the lepton and Higgs fields,
\begin{equation}
L_i\xrightarrow{\gamma} J^{-k_{L_i}}(\gamma, \tau)\left(\rho_{L}(\gamma)\right)_{ij}L_j, ~~~~E^c_i\xrightarrow{\gamma}J^{-k_{E^c_i}}(\gamma, \tau)\left(\rho_{E^c}(\gamma)\right)_{ij}E^c_j\,,~\quad~
H_d\stackrel{\gamma}{\mapsto}J^{-k_{d}}(\gamma, \tau)\rho_{d}(\gamma)H_d\,,
\end{equation}
where $-k_{L_i}$, $-k_{E^c_i}$ and $-k_d$ are the modular weights of the fields $L_i$, $E^c_i$ and $H_d$ respectively. Since $(L_1, L_2, L_3)^T$ are assigned to a irreducible triplet of $\Gamma'_N$, their modular weights should be identical with $k_{L_1}=k_{L_2}=k_{L_3}\equiv k_{L}$. Thus the charged lepton Yukawa couplings $Y^{\ell}_{ij}(\tau)$ transform under the action of modular symmetry as follow,
\begin{eqnarray}
Y^{\ell}_{ij}(\tau)\xrightarrow{\gamma}Y^{\ell}_{ij}(\gamma\tau)=J^{k_{E^c_i}+k_{L_j}+k_d}(\gamma,\tau)\rho^{*}_{d}(\gamma)\left[\rho^*_{E^c}(\gamma) Y^{\ell}(\tau)\rho^{\dagger}_L(\gamma)\right]_{ij}\,,
\end{eqnarray}
The modulus is assumed to be stabilized at the fixed point $\tau_{\ell}$ in the charged lepton sector, thus the charged lepton mass matrix is of the form $M_{\ell}(\tau_{\ell})=Y^{\ell}(\tau_{\ell})v_d$. From the property $\gamma_{\ell}\langle\tau_{\ell}\rangle=\langle\tau_{\ell}\rangle$ of the modular fixed point, we have
\begin{eqnarray}
\left[M_{\ell}(\langle\tau_{\ell}\rangle)\right]_{ij}=J^{k_{E^c_i}+k_{L_j}+k_d}(\gamma_{\ell}, \langle\tau_{\ell}\rangle)\rho^{*}_{d}(\gamma_{\ell})\left[\rho^*_{E^c}(\gamma_{\ell}) M_{\ell}(\langle\tau_{\ell}\rangle)\rho^{\dagger}_L(\gamma_{\ell})\right]_{ij}\,,
\end{eqnarray}
which implies\footnote{Notice $J(\gamma_{\ell}, \langle\tau_{\ell}\rangle)$ is a phase at the modular fixed points.}
\begin{eqnarray}
\label{eq:MldagMl-rho-com}\left[M^{\dagger}_{\ell}(\langle\tau_{\ell}\rangle)M_{\ell}(\langle\tau_{\ell}\rangle), \rho_L(\gamma_{\ell})\right]=0\,.
\end{eqnarray}
Hence $M^{\dagger}_{\ell}(\langle\tau_{\ell}\rangle)M_{\ell}(\langle\tau_{\ell}\rangle)$ and $\rho_L(\gamma_{\ell})$ can be diagonalized by the same unitary transformation $U_{\ell}$, i.e.
\begin{eqnarray}
\label{eq:Ul-daig}U^{\dagger}_{\ell} M^{\dagger}_{\ell}(\langle\tau_{\ell}\rangle)M_{\ell}(\langle\tau_{\ell}\rangle) U_{\ell}=\text{diag}(m^2_e, m^2_{\mu}, m^2_{\tau})\,,~~~U^{\dagger}_{\ell} \rho_L(\gamma_{\ell}) U_{\ell}=\widehat{\rho}_L(\gamma_{\ell})\,,
\end{eqnarray}
where $\widehat{\rho}_L(\gamma_{\ell})$ is a diagonal phase matrix. As a result, the unitary matrix $U_{\ell}$ rotating the left-handed charged lepton fields to their diagonal mass basis is completely fixed by the residual modular symmetry generated by $\gamma_{\ell}$. In practice, it is the diagonalization matrix of $\rho_L(\gamma_{\ell})$. If the eigenvalues of the representation matrix $\rho_L(\gamma_{\ell})$ are non-degenerate, $U_{\ell}$ would be determined up to a permutation matrix $P_{\ell}$ and a diagonal phase matrix $Q_{\ell}$,
\begin{equation}
U_{\ell}\rightarrow U_{\ell}P_{\ell}Q_{\ell}\,.
\end{equation}
The residual modular symmetry can not predict the charged lepton masses, and the permutation matrix $P_{\ell}$ can be used to bring the eigenvalues of $M^{\dagger}_{\ell}(\langle\tau_{\ell}\rangle)M_{\ell}(\langle\tau_{\ell}\rangle)$ into the ordered form $\text{diag}(m^2_e, m^2_{\mu}, m^2_{\tau})$. The permutation $P_{\ell}$ can take the following six possible forms,
\begin{eqnarray}
\nonumber&&P_{123}=\begin{pmatrix}
1    ~&~ 0     ~&~ 0   \,\\
0    ~&~ 1     ~&~ 0   \,\\
0   ~&~ 0     ~&~ 1
\end{pmatrix}\,,
~~
P_{231}=\begin{pmatrix}
0    ~&~ 1    ~&~ 0   \,\\
0    ~&~ 0     ~&~ 1   \,\\
1    ~&~ 0     ~&~ 0
\end{pmatrix} \,,
~~
P_{312}=\begin{pmatrix}
0    ~&~ 0    ~&~ 1   \,\\
1    ~&~ 0     ~&~ 0   \,\\
0   ~&~ 1    ~&~ 0
\end{pmatrix}\,,\\
\label{eq:permu-matr}&&
P_{132}=\begin{pmatrix}
1    ~&~ 0     ~&~ 0   \,\\
0    ~&~ 0    ~&~ 1   \,\\
0   ~&~ 1    ~&~ 0
\end{pmatrix}\,,
~~
P_{213}=\begin{pmatrix}
0    ~&~ 1     ~&~ 0   \,\\
1    ~&~ 0    ~&~ 0   \,\\
0   ~&~ 0    ~&~ 1
\end{pmatrix}\,,
~~
P_{321}=\begin{pmatrix}
0   ~&~ 0     ~&~ 1   \,\\
0   ~&~ 1    ~&~ 0   \,\\
1    ~&~ 0   ~&~ 0
\end{pmatrix}\,.
\end{eqnarray}
We proceed to analyze the neutrino sector. After the electroweak and modular symmetries spontaneous breaking, the two columns of the Dirac neutrino mass matrix $M_D$ are determined by the triplet modular froms $Y_{\text{atm}}(\tau_{\text{atm}})$ and $Y_{\text{sol}}(\tau_{\text{sol}})$ which are aligned along certain directions, since the moduli VEVs $\tau_{\text{atm}}$ and $\tau_{\text{sol}}$ are assumed to be stabilized at modular fixed points $\langle\tau_{\text{atm}}\rangle$ and $\langle\tau_{\text{sol}}\rangle$ respectively, and they preserve the residual modular symmetries generated by $\gamma_{\text{atm}}$ and $\gamma_{\text{sol}}$ respectively. Thus we have
\begin{equation}
M_{D}=\begin{pmatrix}
y_{\text{atm}}Y_{\text{atm}}\left(\langle\tau_{\mathrm{atm}}\rangle\right)v_u,  & y_{\text{sol}}Y_{\text{sol}}\left(\langle\tau_{\mathrm{sol}}\rangle\right)v_u
\end{pmatrix}\,,
\end{equation}
where the Clebsch-Gordan contraction coefficients in the neutrino Yuakwa couplings $y_{\mathrm{atm}}LY_{\mathrm{atm}}N^c_{\mathrm{atm}}H_u$ and $y_{\mathrm{sol}}LY_{\mathrm{sol}}N^c_{\mathrm{sol}}H_u$ are dropped for notation simplicity. The two dimensional mass matrix $M_N$ of the right-handed neutrinos is diagonal,
\begin{equation}
M_{N}=\begin{pmatrix}
M_{\textrm{atm}} ~&~ 0 \\
0  ~& ~ M_{\textrm{sol}}
\end{pmatrix}\,,
\end{equation}
which is enforced by proper assignment of $N^c_{\mathrm{atm}}$ and $N^c_{\mathrm{sol}}$ under modular symmetry. The light neutrino mass matrix is given by the well-known seesaw formula, \begin{equation}
\label{eq:Mnu-ma-ms}M_{\nu}=-M_{D}M_{N}M^T_{D}=m_aY_{\text{atm}}Y^{T}_{\text{atm}}+m_s e^{i\eta}Y_{\text{sol}}Y^{T}_{\text{sol}}\,.
\end{equation}
with
\begin{equation}
m_a=-\frac{y^2_{\text{atm}}v^2_u}{M_{\text{atm}}}\,,~~~m_s=\left|\frac{y^2_{\text{sol}}v^2_u}{M_{\mathrm{sol}}}\right|\,,~~~\eta=\text{arg}\left(\frac{M_{\mathrm{atm}}}{M_{\text{sol}}}\frac{y^2_{\mathrm{sol}}}{y^2_{\text{atm}}}\right)\,.
\end{equation}
We see that the light neutrino mass matrix $M_{\nu}$ only depends on three real parameters $m_a$, $m_s$ and $\eta$, therefore the modular Littlest seesaw models are very predictive. Moreover, we see that the light neutrino mass matrix $M_{\nu}$ admits a vanishing mass eigenvalue and the corresponding eigenvector is proportional to the cross product of $Y_{\text{atm}}$ and $Y_{\text{sol}}$, i.e,
\begin{equation}
\label{eq:fixed-column}M_{\nu} \left[Y_{\text{atm}}\times Y_{\text{sol}}\right]=(0, 0, 0)^T\,.
\end{equation}
Hence the lightest neutrino is massless. We shall focus on neutrino ordering neutrino masses in this work, since the current neutrino oscillation data slightly prefer normal ordering over inverted ordering at $2.2\sigma$ confidence level~\cite{Capozzi:2025wyn,Capozzi:2025ovi,Esteban:2024eli,deSalas:2020pgw}. Furthermore, the left-handed neutrino fields can be rotated to their mass eigenstates by the unitary transformation $U_{\nu}$ satisfying
\begin{equation}
U^T_{\nu} M_{\nu} U_{\nu}=\text{diag}(0, m_2, m_3)\,.
\end{equation}
Accordingly the first column of $U_{\nu}$ is $Y_{\text{atm}}\times Y_{\text{sol}}/\left|Y_{\text{atm}}\times Y_{\text{sol}}\right|$ for normal ordering neutrino mass, and it is fixed by the residual modular symmetry. After including the Clebsch-Gordan rearrangment, the first column of $U_{\nu}$ turns out to be $P_{132}\left(Y_{\text{atm}}\times Y_{\text{sol}}\right)/\left|Y_{\text{atm}}\times Y_{\text{sol}}\right|$ in the basis of Appendices~\ref{app:N4-group-MF} and~\ref{app:N5}. As a consequence, the lepton mixing matrix is determined to be
\begin{equation}
U_{PMNS}=Q^{\dagger}_{\ell} P^{\dagger}_{\ell} U^{\dagger}_{\ell} U_{\nu} \,,
\end{equation}
where the phase factor $Q_{\ell}$ is unphysical and it can be absorbed by redefinition of the charged lepton fields.

In the standard parametrization, the lepton mixing matrix can be written as~\cite{ParticleDataGroup:2024cfk},
\begin{equation}\label{eq:PMNS_def}
U_{PMNS}=\left(\begin{array}{ccc}
c_{12}c_{13}  &   s_{12}c_{13}   &   s_{13}e^{-i\delta_{CP}}  \\
-s_{12}c_{23}-c_{12}s_{13}s_{23}e^{i\delta_{CP}}   &  c_{12}c_{23}-s_{12}s_{13}s_{23}e^{i\delta_{CP}}  &  c_{13}s_{23}  \\
s_{12}s_{23}-c_{12}s_{13}c_{23}e^{i\delta_{CP}}   & -c_{12}s_{23}-s_{12}s_{13}c_{23}e^{i\delta_{CP}}  &  c_{13}c_{23}
\end{array}\right)\text{diag}(e^{i\frac{\alpha}{2}},e^{i\frac{\beta}{2}},1)\,,
\end{equation}
where $c_{ij}=\cos\theta_{ij}$, $s_{ij}=\sin\theta_{ij}$, the mixng angle $\theta_{ij}$ can be taken in the first octant $\theta_{ij}\in[0, \pi/2]$ and the Dirac CP violation phase $\delta_{CP}\in[0, 2\pi)$, and the Majorana phase $\alpha$ is unphysical since the lightest neutrino is massless.

In the approach of modular Littlest seesaw, the neutrino mass matrix and the lepton flavor mixing are fixed by the residual modular symmetries in the charged lepton sector, atmospheric neutrino sector and solar neutrino sectors, which are characterized by the stabilizers  $\gamma_{\ell}$, $\gamma_{\text{atm}}$ and $\gamma_{\text{sol}}$ respectively. If another set of stabilizers $\gamma'_{\ell}$, $\gamma'_{\text{atm}}$, $\gamma'_{\text{sol}}$ are conjugated to $\gamma_{\ell}$, $\gamma_{\text{atm}}$, $\gamma_{\text{sol}}$ under a modular transformation, i.e.
\begin{equation}
\gamma'_{\ell}=\gamma\, \gamma_{\ell} \gamma^{-1}\,,~~\gamma'_{\text{atm}}=\gamma \,\gamma_{\text{atm}} \gamma^{-1}\,,~~\gamma'_{\text{sol}}=\gamma \,\gamma_{\text{sol}} \gamma^{-1}\,,~~\gamma\in SL(2,\mathbb{Z})\,, \label{eq:stabilizers-conjugate}
\end{equation}
accordingly the corresponding charged lepton mass matrix $M^{\dagger}_{\ell}M_{\ell}$ and neutrino mass matrix $M_{\nu}$ would become $\rho_{L}(\gamma)M^{\dagger}_{\ell}M_{\ell}\rho^{\dagger}_{L}(\gamma)$ and $\rho^{*}_{L}(\gamma)M_{\nu}\rho^{\dagger}_{L}(\gamma)$ respectively. Therefore one reaches the same lepton mixing matrix $U_{PMNS}$.

In the following, we perform an exhaustive scan over all possible alignments of triplet modular forms at the fixed points of the finite modular groups $S'_4$ and $A_5$, and then identify the phenomenologically viable modular Littlest Seesaw models by confronting them with the latest global fits to neutrino oscillation data.

\section{\label{sec:MLS-N4}Modular Littlest seesaw models based on finite group $S_4^\prime$}

The homogeneous finite modular group $\Gamma'_4\cong S'_4$ is the double cover of the $S_4$ group. It has four triplet representations $\bm{3}$, $\bm{3'}$, $\widehat{\bm{3}}$ and $\widehat{\bm{3}}^\prime$, two doublet representations $\bm{2},\,\widehat{\bm{2}}$, and four singlet representations $\bm{1}$, $\bm{1}^\prime$, $\widehat{\bm{1}}$, $\widehat{\bm{1}}^{\prime}$. The group theory and the representation matrices of $S'_4$ as well as the Clebsch–Gordan coefficients are listed in Appendix~\ref{app:N4-group-MF}. There are three linearly independent weight one modular forms at level $N=4$, and they can be arranged into a $S'_4$ triplet $\widehat{\bm{3}}^\prime$~\cite{Novichkov:2020eep,Liu:2020akv,Liu:2020msy,Ding:2022nzn}, i.e.
\begin{eqnarray}
\label{eq:MF-w13hp-N4}Y^{(1)}_{\widehat{\bm{3}}^\prime}(\tau)=\begin{pmatrix}\sqrt{2}\vartheta_1\vartheta_2\\-\vartheta_2^2\\\vartheta_1^2\end{pmatrix}\equiv\begin{pmatrix} Y_1(\tau)\\Y_2(\tau)\\Y_3(\tau)\end{pmatrix}\,,
\end{eqnarray}
where $\vartheta_1\left(\tau\right)$ and $\vartheta_2\left(\tau\right)$ are half weight modular forms of level $N=4$~\cite{Liu:2020msy},
\begin{eqnarray}
\nonumber \vartheta_1\left(\tau\right)& = & \sum_{m\in \mathbb{Z}}e^{2\pi i \tau m^2}=1+2q+2q^4+2q^9+2q^{16}+2q^{25}+2q^{36}+2q^{49}+2q^{64}+\cdots\,, \\
\label{eq:vartheta12}\vartheta_2\left(\tau\right) & = & -\sum_{m\in \mathbb{Z}}e^{2\pi i \tau \left(m+1/2\right)^2}=-2q^{1/4}\left(1+q^2+q^6+q^{12}+q^{20}+q^{30}+q^{42}+q^{56}+\cdots\right)\,,
\end{eqnarray}
with $q=e^{2\pi i\tau}$. Therefore the $q$-expansion of the modular forms $Y_{1,2,3}(\tau)$ read as
\begin{eqnarray}
\nonumber Y_1(\tau) & = & -2\sqrt{2}q^{1/4}(1+2q+q^2+2q^3+2q^4+3q^6+2q^7+2q^9+2q^{10}+2q^{11}+q^{12}+\ldots)\,,\\
\nonumber Y_2(\tau) & = &-4q^{1/2}(1+2q^2+q^4+2q^6+2q^8+3q^{12}+\cdots)\,,\\
 Y_3(\tau) & = &1+4q+4q^2+4q^4+8q^5+4q^8+4q^9+8q^{10}+8q^{13}+\cdots\,.
\end{eqnarray}
The higher weight modular forms of level $N=4$ can be constructed from the tensor products of $Y^{(1)}_{\widehat{\bm{3}}^\prime}$, they are homogeneous polynomials of $Y_{1,2,3}(\tau)$, and their explicit expressions are summarized in Appendix~\ref{app:N4-group-MF}. The ratio $\vartheta_2\left(\tau\right)/\vartheta_1\left(\tau\right)$ at the modular fixed points $\tau_S=i$, $\tau_{ST}=e^{i2\pi/3}\equiv\omega$, $\tau_{TS}=e^{i\pi/3}$ and $\tau_{T}=i\infty$ are found to be
\begin{eqnarray}
\frac{\vartheta_2(\tau_S)}{\vartheta_1(\tau_S)}=1-\sqrt{2}\,,~~~\frac{\vartheta_2(\tau_{ST})}{\vartheta_1(\tau_{ST})}=-\frac{1-i}{1+\sqrt{3}}\,,~~~\frac{\vartheta_2(\tau_{TS})}{\vartheta_1(\tau_{TS})}=-\frac{1+i}{1+\sqrt{3}}\,,~~~\frac{\vartheta_2(\tau_{T})}{\vartheta_1(\tau_{T})}=0\,.
\end{eqnarray}
As a result, the alignment of weight 1 modular form triplet $Y^{(1)}_{\widehat{\bm{3}}^\prime}(\tau)$ at the modular fixed points are determined to be
\begin{eqnarray}
\nonumber&& Y^{(1)}_{\widehat{\bm{3}}^\prime}(\tau_S)=Y_S\begin{pmatrix}1\\1-\frac{1}{\sqrt{2}}\\-1-\frac{1}{\sqrt{2}} \end{pmatrix}\,,\qquad Y^{(1)}_{\widehat{\bm{3}}^\prime}(\tau_{ST})=Y_{ST}\begin{pmatrix}1\\\frac{1-\sqrt{3}}{2}\xi^3\\\frac{1+\sqrt{3}}{2}\xi^5 \end{pmatrix}\,, \\
&&
Y^{(1)}_{\widehat{\bm{3}}^\prime}(\tau_{TS})=Y_{TS}\begin{pmatrix}1\\\frac{1-\sqrt{3}}{2}\xi^5\\\frac{1+\sqrt{3}}{2}\xi^3 \end{pmatrix}\,,\qquad Y^{(1)}_{\widehat{\bm{3}}^\prime}(\tau_{T})=\begin{pmatrix}0\\0\\1 \end{pmatrix} \,, \label{eq:Y-w1-3hatp-N4}
\end{eqnarray}
where $Y_S=-0.5902$, $Y_{ST}=-0.5087+0.5087i$, $Y_{TS}=-0.5087-0.5087i$ and $\xi=e^{\pi i/4}=(1+i)/\sqrt{2}$. The modular multiplets of higher weights are tensor products of $Y^{(1)}_{\widehat{\bm{3}}^\prime}(\tau)$, thus theur alignments at the modular fixed points can be determined from Eq.~\eqref{eq:Y-w1-3hatp-N4}, and the results are collected in table~\ref{tab:mf-S4p}.

As shown in Eq.~\eqref{eq:tauf-Hplane}, although there are only four modular fixed points $\tau_0=\tau_S$, $\tau_{ST}$, $\tau_{TS}$ and $\tau_{T}$ in the fundamental domain, there are infinity numbers of modular fixed  points denoted as $\tau_f$ in the upper half complex plane. $\tau_f$ is related to $\tau_0$ through some modular transformation, i.e. $\tau_f=\gamma\tau_0$ where $\gamma$ can be any element of $SL(2,\,\mathbb{Z})$. From Eq.~\eqref{eq:Ytauf}, we know that it is sufficient to focus on $\gamma\in\Gamma'_N$ to obtain all possible independent alignment of modular multiplets at the fixed points. In the framework of modular Littlest  seesaw, both $Y_{\text{atm}}(\tau_{\text{atm}})$ and $Y_{\text{sol}}(\tau_{\text{sol}})$ are modular triplets. We list all the alignments of the modular triplets of level $N=4$ in table~\ref{tab:tb-fp-S4p}. $Y_{\text{atm}}(\tau_{\text{atm}})$ and $Y_{\text{sol}}(\tau_{\text{sol}})$ could be any of these alignments for proper values of $\langle\tau_{\text{atm}}\rangle$ and $\langle\tau_{\text{sol}}\rangle$.

\begin{center}
\setlength\LTcapwidth{\textwidth}
\begin{longtable}{|c|c|c|c|c|}
\caption{\normalsize\label{tab:mf-S4p} The values of the modular form multiplets of level $N=4$ and weights $k=1,\,2,\,3,\,4,\,5,\,6$ at the modular fixed points $\tau_S,\,\tau_{ST},\,\tau_{TS}$ and $\tau_T$, where $Y_S=-0.5902,\,Y_{ST}=-0.5087+0.5087i,\,Y_{TS}=-0.5087-0.5087i$ and $\xi=e^{\pi i/4}$.} \\
\toprule \hline
 & $\tau_S$ &  $\tau_{ST}$ & $\tau_{TS}$ & $\tau_T$\\ \hline
\endfirsthead
\multicolumn{5}{c}{{\bfseries \tablename\ \thetable{} -- continued from previous page}}\\
\hline
 & $\tau_S$ &  $\tau_{ST}$ & $\tau_{TS}$ & $\tau_T$\\
\hline
\endhead

\multicolumn{5}{c}{{\bfseries \tablename\ \thetable{} -- continues on next page}}\\
\endfoot

\hline
\endlastfoot
$Y^{(1)}_{\widehat{\bm{3}}^\prime}$ & $Y_S\begin{pmatrix}1\\1-\frac{1}{\sqrt{2}}\\-1-\frac{1}{\sqrt{2}} \end{pmatrix}$ & $Y_{ST}\begin{pmatrix}1\\\frac{1-\sqrt{3}}{2}\xi^3\\\frac{1+\sqrt{3}}{2}\xi^5 \end{pmatrix} $ & $Y_{TS}\begin{pmatrix}1\\\frac{1-\sqrt{3}}{2}\xi^5\\\frac{1+\sqrt{3}}{2}\xi^3 \end{pmatrix} $ & $\begin{pmatrix}0\\0\\1 \end{pmatrix}$\\
\hline
$Y^{(2)}_{\bm{2}}$ & $3Y^2_S\begin{pmatrix}1\\-\frac{1}{\sqrt{3}} \end{pmatrix} $ & $\sqrt{3}\,iY^2_{ST}\begin{pmatrix}1\\i \end{pmatrix}$ & $-\sqrt{3}\,iY^2_{TS}\begin{pmatrix}1\\-i \end{pmatrix} $ & $\begin{pmatrix}1\\0 \end{pmatrix} $\\
\hline
$Y^{(2)}_{\bm{3}}$ & $2\sqrt{2}\,Y^2_S\begin{pmatrix}1\\-\frac{1}{2}-\frac{1}{\sqrt{2}}\\\frac{1}{2}-\frac{1}{\sqrt{2}} \end{pmatrix} $ & $2iY^2_{ST}\begin{pmatrix}1\\\frac{1+\sqrt{3}}{2}\xi^3\\\frac{1-\sqrt{3}}{2}\xi^5 \end{pmatrix} $ & $-2iY^2_{TS}\begin{pmatrix}1\\\frac{1+\sqrt{3}}{2}\xi^5\\\frac{1-\sqrt{3}}{2}\xi^3 \end{pmatrix} $ & $\begin{pmatrix}1\\0\\0 \end{pmatrix} $ \\
\hline
$Y^{(3)}_{\widehat{\bm{1}}^\prime}$ & $2Y^3_S $ & $\sqrt{2}\,iY^3_{ST} $ & $-\sqrt{2}\,iY^3_{TS} $ & $0 $ \\
\hline
$Y^{(3)}_{\widehat{\bm{3}}}$ & $2Y^3_S\begin{pmatrix}1\\-2-\frac{1}{\sqrt{2}}\\2-\frac{1}{\sqrt{2}} \end{pmatrix} $ & $2Y_{ST}^3\begin{pmatrix} 1\\\xi^7\\\xi\end{pmatrix} $ & $2Y_{TS}^3\begin{pmatrix} 1\\\xi\\\xi^7\end{pmatrix} $ & $ \begin{pmatrix}0\\1\\0 \end{pmatrix}$ \\
\hline
$Y^{(3)}_{\widehat{\bm{3}}^\prime}$ & $6Y_S^3\begin{pmatrix}1\\\frac{1}{\sqrt{2}}\\\frac{1}{\sqrt{2}} \end{pmatrix} $ & $2\sqrt{3}\,iY^3_{ST}\begin{pmatrix}1\\\xi^7\\\xi \end{pmatrix}  $ & $-2\sqrt{3}\,iY^3_{TS}\begin{pmatrix}1\\\xi\\\xi^7 \end{pmatrix}$ & $ \begin{pmatrix}0\\0\\-1 \end{pmatrix}$ \\
\hline
$Y^{(4)}_{\bm{1}}$ & $12Y^4_S $ & $0 $ & $0 $& $1 $ \\
\hline
$Y^{(4)}_{\bm{2}}$ & $6Y^4_S\begin{pmatrix}1\\\sqrt{3} \end{pmatrix} $ & $-6Y^4_{ST}\begin{pmatrix}1\\-i \end{pmatrix} $ & $-6Y^4_{TS}\begin{pmatrix}1\\i \end{pmatrix} $ & $\begin{pmatrix}1\\0 \end{pmatrix} $ \\
\hline
$Y^{(4)}_{\bm{3}}$ & $-6\sqrt{2}\,Y^4_S\begin{pmatrix} 1\\\frac{1}{\sqrt{2}}\\\frac{1}{\sqrt{2}}\end{pmatrix} $ & $2\sqrt{3}\,Y^4_{ST}\begin{pmatrix}1\\\frac{1-\sqrt{3}}{2}\xi^3\\\frac{1+\sqrt{3}}{2}\xi^5 \end{pmatrix} $ & $2\sqrt{3}\,Y^4_{TS}\begin{pmatrix}1\\\frac{1-\sqrt{3}}{2}\xi^5\\\frac{1+\sqrt{3}}{2}\xi^3 \end{pmatrix} $ & $\begin{pmatrix}-1\\0\\0 \end{pmatrix} $\\
\hline
$Y^{(4)}_{\bm{3}^\prime}$ & $2Y^4_S\begin{pmatrix}1\\1-\frac{1}{\sqrt{2}}\\-1-\frac{1}{\sqrt{2}} \end{pmatrix} $ & $\sqrt{2}\,iY^4_{ST}\begin{pmatrix}1\\\frac{1-\sqrt{3}}{2}\xi^3\\\frac{1+\sqrt{3}}{2}\xi^5 \end{pmatrix} $ & $-\sqrt{2}\,iY^4_{TS}\begin{pmatrix}1\\\frac{1-\sqrt{3}}{2}\xi^5\\\frac{1+\sqrt{3}}{2}\xi^3 \end{pmatrix} $ & $\begin{pmatrix}0\\0\\0 \end{pmatrix} $\\
\hline
$Y^{(5)}_{\widehat{\bm{2}}}$ & $2\sqrt{3}\,Y^5_S\begin{pmatrix}1\\\sqrt{3} \end{pmatrix} $ & $\sqrt{6}\,iY^5_{ST}\begin{pmatrix}1\\i \end{pmatrix} $ & $-\sqrt{6}\,iY^5_{TS}\begin{pmatrix}1\\-i \end{pmatrix} $ & $ \begin{pmatrix}0\\0 \end{pmatrix}$ \\
\hline
$Y^{(5)}_{\widehat{\bm{3}}}$ & $-12\,Y^5_S\begin{pmatrix}1\\\frac{1}{\sqrt{2}}\\\frac{1}{\sqrt{2}} \end{pmatrix} $ & $-4\sqrt{3}\,iY^5_{ST}\begin{pmatrix}1\\\frac{1+\sqrt{3}}{2}\xi^3\\\frac{1-\sqrt{3}}{2}\xi^5 \end{pmatrix} $ & $4\sqrt{3}\,iY^5_{TS}\begin{pmatrix}1\\\frac{1+\sqrt{3}}{2}\xi^5\\\frac{1-\sqrt{3}}{2}\xi^3 \end{pmatrix} $ & $\begin{pmatrix}0\\1\\0 \end{pmatrix} $ \\
\hline
$Y^{(5)}_{\widehat{\bm{3}}^\prime I}$ & $12Y_S^5\begin{pmatrix}1\\-2-\frac{1}{\sqrt{2}}\\2-\frac{1}{\sqrt{2}} \end{pmatrix} $ & $-12Y^5_{ST}\begin{pmatrix}1\\\frac{1+\sqrt{3}}{2}\xi^3\\\frac{1-\sqrt{3}}{2}\xi^5 \end{pmatrix} $ & $-12Y^5_{TS}\begin{pmatrix}1\\\frac{1+\sqrt{3}}{2}\xi^5\\\frac{1-\sqrt{3}}{2}\xi^3 \end{pmatrix}  $ & $ \begin{pmatrix}0\\0\\-1 \end{pmatrix}$ \\
\hline
$Y^{(5)}_{\widehat{\bm{3}}^\prime II}$ & $12Y_S^5\begin{pmatrix}1\\1-\frac{1}{\sqrt{2}} \\-1-\frac{1}{\sqrt{2}}\end{pmatrix} $ & $\begin{pmatrix}0\\0\\0 \end{pmatrix} $ & $\begin{pmatrix}0\\0\\0 \end{pmatrix} $ & $\begin{pmatrix}0\\0\\1 \end{pmatrix} $ \\
\hline
$Y^{(6)}_{\bm{1}}$ & $ 0$ & $-12\sqrt{3}\,iY^6_{ST} $ & $12\sqrt{3}\,iY^6_{TS} $ & $1 $ \\
\hline
$Y^{(6)}_{\bm{1}^\prime}$ & $4Y^6_S $ & $-2Y^6_{ST} $ & $-2Y^6_{TS}  $ & $0 $ \\
\hline
$Y^{(6)}_{\bm{2}}$ & $36\,Y_S^6\begin{pmatrix}1\\-\frac{1}{\sqrt{3}} \end{pmatrix} $ & $\begin{pmatrix}0\\0 \end{pmatrix} $ & $\begin{pmatrix}0\\0 \end{pmatrix} $ & $\begin{pmatrix}1\\0 \end{pmatrix} $ \\
\hline
$Y^{(6)}_{\bm{3} I}$ & $12\sqrt{2}\,Y_S^6\begin{pmatrix}1\\1-\frac{1}{\sqrt{2}}\\-1-\frac{1}{\sqrt{2}} \end{pmatrix} $ & $-12iY^6_{ST}\begin{pmatrix}1\\\xi^7\\\xi \end{pmatrix} $ & $12iY^6_{TS}\begin{pmatrix}1\\\xi\\\xi^7 \end{pmatrix} $ & $\begin{pmatrix} 1\\0\\0\end{pmatrix} $ \\
\hline
$Y^{(6)}_{\bm{3} II}$ & $24\sqrt{2}\,Y_S^6\begin{pmatrix}1\\-\frac{1}{2}-\frac{1}{\sqrt{2}}\\\frac{1}{2}-\frac{1}{\sqrt{2}} \end{pmatrix} $ & $\begin{pmatrix}0\\0\\0 \end{pmatrix} $ & $\begin{pmatrix}0\\0\\0 \end{pmatrix} $ & $\begin{pmatrix}1\\0\\0 \end{pmatrix} $ \\
\hline
$Y^{(6)}_{\bm{3}^\prime}$ & $ 12Y^6_S\begin{pmatrix}1\\\frac{1}{\sqrt{2}}\\\frac{1}{\sqrt{2}} \end{pmatrix}$ & $-2\sqrt{6}\,Y^6_{ST}\begin{pmatrix}1\\\xi^7\\\xi \end{pmatrix} $ & $-2\sqrt{6}\,Y^6_{TS}\begin{pmatrix}1\\\xi\\\xi^7 \end{pmatrix} $ & $\begin{pmatrix}0\\0\\0 \end{pmatrix} $ \\
\hline \bottomrule
\end{longtable}
\end{center}

\subsection{Modular Littlest seesaw models in $S'_4$ }

The three generations of the left-handed lepton doublets are assumed to transform as an irreducible triplet $\bm{3}$, $\bm{3}^\prime$,$\widehat{\bm{3}}$ or $\widehat{\bm{3}}^\prime$ under $S'_4$, the two right-handed neutrinos $N^c_{\mathrm{atm}}$ and $N^c_{\mathrm{sol}}$ are assigned to be two different singlets of $S'_4$ which are $\bm{1}$,$\bm{1}^\prime$, $\widehat{\bm{1}}$, $\widehat{\bm{1}}^\prime$. As explained in Eq.~\eqref{eq:stabilizers-conjugate} and section~\ref{sec:modular-LSS},  the same predictions for lepton masses and flavor mixing would be reached if two pairs of residual symmetries in the charged lepton, solar neutrino and atmospheric neutrino sectors are conjugate to each other. As a consequence, it is sufficient to focus on $\gamma_{\ell}=ST$ or $\gamma_{\ell}=T$\footnote{If $\gamma_{\ell}=S$, the eigenvalues of $\rho_{\bm{3}}(\gamma_{\ell})$ are degenerate so that the three generations of charged leptons can not be fully distinguished and only one column of the unitary diagonalization matrix $U_{\ell}$ is fixed. Thus more free parameters would be involved and the predictive power of Littlest seesaw would be reduced. Moreover, $TS=T(ST)T^{-1}$ is conjugate to $ST$.}, the corresponding unitary diagonalization matrix $U_{\ell}$ is given by
\begin{eqnarray}
\label{eq:Uell-ST} \gamma_{\ell}&=&ST ~:~U_{\ell}=\frac{1}{2\sqrt{3}}\left(
\begin{array}{ccc}
 2e^{-\pi i/4} ~&~  2e^{-\pi i/4} ~&~  2e^{3\pi i/4}  \\
 -2i   ~&~  \left(1+\sqrt{3}\right)i ~~&~~ \left(\sqrt{3}-1\right)i \\
2  ~&~ \sqrt{3}-1  ~&~ 1+\sqrt{3}
\end{array}
\right) \,,\\
\label{eq:Uell-T} \gamma_{\ell}&=&T~:~U_{\ell}=\left(
\begin{array}{ccc}
 1 ~&~ 0 ~&~ 0 \\
 0 ~&~ 1 ~&~ 0 \\
 0 ~&~ 0 ~&~ 1 \\
\end{array}
\right)\,,
\end{eqnarray}
up to permutations and phases of the column vectors. $\tau_{\text{atm}}$ and $\tau_{\text{sol}}$ could be any fixed points in the upper half complex plane, so that the alignment of the modular triplets $Y_{\text{atm}}\left(\langle\tau_{\mathrm{atm}}\rangle\right)$ and $Y_{\text{sol}}\left(\langle\tau_{\mathrm{sol}}\rangle\right)$ could take the values listed in table~\ref{tab:tb-fp-S4p}. By considering all possible values of $\langle\tau_{\mathrm{atm}}\rangle$ and $\langle\tau_{\mathrm{sol}}\rangle$, we find four phenomenologically viable Littlest seesaw models denoted as case A, case B, case C and case D, which are listed in table~\ref{tab:4cases-S4p}. The first two cases A and B are for $\gamma_{\ell}=ST$, and the latter two cases C and D are for $\gamma_{\ell}=T$. We discuss the structure and the phenomenological predictions of these four models one by one in the following.

\begin{table}[t!]
\begin{center}
\resizebox{0.98\textwidth}{!}{
\begin{tabular}{|c|c|c|c|c|c|}\hline\hline
Case & $\gamma_{\ell}$ & $\langle\tau_{\mathrm{atm}}\rangle$ & $\langle\tau_{\mathrm{sol}}\rangle$ & $Y_{\text{atm}}\left(\langle\tau_{\mathrm{atm}}\rangle\right)$ & $Y_{\text{sol}}\left(\langle\tau_{\mathrm{sol}}\rangle\right)$ \\
\hline
A & $ST$ & $T^2\tau_S=2+i$ & $\tau_S=i$ & $\begin{matrix} & \\Y^{(3)}_{\widehat{\bm{3}}^\prime}\left(\langle\tau_{\mathrm{atm}}\rangle\right),& Y^{(4)}_{\bm{3}}\left(\langle\tau_{\mathrm{atm}}\rangle\right), \\Y^{(5)}_{\widehat{\bm{3}}}\left(\langle\tau_{\mathrm{atm}}\rangle\right), &Y^{(6)}_{\bm{3}^\prime}\left(\langle\tau_{\mathrm{atm}}\rangle\right)\\ & \end{matrix}$ & $Y^{(2)}_{\bm{3}}\left(\langle\tau_{\mathrm{sol}}\rangle\right)$  \\
\hline
B & $ST$& $T^2\tau_S=2+i$  & $T^2ST^2\tau_S=\frac{8}{5}+\frac{i}{5}$ & $\begin{matrix} & \\Y^{(3)}_{\widehat{\bm{3}}^\prime}\left(\langle\tau_{\mathrm{atm}}\rangle\right), & Y^{(4)}_{\bm{3}}\left(\langle\tau_{\mathrm{atm}}\rangle\right),\\ Y^{(5)}_{\widehat{\bm{3}}}\left(\langle\tau_{\mathrm{atm}}\rangle\right), & Y^{(6)}_{\bm{3}^\prime}\left(\langle\tau_{\mathrm{atm}}\rangle\right) \\ & \end{matrix}$ &  $Y^{(2)}_{\bm{3}}\left(\langle\tau_{\mathrm{sol}}\rangle\right)$  \\
\hline
\multirow{2}{*}[-5ex]{C} & \multirow{2}{*}[-5ex]{$T$}& \multirow{2}{*}[-5ex]{$TS^3T^2\tau_S=\frac{3}{5}+\frac{i}{5}$} & $T^2S\tau_T=2$& \multirow{2}{*}[-5ex]{$\begin{matrix}Y^{(2)}_{\bm{3}}\left(\langle\tau_{\mathrm{atm}}\rangle\right) \end{matrix}$} & $
\begin{matrix} & \\Y^{(3)}_{\widehat{\bm{3}}}\left(\langle\tau_{\mathrm{sol}}\rangle\right), & Y^{(5)}_{\widehat{\bm{3}}}\left(\langle\tau_{\mathrm{sol}}\rangle\right)\\ &  \end{matrix}$  \\
\cline{4-4}\cline{6-6}
 &   & &$S\tau_T=0$ & &$\begin{matrix} & \\Y^{(1)}_{\widehat{\bm{3}}^\prime}\left(\langle\tau_{\mathrm{sol}}\rangle\right), & Y^{(3)}_{\widehat{\bm{3}}^\prime}\left(\langle\tau_{\mathrm{sol}}\rangle\right),\\ Y^{(5)}_{\widehat{\bm{3}}^\prime I}\left(\langle\tau_{\mathrm{sol}}\rangle\right), & Y^{(5)}_{\widehat{\bm{3}}^{\prime} II}\left(\langle\tau_{\mathrm{sol}}\rangle\right)\\ & \end{matrix}$  \\
\hline
\multirow{2}{*}[-5ex]{D}  & \multirow{2}{*}[-5ex]{$T$}& \multirow{2}{*}[-5ex]{$T\tau_S=1+i$} & $T^2S\tau_T=2$  & \multirow{2}{*}[-5ex]{$\begin{matrix}Y^{(2)}_{\bm{3}}\left(\langle\tau_{\mathrm{atm}}\rangle\right) \end{matrix}$} & $\begin{matrix}& \\ Y^{(3)}_{\widehat{\bm{3}}}\left(\langle\tau_{\mathrm{sol}}\rangle\right), & Y^{(5)}_{\widehat{\bm{3}}}\left(\langle\tau_{\mathrm{sol}}\rangle\right)\\ &  \end{matrix}$\\
\cline{4-4}\cline{6-6}
 & & &$S\tau_T=0$& &$\begin{matrix}& \\Y^{(1)}_{\widehat{\bm{3}}^\prime}\left(\langle\tau_{\mathrm{sol}}\rangle\right), & Y^{(3)}_{\widehat{\bm{3}}^\prime}\left(\langle\tau_{\mathrm{sol}}\rangle\right),\\ Y^{(5)}_{\widehat{\bm{3}}^\prime I}\left(\langle\tau_{\mathrm{sol}}\rangle\right), & Y^{(5)}_{\widehat{\bm{3}}^{\prime}II}\left(\langle\tau_{\mathrm{sol}}\rangle\right)\\ & \end{matrix}$    \\
\hline \hline
\end{tabular}}
\caption{\label{tab:4cases-S4p} The stabilizer $\gamma_{\ell}$, the VEVs of the atmospheric modulus $\langle\tau_{\mathrm{atm}}\rangle$ and the solar modulus $\langle\tau_{\mathrm{sol}}\rangle$ as well the assignments of the modular forms $Y_{\text{atm}}$ and $Y_{\text{sol}}$ for the four phenomenologically viable modular Littlest Seesaw based on $S'_4$. }
\end{center}
\end{table}

\subsubsection{$\gamma_{\ell}=ST$ }

For both case A and case B, the three generations of left-handed lepton doublets can be assigned to be a triplet $\bm{3}$ of $S'_4$, $N^c_{\mathrm{atm}}$ and $N^c_{\mathrm{sol}}$ are $S'_4$ singlets $\bm{1}$ and $\bm{1'}$ respectively. The representation and modular weight assignments of the lepton fields are given by\footnote{The modular transformations of lepton fields can also be chosen as $L\sim(\bm{3}, 0)\,,~ N^c_{\mathrm{atm}}\sim(\widehat{\bm{1}}, 3)\,,~ N^c_{\mathrm{sol}}\sim(\bm{1}, 2)$ or $L\sim(\bm{3}, 2)\,,~ N^c_{\mathrm{atm}}\sim(\bm{\widehat{1}'}, 3)\,,~ N^c_{\mathrm{sol}}\sim(\bm{1}, 0)$ or $L\sim(\bm{3}, 2)\,,~ N^c_{\mathrm{atm}}\sim(\bm{1}, 2)\,,~ N^c_{\mathrm{sol}}\sim(\bm{1}, 0)$.}
\begin{eqnarray}
L\sim(\bm{3}, 2)\,,\quad N^c_{\mathrm{atm}}\sim(\bm{1'}, 4)\,,\quad N^c_{\mathrm{sol}}\sim(\bm{1}, 0)\,,
\end{eqnarray}
where the first number in the parentheses refers to the representation under $S'_4$ and the second number stands for the modular weight. Modular invariance requires that the cross term $N^c_{\text{atm}}N^{c}_{\text{sol}}$ couples with a single modular form $Y^{(4)}_{\bm{1}'}(\tau)$ which is
is absent, as can be seen from Appendix~\ref{app:N4-group-MF}. The three generations of right-handed charged leptons may be assigned to singlet representations of $S'_4$, as in conventional constructions, and can simultaneously carry different modular weights. For instance,
\begin{eqnarray}
e^{c}\sim(\bm{1}, 0),~~\mu^{c}\sim(\bm{1'}, 2),~~\tau^{c}\sim(\widehat{\bm{1}}, -1)\,,
\end{eqnarray}
and the down-type Higgs $H_d$ is an invariant singlet of $S'_4$ with zero modular weight. Then one can read out the modular invariant superpotential for the charged lepton mass
\begin{eqnarray}
\mathcal{W}_{\ell}=y_{e} e^c \left(LY^{(2)}_{\bm{3}}\right)_{\bm{1}}
H_d+y_{\mu} \mu^c \left(LY^{(4)}_{\bm{3'}}\right)_{\bm{1}}
H_d+y_{\tau} \tau^c \left(LY^{(1)}_{\widehat{\bm{3}}^\prime}\right)_{\widehat{\bm{1}}'}
H_d\,.
\end{eqnarray}
Note that the modular transformations of right-handed charged leptons are not unique. The VEV of $\tau_{\ell}$ invariant under $\gamma_{\ell}=ST$ is $\langle\tau_{\ell}\rangle=\omega$, thus the modular symmetry is broken down to the $Z^{ST}_3$ subgroup in the charged lepton sector. The charged lepton mass matrix is diagonalized by the unitary matrix $U_{\ell}$ in Eq.~\eqref{eq:Uell-ST} with the choice  $P_{\ell}=\mathbb{1}_3$. For case A and case B, the VEVs of the atmospheric and solar moduli are $\left(\langle\tau_{\text{atm}}\rangle, \langle\tau_{\text{sol}}\rangle\right)=(2+i,i )$ and $(2+i, 8/5+i/5)$ respectively. From table~\ref{tab:4cases-S4p} we can read off the alignments of $Y_{\text{atm}}$ and $Y_{\text{sol}}$ as follows~\footnote{The case B can also be obtained from the alignment of case A in Eq.~\eqref{eq:Yatm-Ysol-caseA} by choosing the permutation $P_{\ell}=P_{132}$. },
\begin{eqnarray}
\label{eq:Yatm-Ysol-caseA}\text{case A}: && Y_{\text{atm}}=Y^{(6)}_{\bm{3}'}\left(\langle\tau_{\text{atm}}\rangle=2+i\right)\propto\begin{pmatrix}1\\-\frac{1}{\sqrt{2}}\\-\frac{1}{\sqrt{2}} \end{pmatrix},~~ Y_{\text{sol}}=Y^{(2)}_{\bm{3}}\left(\langle\tau_{\text{sol}}\rangle=i\right)\propto\begin{pmatrix} 1\\
-\frac{1}{2}-\frac{1}{\sqrt{2}} \\
\frac{1}{2}-\frac{1}{\sqrt{2}}
\end{pmatrix}\,, \\
\text{case B}: && Y_{\text{atm}}=Y^{(6)}_{\bm{3}'}\left(\langle\tau_{\text{atm}}\rangle=2+i\right)\propto\begin{pmatrix}1\\-\frac{1}{\sqrt{2}}\\-\frac{1}{\sqrt{2}} \end{pmatrix},~~ Y_{\text{sol}}=Y^{(2)}_{\bm{3}}\left(\langle\tau_{\text{sol}}\rangle=8/5+i/5\right)\propto\begin{pmatrix} 1\\
\frac{1}{2}-\frac{1}{\sqrt{2}}\\
-\frac{1}{2} -\frac{1}{\sqrt{2}}
\end{pmatrix} \,.~~~~
\end{eqnarray}
It is convenient to switch to the charged lepton diagonal basis through the similarity transformation $V= \dfrac{1}{2\sqrt{3}}
\begin{pmatrix}
2\,i ~&~ 2\,e^{\frac{3\pi i}{4}} ~&~ 2\,e^{\frac{\pi i}{4}}\\
2\,e^{-\frac{\pi i}{6}} ~&~ (\sqrt{3}+1)\,e^{-\frac{11\pi i}{12}} ~&~ (\sqrt{3}-1)\,e^{-\frac{5\pi i}{12}}\\
2\,e^{-\frac{5\pi i}{6}} ~&~ (\sqrt{3}-1)\,e^{-\frac{7\pi i}{12}} ~&~ (\sqrt{3}+1)\,e^{-\frac{\pi i}{12}}
\end{pmatrix}$, then $U_{\ell}$ is a diagonal unitary matrix and the lepton mixing completely arises from the neutrino sector. The alignments of $Y_{\text{atm}}$ and $Y_{\text{sol}}$ in such basis would be of the following form,
\begin{eqnarray}
\text{case A}: && Y_{\text{atm}}\left(\langle\tau_{\text{atm}}\rangle=2+i\right)\propto\begin{pmatrix}0\\1\\-1\end{pmatrix},\quad Y_{\text{sol}}\left(\langle\tau_{\text{sol}}\rangle=i\right)\propto\begin{pmatrix} 1\\1+\sqrt{6}\\1-\sqrt{6}\end{pmatrix}\,, \\
\text{case B}: && Y_{\text{atm}}\left(\langle\tau_{\text{atm}}\rangle=2+i\right)\propto\begin{pmatrix}0\\1\\-1\end{pmatrix},\quad Y_{\text{sol}}\left(\langle\tau_{\text{sol}}\rangle=8/5+i/5\right)\propto\begin{pmatrix} 1\\1-\sqrt{6}\\1+\sqrt{6}\end{pmatrix} \,.
\end{eqnarray}
Notice that the above alignments can also be achieved from the $S_4$ modular group, and they are known as original modular Littlest seesaw and flipped modular Littlest seesaw~\cite{Ding:2019gof,Ding:2023htn}. Accordingly the light neutrino mass matrix is given by
\begin{eqnarray}
\label{eq:mnu-caseA}\text{case A}:\,\,\,\,M_{\nu}=m_a\begin{pmatrix}
0 ~&~ 0 ~&~ 0\\
0 ~&~ 1 ~&~ -1 \\
0 ~&~ -1 ~&~ 1
\end{pmatrix}
+m_se^{i\eta}\begin{pmatrix} 1 ~&~ 1-\sqrt{6} ~&~ 1+\sqrt{6}\\
1-\sqrt{6} ~&~ 7-2\sqrt{6} ~&~ -5\\
1+\sqrt{6} ~~&~~ -5 ~~&~~ 7+2\sqrt{6} \end{pmatrix}\,,\\
\label{eq:mnu-caseB}\text{case B}:\,\,\,\,M_{\nu}=m_a\begin{pmatrix}
0 ~&~ 0 ~&~ 0\\
0 ~&~ 1 ~&~ -1 \\
0 ~&~ -1 ~&~ 1 \end{pmatrix}
+m_se^{i\eta}\begin{pmatrix}
1 ~&~ 1+\sqrt{6} ~&~ 1-\sqrt{6}\\
1+\sqrt{6} ~~&~~ 7+2\sqrt{6} ~~&~~ -5\\
1-\sqrt{6} ~&~ -5 ~&~ 7-2\sqrt{6}
\end{pmatrix}\,.
\end{eqnarray}
The light neutrino mass matrices in the two cases are related through the exchange of the second and the third rows and columns. It is notable that the modular Littlest seesaw model of case B can also be obtained from the alignment of case A by choosing another charged lepton permutation $P_{\ell}=P_{132}$. We see that the models are very predictive, all the neutrino masses and mixing angles and CP violation phases are determined by three parameters $m_a$, $m_s$ and $\eta$. From Eq.~\eqref{eq:fixed-column}, we know that one column of the lepton mixing matrix is $\frac{1}{\sqrt{6}}(2, -1,-1)^T$ corresponding to the massless light neutrino. Consequently the lepton mixing is predicted to be of the following form
\begin{eqnarray}
\label{eq:PMNS-caseA}U_{PMNS}=\begin{pmatrix}
\sqrt{\frac{2}{3}} ~&~ \times ~&~ \times \\
-\frac{1}{\sqrt{6}} ~&~ \times ~&~ \times \\
-\frac{1}{\sqrt{6}} ~&~ \times ~&~ \times \\
\end{pmatrix}\,,
\end{eqnarray}
where the second and third columns depend on the ratio $r\equiv m_s/m_a$ and the phase $\eta$, and they are orthogonal to first column vector $(\sqrt{\frac{2}{3}}, -\frac{1}{\sqrt{6}}, -\frac{1}{\sqrt{6}})^T$. Comparing the lepton mixing matrix in Eq.~\eqref{eq:PMNS-caseA} with the standard parametrization of Eq.~\eqref{eq:PMNS_def}, we find that the lepton mixing angles and CP violation phases are strongly correlated with each other as follow,
\begin{eqnarray}
\nonumber&& \quad 3 \cos^{2}\theta_{12}\cos^{2}\theta_{13}= 2\,, \\
\label{eq:cos-deltaCP-sumR-AB}&&\cos\delta_{CP}=-\frac{\cot2\theta_{23} (1-5\sin^{2}\theta_{13})} {2\sin\theta_{13}\sqrt{2-6\sin^{2}\theta_{13}}} \,.
\end{eqnarray}
As a leading order sequential dominance approximation if we only keep the $m_a$ term and neglect $m_s$ term in the light neutrino mass matrix $M_{\nu}$ in Eqs.~(\ref{eq:mnu-caseA}, \ref{eq:mnu-caseB}), then two neutrinos would be massless and a third neutrino would have mass $\sqrt{2}m_a$. Accordingly the third column of the lepton mixing matrix would be $(0, 1/\sqrt{2}, -1/\sqrt{2})^T$ which would give $\theta_{13}=0$ and $\theta_{23}=45^{\circ}$. Including the subleading term proportional to $m_s$ which is generated by integrating out the right-handed neutrino $N^c_{\text{sol}}$, the second neutrino mass, the solar mixing angle $\theta_{12}$ and non-zero reactor angle $\theta_{13}$ are produced.

\begin{figure}[t!]
\begin{center}
\includegraphics[width=0.99\textwidth]{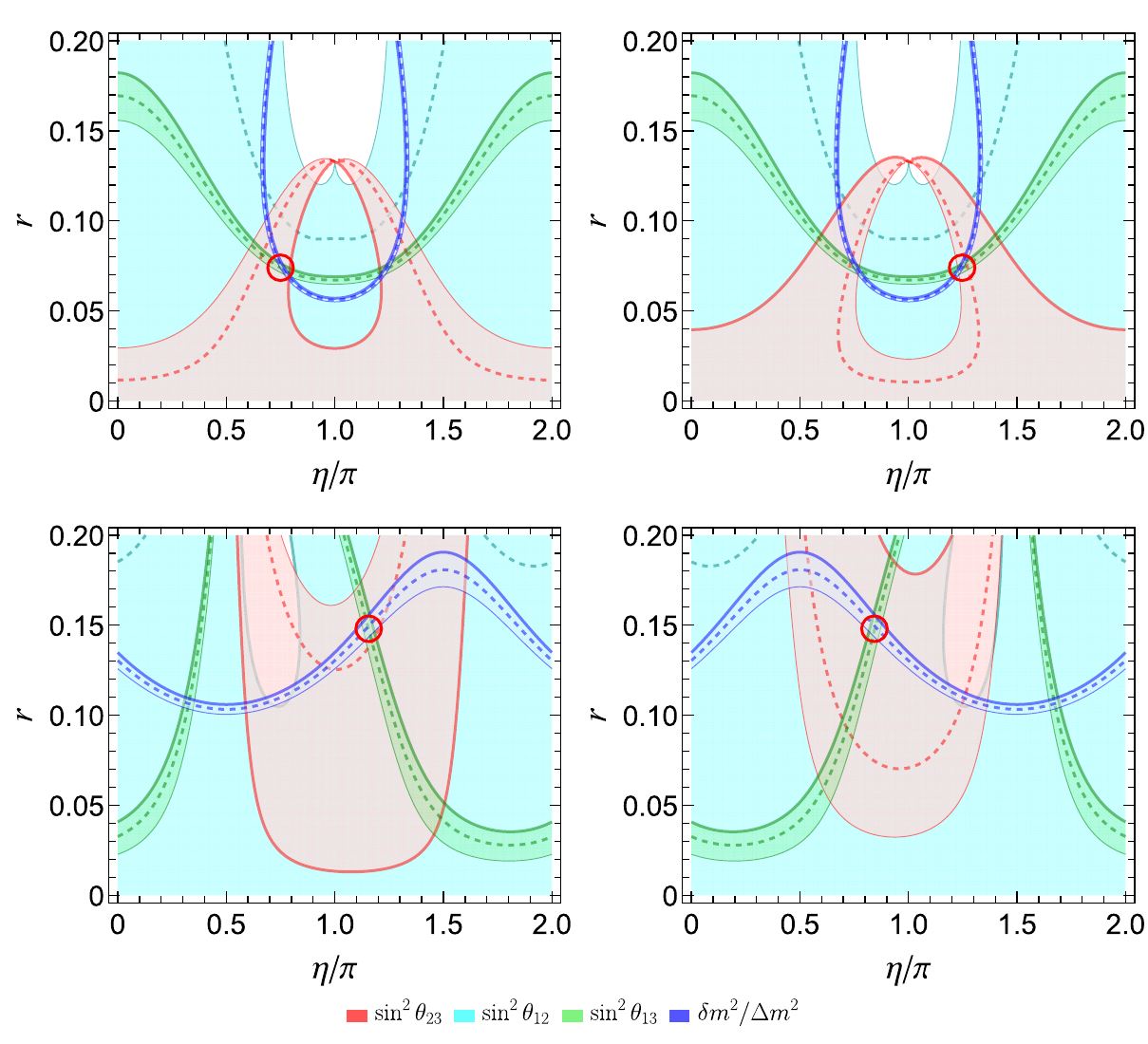}
\end{center}
\caption{\label{fig:mixing-pars-S4pLSS} The contour plots of $\sin^2\theta_{12}$, $\sin^2\theta_{13}$, $\sin^{2}\theta_{23}$ and $\delta m^2/\Delta m^2$ in the plane of $r\equiv m_s/m_a$ versus $\eta$ for the four possible modular Littlest seesaw models in $S'_4$ modular symmetry. The panels in the upper-left, upper-right, lower- left, lower-right are for case A, case B, case C and case D respectively.  The cyan, red, green and blue areas denote the $3\sigma$ regions of $\sin^{2}\theta_{23}$, $\sin^{2}\theta_{13}$ and $\delta m^2/\Delta m^2$ respectively. The solid lines denote the $3\sigma$ upper bounds, the thin lines denote the $3\sigma$ lower bounds and the dashed lines refer to their best fit values, as adopted from latest neutrino global fit~\cite{Capozzi:2025wyn}. The red circle indicates the phenomenologically viable region of parameter space. }
\end{figure}

We show the contour plots of $\sin^2\theta_{12}$, $\sin^2\theta_{13}$, $\sin^2\theta_{23}$ and $\delta m^2/\Delta m^2$ in the plane of $r$ with respect to $\eta$ in figure~\ref{fig:mixing-pars-S4pLSS}, where $\delta m^2\equiv m^2_2-m^2_1$ and $\Delta m^2=m^2_3-\frac{1}{2}(m^2_1+m^2_2)$ are the two squared mass gaps measured by neutrino oscillations. We would like to remind that the lightest neutrino is massless with $m_1=0$ in our models.  One can see that the experimental data can be well accommodated by these Littlest seesaw models in a small parameter space which is dominantly determined by the precisely measured reactor angle $\sin^2\theta_{13}$ and $\delta m^2/\Delta m^2$. Furthermore, we perform a conventional $\chi^2$ analysis~\footnote{The contribution of $\sin^2\theta_{12}$, $\sin^2\theta_{13}$, $\sin^2\theta_{23}$, $\delta_{CP}$, $\delta m^2$ and $\Delta m^2$ are included in the $\chi^2$ analysis, their central values and $1\sigma$ uncertainties are adopted from~\cite{Capozzi:2025wyn} when constructing the $\chi^2$ function.}, the best fit values of $m_a$, $m_s$ and $\eta$ and the corresponding predictions for the light neutrino masses and mixing parameters are listed in table~\ref{tab:BF-results-S4p}. Here $m_{\beta\beta}$ denotes the effective neutrino mass in neutrinoless double decay, i.e.
\begin{eqnarray}
m_{\beta\beta}=\left|m_{1}\cos^{2}\theta_{12}\cos^{2}\theta_{13}e^{i\alpha}+m_{2}\sin^{2}\theta_{12}\cos^{2}\theta_{13}e^{i\beta}+m_{3}\sin^{2}\theta_{13}e^{-2i\delta_{CP}}\right|\,.
\end{eqnarray}
We have $m_1=0$ in Littlest seesaw model, consequently the expression of $m_{\beta\beta}$ simplifies into
\begin{eqnarray}
m_{\beta\beta}=\left|m_{2}\sin^{2}\theta_{12}\cos^{2}\theta_{13}+m_{3}\sin^{2}\theta_{13}e^{-i(\beta+2\delta_{CP})}\right|\,.
\end{eqnarray}
It is notable that $m_{\beta\beta}$ is predicted to be around 2.3 meV, which is below the current experimental bound~\cite{KamLAND-Zen:2024eml} as well as the sensitivities of future ton-scale neutrinoless double beta decay experiments~\cite{Agostini:2022zub,Cirigliano:2022oqy}. Hence our model would be ruled out if the signal of neutrinoless double decay is observed in future.

\begin{table}[t!]
\centering
\resizebox{1.00\textwidth}{!}{
\begin{tabular}{|c|c|c|c|c|c|c|c|c|c|c|c|c|c|c|} \hline \hline
Case & $\chi^2_{\text{min}}$ & $m_{a}(\text{meV})$ & $m_s/m_a$  & $\eta/\pi$  &$\sin^2\theta_{12}$ &  $\sin^2\theta_{13}$ & $\sin^2\theta_{23}$  & $\delta_{CP}/\pi$ &  $\beta/\pi$ & $m_2(\text{meV})$ & $m_3(\text{meV})$ & $m_{\beta\beta}(\text{meV})$ \\
\hline
A & $16.033$ & $30.991$ & $0.074$ & $0.750$ & $0.318$ & $0.02260$ & $0.544$ & $1.561$ & $1.544$ & $8.520$ & $50.362$ & $2.308$\\
\hline
B & $5.639$ & $31.092$ & $0.074$ & $1.247$ & $0.318$ & $0.02250$ & $0.454$ & $1.436$ & $0.461$ & $8.542$ & $50.347$ & $2.305$\\
\hline
C & $7.602$ & $26.626$ & $0.148$ & $1.156$ & $0.333$ & $0.02232$ & $0.465$ & $1.481$ & $1.454$ & $8.578$ & $50.322$ & $3.267$\\
\hline
D & $15.244$ & $26.663$ & $0.148$ & $0.843$ & $0.333$ & $0.02233$ & $0.535$ & $1.518$ & $0.547$ & $8.561$ & $50.334$ & $3.260$\\ \hline \hline
\end{tabular}}
\caption{\label{tab:BF-results-S4p}Results of the $\chi^2$ analysis for the modular Littlest seesaw based on the $S'_4$ modular group in the charged lepton diagonal basis. Note that the lightest neutrino mass is zero with $m_1=0$. }
\end{table}

\subsubsection{$\gamma_{\ell}=T$ }

For both case C and case D, the modular transformations of the lepton fields $L$, $N^c_{\mathrm{atm}}$ and $N^c_{\mathrm{sol}}$ can be chosen as\footnote{We can also assign $L\sim(\bm{3}, 2)\,,~ N^c_{\mathrm{atm}}\sim(\bm{1}, 0)\,,~ N^c_{\mathrm{sol}}\sim(\bm{\widehat{1}'}, 3)$ or $L\sim(\bm{\widehat{3}}, -1)\,,~ N^c_{\mathrm{atm}}\sim(\bm{\widehat{1}'}, 3)\,,~ N^c_{\mathrm{sol}}\sim(\bm{1}, 2)$ or $L\sim(\bm{3}, 0)\,,~ N^c_{\mathrm{atm}}\sim(\bm{1}, 2)\,,~ N^c_{\mathrm{sol}}\sim(\bm{\widehat{1}}, 3)$. They can give rise to the same modular Littlest seesaw model, as can be seen from table~\ref{tab:4cases-S4p}.}
\begin{eqnarray}
L\sim(\bm{3}, 0)\,,\quad N^c_{\mathrm{atm}}\sim(\bm{1}, 2)\,,\quad N^c_{\mathrm{sol}}\sim(\bm{\widehat{1}'}, 3)\,.
\end{eqnarray}
The term $N^{c}_{\text{atm}}N^c_{\text{sol}}Y^{(5)}_{\widehat{\bm{1}}}(\tau)$ is forbidden by modular symmetry for the above assignment because there doesn't exist the modular form singlet $Y^{(5)}_{\widehat{\bm{1}}}(\tau)$. The atmospheric modulus $\tau_{\text{atm}}$ and solar modulus $\tau_{\text{sol}}$ are assumed to be stabilized at the modular fixed points $\left(\langle\tau_{\text{atm}}\rangle, \langle\tau_{\text{sol}}\rangle\right)=\left(3/5+i/5, 2\right)$ or $(3/5+i/5, 0)$ for case C and $\left(\langle\tau_{\text{atm}}\rangle, \langle\tau_{\text{sol}}\rangle\right)=\left(1+i, 2\right)$ or $(1+i, 0)$ for case D. As shown in table~\ref{tab:tb-fp-S4p}, the modular forms $Y_{\text{atm}}$ and $Y_{\text{sol}}$ at these fixed points are along the following directions,
\begin{eqnarray}
\text{case C}: && Y_{\text{atm}}=Y^{(2)}_{\bm{3}}\left(\langle\tau_{\text{atm}}\rangle=3/5+i/5\right)\propto\begin{pmatrix}1\\ i\left(-\frac{1}{2}+\frac{1}{\sqrt{2}}\right)\\  i\left(-\frac{1}{2}-\frac{1}{\sqrt{2}}\right)\end{pmatrix}\,,~Y_{\text{sol}}=Y^{(3)}_{\widehat{\bm{3}}}\left(\langle\tau_{\text{sol}}\rangle=2\right)\propto\begin{pmatrix}1\\\frac{1}{\sqrt{2}}\\-\frac{1}{\sqrt{2}} \end{pmatrix}\,, ~~~~\\
\text{case D}: && Y_{\text{atm}}=Y^{(2)}_{\bm{3}}\left(\langle\tau_{\text{atm}}\rangle=1+i\right)\propto\begin{pmatrix}1\\i\left(-\frac{1}{2}-\frac{1}{\sqrt{2}}\right)\\ i\left(-\frac{1}{2}+\frac{1}{\sqrt{2}}\right) \end{pmatrix}\,,~ Y_{\text{sol}}=Y^{(3)}_{\widehat{\bm{3}}}\left(\langle\tau_{\text{sol}}\rangle=2\right)\propto\begin{pmatrix}1\\\frac{1}{\sqrt{2}}\\-\frac{1}{\sqrt{2}} \end{pmatrix}\,.
\end{eqnarray}
In this scenario, the VEV of $\tau_{\ell}$ is $\langle\tau_{\ell}\rangle=i\infty$ and a residual symmetry $Z^T_4$ is preserved by the charged lepton Yukawa couplings. As a consequence, the unitary transformation $U_{\ell}$ is a permutation matrix in the representation basis in Appendix~\ref{app:N4-group-MF} since the residual symmetry doesn't allow to predict charged lepton masses. It turns out that the modular Littlest seesaw models of case C and case D can be compatible with experimental data, if one takes the permutation matrices $P_{\ell}=P_{321}$ and $P_{\ell}=P_{213}$ for the case C and case D respectively. Performing basis transformation by the similarity transformation $V=\begin{pmatrix} 0~&~0~&~1\\0~&~1~&~0\\1~&~0~&~0\end{pmatrix}$ for case C and  $V=\begin{pmatrix}0~&~1~&~0\\1~&~0~&~0\\0~&~0~&~1 \end{pmatrix}$ for case D, we can go to the flavor basis in which $U_{\ell}$ is the unit matrix exactly, and then the alignments of $Y_{\text{atm}}$ and $Y_{\text{sol}}$ in the flavor basis are given by
\begin{eqnarray}
\text{case C}: && Y_{\text{atm}}\left(\langle\tau_{\text{atm}}\rangle=3/5+i/5\right)\propto\begin{pmatrix}1\\2\sqrt{2}-3\\ (2\sqrt{2}-2)i \end{pmatrix}\,,
\quad Y_{\text{sol}}\left(\langle\tau_{\text{sol}}\rangle=2\right)\propto\begin{pmatrix}1\\-1\\-\sqrt{2} \end{pmatrix}\,, \\
\text{case D}: && Y_{\text{atm}}\left(\langle\tau_{\text{atm}}\rangle=1+i\right)\propto\begin{pmatrix}1 \\ (2\sqrt{2}-2)i\\ 2\sqrt{2}-3 \end{pmatrix}\,,
\quad Y_{\text{sol}}\left(\langle\tau_{\text{sol}}\rangle=2\right)\propto\begin{pmatrix}1\\\sqrt{2}\\-1 \end{pmatrix}\,.
\end{eqnarray}
The Clebsch–Gordan matrices corresponding to the contraction of two triplets into a singlet are given by $P_{213}$ for case C  and $P_{321}$ for case D respectively. The light neutrino mass matrix in the flavor basis is of the following form
\begin{eqnarray}
\text{case C}:~ M_{\nu}&=&m_a
\begin{pmatrix}
17-12\sqrt{2} ~& 2\sqrt{2}-3 ~&  (14-10\sqrt{2})\,i\\
2\sqrt{2}-3 ~& 1 ~& (2\sqrt{2}-2)\,i\\
(14-10\sqrt{2})\,i ~&~ (2\sqrt{2}-2)\,i ~& 8\sqrt{2}-12
\end{pmatrix}+m_se^{i\eta}
\begin{pmatrix}
1~&-1 ~& \sqrt{2}\\
-1 ~& 1 ~& -\sqrt{2} \\
\sqrt{2} ~& -\sqrt{2} ~& 2 \end{pmatrix}\,,\\
\text{case D}:~ M_{\nu}&=&m_a
\begin{pmatrix}
17-12\sqrt{2} ~& (14-10\sqrt{2})\,i ~& 2\sqrt{2}-3\\
(14-10\sqrt{2})\,i ~& 8\sqrt{2}-12 ~& (2\sqrt{2}-2)\,i\\
2\sqrt{2}-3 ~& (2\sqrt{2}-2)\,i ~& 1
\end{pmatrix}+m_se^{i\eta}
\begin{pmatrix}
1~&-\sqrt{2} ~& -1\\
-\sqrt{2} ~& 2 ~& \sqrt{2} \\
-1 ~& \sqrt{2} ~& 1 \end{pmatrix}\,.
\end{eqnarray}
We see that the neutrino mass matrix $M_{\nu}$ in cases C and D are related each other through the exchange of the second and third rows and columns. In the limit of atmospheric neutrino $N^{c}_{\text{atm}}$ dominance, the third column of the lepton mixing matrix is $(-0.131, 0.763, -0.632i)^T$ in case C and $(-0.131, -0.632i, 0.763)^T$ in case D. Thus we have $(\theta_{13}, \theta_{23})\approx(7.527^{\circ}, 50.361^{\circ})$ and $(7.527^{\circ}, 39.639^{\circ})$ at leading order for cases C and D respectively. The contribution of $N^{c}_{\text{sol}}$ would generate the solar mixing angle $\theta_{12}$ and small corrections to the above leading order values of $\theta_{13}$ and $\theta_{23}$.

The contour plots of $\sin^2\theta_{12}$, $\sin^2\theta_{13}$, $\sin^{2}\theta_{23}$ and $\delta m^2/\Delta m^2$ in the $\eta-r$ plane are displayed in figure~\ref{fig:mixing-pars-S4pLSS}. It can be seen that the experimental data can be well accommodated. We collect the best fit values of $m_a$, $m_s$, $\eta$ and the predictions for the lepton mixing parameters in table~\ref{tab:BF-results-S4p}. The effective Majorana neutrino mass $m_{\beta\beta}$ is predicted to be around 3.3 meV which is out of the reach of future neutrinoless double decay experiments. In case C, the lepton mixing matrix is of the following form,
\begin{eqnarray}
\label{eq:PMNS-caseC} U_{PMNS}=\begin{pmatrix}\frac{1}{2}\sqrt{\frac{1}{3}(5+2\sqrt{2})} ~&~ \times ~&~ \times \\
\frac{1}{2}\sqrt{\frac{1}{3}(5-2\sqrt{2})} ~&~ \times ~&~ \times \\
\frac{1}{\sqrt{6}} ~&~ \times ~&~ \times \end{pmatrix}
\simeq\begin{pmatrix}0.808 ~&~ \times ~&~ \times\\ 0.425 ~&~ \times ~&~ \times\\ 0.408 ~&~ \times ~&~ \times \end{pmatrix}
\end{eqnarray}
Therefore we can reach the sum rules between the lepton mixing angles and Dirac CP phase $\delta_{CP}$ as follow,
\begin{eqnarray}
\nonumber &&\qquad \qquad\qquad \qquad  12\cos^2\theta_{12}\cos^2\theta_{13}=5+2\sqrt{2}\,,\\[0.1in]
&& \cos\delta_{CP}=\dfrac{\left(6-4\sqrt{2}\right)\cos^2\theta_{13}+\left[3+6\sqrt{2}-\left(17+2\sqrt{2}\right)\cos 2\theta_{13}\right]\cos 2\theta_{23}}{4\sin\theta_{13}\sin  2\theta_{23}\sqrt{\left(5+2\sqrt{2}\right)\left(1-2\sqrt{2}+6\cos 2\theta_{13}\right)}}\,.
\end{eqnarray}
In case D, the lepton mixing matrix is of the following form,
\begin{eqnarray}
\label{eq:PMNS-caseD} U_{PMNS}=\begin{pmatrix}\frac{1}{2}\sqrt{\frac{1}{3}(5+2\sqrt{2})} ~&~ \times ~&~ \times \\
\frac{1}{\sqrt{6}} ~&~ \times ~&~ \times \\
\frac{1}{2}\sqrt{\frac{1}{3}(5-2\sqrt{2})} ~&~ \times ~&~ \times
 \end{pmatrix}
\simeq\begin{pmatrix}0.808 ~&~ \times ~&~ \times\\ 0.408 ~&~ \times ~&~ \times\\ 0.425 ~&~ \times ~&~ \times \end{pmatrix}\,.
\end{eqnarray}
The corresponding sum rule is found to be
\begin{eqnarray}
\nonumber &&\qquad \qquad\qquad \qquad  12\cos^2\theta_{12}\cos^2\theta_{13}=5+2\sqrt{2}\,,\\[0.1in]
&& \cos\delta_{CP}=\dfrac{\left(4\sqrt{2}-6\right)\cos^2\theta_{13}+\left[3+6\sqrt{2}-\left(17+2\sqrt{2}\right)\cos 2\theta_{13}\right]\cos 2\theta_{23}}{4\sin\theta_{13}\sin  2\theta_{23}\sqrt{\left(5+2\sqrt{2}\right)\left(1-2\sqrt{2}+6\cos 2\theta_{13}\right)}}\,.
\end{eqnarray}
Using for $\sin^2\theta_{13}$ its global best fit value $(\sin^2\theta_{13})^{\text{bf}}=0.0223$~\cite{Capozzi:2025wyn}, we find for the solar mixing angle $\sin^2\theta_{12}\approx0.333$ which is compatible with the numerical results in table~\ref{tab:BF-results-S4p}. It is notable that the prediction for $\sin^2\theta_{12}$ is different from that of case A and case B. The JUNO collaboration is expected to perform very high precision measurement of $\sin^2\theta_{12}$ which will be determined to better than $0.5\%$ precision in six years of data collection~\cite{JUNO:2022mxj}. Therefore JUNO can test the above four Littlest seesaw models in $S'_4$.

\section{\label{sec:MLS-N5}Modular Littlest seesaw models based on finite group $A_5$ and $A'_5$ }

The finite modular group $\Gamma'_5\cong A'_5$ which is the double covering of the icosahedral group $A_5$. It has one singlet representation $\bm{1}$, two double representations $\bm{\widehat{2}}$ and $\bm{\widehat{2}'}$, two triplet representations $\bm{3}$ and $\bm{3'}$, two quartet representations $\bm{4}$ and $\widehat{\bm{4}}$, one quintuplet representation $\bm{5}$, and one sextet representation $\bm{\widehat{6}}$. The representation matrices of $S$ and $T$ in these irreducible representations are collected in Eq.~\eqref{eq:Irr-matrix-N5}. The modular forms of integer weight $k$ at level $N=5$ span a linear space of dimension $5k+1$, and each modular form can be written as a polynomial of degree $5k$ in $F_1(\tau)$ and $F_2(\tau)$~\cite{Yao:2020zml}, where $F_1(\tau)$ and $F_2(\tau)$ are weight $1/5$ modular forms and they can be expressed in terms of Dedekind eta-function $\eta(\tau)$ and
the Klein form $\mathfrak{k}_{r_1,r_2}(\tau)$ as follows,
\begin{eqnarray}
F_1(\tau)=\dfrac{\eta^3(5\tau)}{\eta^{\frac{3}{5}}(\tau)}\mathfrak{k}_{\frac{2}{5},0}(5\tau),\qquad  F_2(\tau)=\dfrac{\eta^3(5\tau)}{\eta^{\frac{3}{5}}(\tau)}\mathfrak{k}_{\frac{1}{5},0}(5\tau)\,.
\end{eqnarray}
Here $\eta(\tau)$ and $\mathfrak{k}_{r_1,r_2}(\tau)$ are defined as
\begin{eqnarray}
\nonumber&\eta(\tau)=q^{\frac{1}{24}}\prod^{\infty}_{n=1}(1-q^n),\,\,\,\,q\equiv e^{2\pi i\tau}\,,\\[0.1in]
&\mathfrak{k}_{r_1,r_2}(\tau)=q_z^{(r_1-1)/2}(1-q_z)\prod^{\infty}_{n=1}(1-q^nq_z)(1-q^nq_z^{-1})(1-q^n)^{-2}\,,
\end{eqnarray}
with $q_z=e^{2\pi iz}$ and $z=r_1\tau+r_2$. As a consequence, the $q$-expansion of $F_1(\tau)$ and $F_2(\tau)$ reads as
\begin{eqnarray}
\nonumber F_1(\tau) & = & 1 +\frac{3}{5}q+\frac{2}{25}q^2-\frac{28}{125}q^3+\frac{264}{625}q^4+\frac{532}{15625}q^5+\frac{20916}{78125}q^6-\frac{88416}{390625}q^7+\cdots\,,\\
F_2(\tau) & = & q^{\frac{1}{5}}\left(1-\frac{2}{5}q+\frac{12}{25}q^2+\frac{37}{125}q^3-\frac{171}{625}q^4+\frac{3318}{15625}q^5+\frac{13756}{78125}q^6+\frac{95304}{390625}q^7+\cdots\right)
\end{eqnarray}
At the lowest integer weight $k=1$ and level $N=5$, there are six linearly independent modular forms including $F_1^5(\tau)$, $F^4_1(\tau)F_2(\tau)$, $F^3_1(\tau)F^2_2(\tau)$, $F^2_1(\tau)F^3_2(\tau)$, $F_1(\tau)F^4_2(\tau)$ and $F^5_2(\tau)$. They can be arranged into a sextet $Y^{(1)}_{\widehat{\bm{6}}}$ in the $\Gamma^\prime_5\cong A^\prime_5$ representation $\widehat{\bm{6}}$,
\begin{eqnarray}
Y^{(1)}_{\widehat{\bm{6}}}(\tau)=\begin{pmatrix}F_1^5+2F_2^5\\2F_1^5-F_2^5\\5F_1^4F_2\\5\sqrt{2}\,F_1^3F_2^2\\-5\sqrt{2}\,F_1^2F_2^3\\5F_1F_2^4 \end{pmatrix}\equiv\begin{pmatrix}Y_1(\tau)\\Y_2(\tau)\\Y_3(\tau)\\Y_4(\tau)\\Y_5(\tau)\\Y_6(\tau) \end{pmatrix}\label{eq-sixletsF1F2}
\end{eqnarray}
The $q$-expansion of the modular forms $Y_{1,,\,2,\,3,\,4,\,5,\,6}(\tau)$ reads as
\begin{eqnarray}
\nonumber&&Y_1(\tau)=1+5q+10q^3-5q^4+5q^5+10q^6+5q^9+\ldots\,, \\
\nonumber&&Y_2(\tau)=2+5 q+10 q^2+5 q^4+5 q^5+10 q^6+10q^7-5q^9+\ldots\,, \\
\nonumber&&Y_3(\tau)=5q^{1/5}\left(1+2q+2q^2+q^3+2q^4+2q^5+2q^6+q^7+2q^8+2q^9+\ldots\right)\,,\\
\nonumber&&Y_4(\tau)=5\sqrt{2}q^{2/5}\left(1+q+q^2+q^3+2q^4+q^6+q^7+2 q^8+q^9+\ldots\right)\,,\\
\nonumber&&Y_5(\tau)=-5\sqrt{2}q^{3/5}\left(1+q^2+q^3+q^4-q^5+2 q^6+q^8+q^9+\ldots\right)\,,\\
\label{eq:q_series_wt1}&&Y_6(\tau)=5q^{4/5}\left(1-q+2q^2+2q^6-2q^7+2q^8+q^9+\ldots\right)\,.
\end{eqnarray}
At the fixed points $\tau_S=i$, $\tau_{ST}=-\frac{1}{2}+i\frac{\sqrt{3}}{2}$, $\tau_{TS}=\frac{1}{2}+i\frac{\sqrt{3}}{2}$ and $\tau_T=i\infty$, the ratio $F_2(\tau)/F_1(\tau)$ is determined to be
\begin{eqnarray}
\nonumber \dfrac{F_2(\tau_S)}{F_1(\tau_S)} & = & \frac{1}{2}\left(\sqrt{10+2\sqrt{5}}-1-\sqrt{5}\right) \approx  0.284\,,\\
\nonumber\dfrac{F_2(\tau_{ST})}{F_1(\tau_{ST})} & = & \frac{1}{4}\left(\sqrt{30+6\sqrt{5}}-3-\sqrt{5}\right)e^{-\frac{\pi i}{5}} \approx 0.338\,e^{-\frac{\pi i}{5}} \,,\\
\nonumber\dfrac{F_2(\tau_{TS})}{F_1(\tau_{TS})} & = & \frac{1}{4}\left(\sqrt{30+6\sqrt{5}}-3-\sqrt{5}\right)e^{\frac{\pi i}{5}} \approx0.338\,e^{\frac{\pi i}{5}} \,,\\
\dfrac{F_2(\tau_T)}{F_1(\tau_T)} & = & 0\,.
\end{eqnarray}
Hence the modular form sextet $Y^{(1)}_{\widehat{\bm{6}}}(\tau)$ is aligned along the following directions at the fixed points,
\begin{eqnarray}
Y^{(1)}_{\widehat{\bm{6}}}(\tau_S) & \propto &
\begin{pmatrix}
1\\
\frac{1}{118}\left(99-5\sqrt{5}+3\sqrt{2650-110\sqrt{5}}\right)\\
\frac{1}{118}\left(-43+57\sqrt{5}+\sqrt{23050-7262\sqrt{5}}\right)\\
\frac{1}{59}\left(37\sqrt{2}-12\sqrt{10}+2\sqrt{425-149\sqrt{5}}\right)\\
-\frac{1}{118}\left(3\sqrt{2}+7\sqrt{10}+(31\sqrt{5}-71)\sqrt{10+4\sqrt{5}}\right)\\
\frac{1}{59} \left(18-17 \sqrt{5}+\sqrt{530-22 \sqrt{5}}\right)
\end{pmatrix} \approx \begin{pmatrix}1\\1.991 \\ 1.415 \\ 0.569\\ -0.162\\ 0.0324 \end{pmatrix}\,, \\
Y^{(1)}_{\widehat{\bm{6}}}(\tau_{ST}) & \propto &
\begin{pmatrix}
1\\
3-\sqrt{5}+\sqrt{15-6\sqrt{5}}\\
\frac{1}{4}\left(19-7\sqrt{5}+\sqrt{750-330\sqrt{5}}\right)\,e^{-\frac{\pi i}{5}}\\
\frac{1}{2\sqrt{2}}\left(17-7\sqrt{5}+\sqrt{390-174\sqrt{5}}\right)\,e^{-\frac{2\pi i}{5}}\\
\frac{1}{2\sqrt{2}}\left(11-5\sqrt{5}+\sqrt{390-174\sqrt{5}}\right)\,e^{\frac{2\pi i}{5}}\\
\frac{1}{4}\left(13-5\sqrt{5}-\sqrt{150-66\sqrt{5}}\right)\,e^{-\frac{4\pi i}{5}}
\end{pmatrix}
\approx \begin{pmatrix}1\\ 2.022\\ 1.706 \,e^{-\frac{\pi i}{5}}\\ 0.816\,e^{-\frac{2\pi i}{5}}\\ 0.276\,e^{\frac{2\pi i}{5}}\\ 0.066\,e^{-\frac{4\pi i}{5}} \end{pmatrix} \,,~~~\quad \\
Y^{(1)}_{\widehat{\bm{6}}}(\tau_{TS}) & \propto &\begin{pmatrix}
1\\
3-\sqrt{5}+\sqrt{15-6\sqrt{5}}\\
\frac{1}{4}\left(19-7\sqrt{5}+\sqrt{750-330\sqrt{5}}\right)\,e^{\frac{\pi i}{5}}\\
\frac{1}{2\sqrt{2}}\left(17-7\sqrt{5}+\sqrt{390-174\sqrt{5}}\right)\,e^{\frac{2\pi i}{5}}\\
\frac{1}{2\sqrt{2}}\left(11-5\sqrt{5}+\sqrt{390-174\sqrt{5}}\right)\,e^{-\frac{2\pi i}{5}}\\
\frac{1}{4}\left(13-5\sqrt{5}-\sqrt{150-66\sqrt{5}}\right)\,e^{\frac{4\pi i}{5}}
\end{pmatrix}\approx  \begin{pmatrix}1\\ 2.022\\ 1.706\,e^{\frac{\pi i}{5}}\\ 0.816\,e^{\frac{2\pi i}{5}}\\ 0.276\,e^{-\frac{2\pi i}{5}}\\ 0.066\,e^{\frac{4\pi i}{5}} \end{pmatrix}\,,~~~\quad \\
Y^{(1)}_{\widehat{\bm{6}}}(\tau_{T}) & \propto &
\begin{pmatrix}
1\\
2\\
0\\
0\\
0\\
0
\end{pmatrix}\,.
\end{eqnarray}
The higher weight modular forms of level $N=5$ can be constructed from the tensor products of $Y^{(1)}_{\widehat{\bm{6}}}(\tau)$, and their explicit expressions are collected in Appendix~\ref{app:N5}. Plugging the above results into the expressions of higher weight modular forms, we can straightforwardly obtain their values at the fixed points $\tau_S$,\, $\tau_{ST}$,\, $\tau_{TS}$ and $\tau_T$. We are concerned with the triplet modular multiplets to which the neutrino Yukawa couplings $Y_{\text{atm}}$ and $Y_{\text{sol}}$ belong, the alignments of triplet modular multiplets of level $N=5$ at the fixed points $\tau_S$,\, $\tau_{ST}$,\,$\tau_{TS}$ \, $\tau_T$ are presented in table~\ref{tab:MF-FPs-N5}. Notice that the modular form triplets at $\tau_{TS}$ and $\tau_{ST}$ are related as $Y^{(k)}_{\bm{3}}(\tau_{TS})=\rho_{\bm{3}}(T)Y^{(k)}_{\bm{3}}(\tau_{ST})$ and $Y^{(k)}_{\bm{3}'}(\tau_{TS})=\rho_{\bm{3}'}(T)Y^{(k)}_{\bm{3}'}(\tau_{ST})$ because of $\tau_{TS}=T\tau_{ST}$. Furthermore, the alignments of triplet modular forms at other fixed points $\tau_f=\gamma\tau_S$, $\gamma\tau_{ST}$, $\gamma\tau_{TS}$, $\gamma\tau_{T}$ in the upper half plane, can be obtained by multiplying the vectors in table~\ref{tab:MF-FPs-N5} with the representation matrix $\rho_{\bm{3}}(\gamma)$ or $\rho_{\bm{3}^\prime}(\gamma)$, as shown in Eq.~\eqref{eq:Ytauf}. The resulting expressions are too lengthy to be included here.

\begin{small}
\begin{center}
\setlength\LTcapwidth{\textwidth}
\begin{longtable}{|c|c|c|}
\caption{\label{tab:MF-FPs-N5}
The alignments of modular triplets of level $N=5$ at the modular fixed points $\tau_S=i$, $\tau_{ST}=-\frac{1}{2}+i\frac{\sqrt{3}}{2}$, $\tau_{TS}=\frac{1}{2}+i\frac{\sqrt{3}}{2}$ and $\tau_T=i\infty$, where modular forms up to weight 6 are considered. }
\endfirsthead
\multicolumn{3}{c}{{\bfseries \tablename\ \thetable{} -- continued from previous page}}
\endhead

\multicolumn{3}{c}{{\bfseries \tablename\ \thetable{} -- continues on next page}}\\
\endfoot
\hline

\endlastfoot

\toprule
  \hline
 & $\tau_S$ & $\tau_T$\\
\hline
$Y^{(2)}_{\bm{3}}\propto$ & $\begin{pmatrix}1\\[1ex]\frac{1}{22\sqrt{2}}\left(11+11\sqrt{5}+\sqrt{1090-58\sqrt{5}}\right)\\[1ex] \frac{1}{22\sqrt{2}}\left(11+11\sqrt{5}-\sqrt{1090-58\sqrt{5}}\right)  \end{pmatrix}\approx \begin{pmatrix}1\\2.140\\0.148 \end{pmatrix}$ & $\begin{pmatrix}1\\[1ex]0\\[1ex]0 \end{pmatrix}$\\
\hline
$Y^{(2)}_{\bm{3}^\prime}\propto$ & $\begin{pmatrix}1\\[1ex]\frac{1}{2\sqrt{2}}\left(1-\sqrt{5}-2\sqrt{85-38\sqrt{5}}\right)\\[1ex]\frac{1}{2\sqrt{2}}\left(1-\sqrt{5}+2\sqrt{85-38\sqrt{5}}\right)\end{pmatrix}\approx\begin{pmatrix} 1\\-0.558\\-0.316\end{pmatrix}$ & $\begin{pmatrix} 1\\[1ex]0\\[1ex]0\end{pmatrix}$\\
\hline
$Y^{(4)}_{\bm{3}}\propto$ & $\begin{pmatrix}1\\[1ex]\frac{1-\sqrt{5}}{2\sqrt{2}}\\[1ex]\frac{1-\sqrt{5}}{2\sqrt{2}} \end{pmatrix}\approx\begin{pmatrix}1\\-0.437\\-0.437 \end{pmatrix}$ & $\begin{pmatrix}1\\[1ex]0\\[1ex]0 \end{pmatrix}$\\
\hline
$Y^{(4)}_{\bm{3}^\prime}\propto$ & $\begin{pmatrix}1\\[1ex]\frac{1+\sqrt{5}}{2\sqrt{2}}\\[1ex]\frac{1+\sqrt{5}}{2\sqrt{2}} \end{pmatrix}\approx\begin{pmatrix}1\\1.144\\1.144 \end{pmatrix}$ & $\begin{pmatrix}1\\[1ex]0\\[1ex]0 \end{pmatrix}$\\
\hline
$Y^{(6)}_{\bm{3}I}\propto$ & $\begin{pmatrix}1\\[1ex]\frac{1}{22\sqrt{2}}\left(11+11\sqrt{5}+\sqrt{1090-58\sqrt{5}}\right)\\[1ex]\frac{1}{22\sqrt{2}}\left(11+11\sqrt{5}-\sqrt{1090-58\sqrt{5}}\right)\end{pmatrix}\approx\begin{pmatrix}1\\2.140\\0.148 \end{pmatrix}$ & $\begin{pmatrix}1\\[1ex]0\\[1ex]0 \end{pmatrix}$\\
\hline
$Y^{(6)}_{\bm{3}II}\propto$ & $\begin{pmatrix}1\\[1ex]\frac{1}{38\sqrt{2}}\left(19+19\sqrt{5}-2\sqrt{4285-1882\sqrt{5}}\right)\\[1ex]\frac{1}{38\sqrt{2}}\left(19+19\sqrt{5}+2\sqrt{4285-1882\sqrt{5}}\right)\end{pmatrix}\approx\begin{pmatrix} 1\\0.818\\1.470\end{pmatrix}$ & $\begin{pmatrix}1\\[1ex]0\\[1ex]0 \end{pmatrix}$\\
\hline
$Y^{(6)}_{\bm{3}^\prime I}\propto$ & $\begin{pmatrix}1\\[1ex]\frac{1}{2\sqrt{2}}\left(1-\sqrt{5}-2\sqrt{85-38\sqrt{5}}\right)\\[1ex]\frac{1}{2\sqrt{2}}\left(1-\sqrt{5}+2\sqrt{85-38\sqrt{5}}\right)\end{pmatrix}\approx\begin{pmatrix}1\\[1ex]-0.558\\[1ex]-0.316
\end{pmatrix}$  & $\begin{pmatrix} 1\\[1ex]0\\[1ex]0\end{pmatrix}$\\
\hline
$Y^{(6)}_{\bm{3}^\prime II}\propto$ & $\begin{pmatrix}
1\\[1ex]
\frac{1}{2\sqrt{2}}\left(1-\sqrt{5}-\sqrt{130+58\sqrt{5}}\right)\\[1ex]
\frac{1}{2\sqrt{2}}\left(1-\sqrt{5}+\sqrt{130+58\sqrt{5}}\right)
\end{pmatrix}\approx\begin{pmatrix}1\\[1ex]-6.135\\[1ex]5.260 \end{pmatrix}$ &  $\begin{pmatrix} 0\\[1ex]0\\[1ex]0\end{pmatrix}$\\
\hline
 & \multicolumn{2}{|c|}{$\tau_{ST}$}\\
\hline
$Y^{(2)}_{\bm{3}}\propto$ & \multicolumn{2}{|c|}{$\begin{pmatrix}1\\[1ex]\frac{1}{4\sqrt{2}}\left(3+\sqrt{5}+\sqrt{30+6\sqrt{5}}\right)\,e^{-\frac{i\pi}{5}}\\[1ex]\frac{1}{4\sqrt{2}}\left(-3-\sqrt{5}+\sqrt{30+6\sqrt{5}}\right)\,e^{-\frac{4i\pi}{5}}\end{pmatrix}\approx\begin{pmatrix} 1\\2.090\,e^{-\frac{i\pi}{5}}\\0.239\,e^{-\frac{4i\pi}{5}}\end{pmatrix}$}\\
\hline
$Y^{(2)}_{\bm{3}^\prime}\propto$ & \multicolumn{2}{|c|}{$\begin{pmatrix}1\\[1ex]\frac{1}{4\sqrt{2}}\left(3-\sqrt{5}+\sqrt{30-6\sqrt{5}}\right)\,e^{\frac{3i\pi}{5}}\\[1ex]\frac{1}{4\sqrt{2}}\left(-3+\sqrt{5}+\sqrt{30-6\sqrt{5}}\right)\,e^{\frac{2i\pi}{5}}\end{pmatrix}\approx\begin{pmatrix} 1\\0.855\,e^{\frac{3i\pi}{5}}\\0.585\,e^{\frac{2i\pi}{5}}\end{pmatrix}$}\\
\hline
$Y^{(4)}_{\bm{3}}\propto$ & \multicolumn{2}{|c|}{$\begin{pmatrix}1\\[1ex]\frac{1}{4\sqrt{2}}\left(-3-\sqrt{5}+\sqrt{30+6\sqrt{5}}\right)\,e^{\frac{4i\pi}{5}}\\[1ex]\frac{1}{4\sqrt{2}}\left(3+\sqrt{5}+\sqrt{30+6\sqrt{5}}\right)\,e^{\frac{i\pi}{5}}\end{pmatrix}\approx\begin{pmatrix} 1\\0.239\,e^{\frac{4i\pi}{5}}\\2.090\,e^{\frac{i\pi}{5}}\end{pmatrix}$}\\
\hline
$Y^{(4)}_{\bm{3}^\prime}\propto$ & \multicolumn{2}{|c|}{$\begin{pmatrix}1\\[1ex]\frac{1}{4\sqrt{2}}\left(-3+\sqrt{5}+\sqrt{30-6\sqrt{5}}\right)\,e^{-\frac{2i\pi}{5}}\\[1ex]\frac{1}{4\sqrt{2}}\left(3-\sqrt{5}+\sqrt{30-6\sqrt{5}}\right)\,e^{-\frac{3i\pi}{5}}\end{pmatrix}\approx\begin{pmatrix} 1\\0.585\,e^{-\frac{2i\pi}{5}}\\0.855\,e^{-\frac{3i\pi}{5}}\end{pmatrix}$}\\
\hline
$Y^{(6)}_{\bm{3} I}\propto$ & \multicolumn{2}{|c|}{$\begin{pmatrix}0\\0\\0 \end{pmatrix}$}\\
\hline
$Y^{(6)}_{\bm{3} II}\propto$ & \multicolumn{2}{|c|}{$\begin{pmatrix}1\\[1ex]\frac{1}{\sqrt{2}}\left(3-\sqrt{5}\right)\,e^{\frac{4i\pi}{5}}\\[2ex]\frac{1}{\sqrt{2}}\left(3-\sqrt{5}\right)\,e^{-\frac{4i\pi}{5}} \end{pmatrix}\approx\begin{pmatrix} 1\\0.540\,e^{\frac{4i\pi}{5}}\\0.540\,e^{-\frac{4i\pi}{5}}\end{pmatrix}$}\\
\hline
$Y^{(6)}_{\bm{3}^\prime I}\propto$ & \multicolumn{2}{|c|}{$\begin{pmatrix}0\\0\\0 \end{pmatrix}$}\\
\hline
$Y^{(6)}_{\bm{3}^\prime II}\propto$ & \multicolumn{2}{|c|}{$\begin{pmatrix}1\\[1ex]\frac{1}{\sqrt{2}}\left(3+\sqrt{5}\right)\,e^{-\frac{2i\pi}{5}}\\[2ex]\frac{1}{\sqrt{2}}\left(3+\sqrt{5}\right)\,e^{\frac{2i\pi}{5}} \end{pmatrix}\approx\begin{pmatrix}1\\3.702\,e^{-\frac{2i\pi}{5}}\\3.702\,e^{\frac{2i\pi}{5}}\end{pmatrix}$}\\
\hline
 & \multicolumn{2}{|c|}{$\tau_{TS}$}\\
\hline
$Y^{(2)}_{\bm{3}}\propto$ & \multicolumn{2}{|c|}{$\begin{pmatrix}1\\[1ex]\dfrac{1}{4\sqrt{2}}\left(3+\sqrt{5}+\sqrt{30+6\sqrt{5}}\right)\,e^{\frac{i\pi}{5}}\\[1ex]\dfrac{1}{4\sqrt{2}}\left(-3-\sqrt{5}+\sqrt{30+6\sqrt{5}}\right)\,e^{\frac{4i\pi}{5}}\end{pmatrix}\approx\begin{pmatrix} 1\\2.090\,e^{\frac{i\pi}{5}}\\0.239\,e^{\frac{4i\pi}{5}}\end{pmatrix}$}\\
\hline
$Y^{(2)}_{\bm{3}^\prime}\propto$ & \multicolumn{2}{|c|}{$\begin{pmatrix}1\\[1ex]\dfrac{1}{4\sqrt{2}}\left(3-\sqrt{5}+\sqrt{30-6\sqrt{5}}\right)\,e^{-\frac{3i\pi}{5}}\\[1ex]\dfrac{1}{4\sqrt{2}}\left(-3+\sqrt{5}+\sqrt{30-6\sqrt{5}}\right)\,e^{-\frac{2i\pi}{5}}\end{pmatrix}\approx\begin{pmatrix} 1\\0.855\,e^{-\frac{3i\pi}{5}}\\0.585\,e^{-\frac{2i\pi}{5}}\end{pmatrix}$}\\
\hline
$Y^{(4)}_{\bm{3}}\propto$ & \multicolumn{2}{|c|}{$\begin{pmatrix}1\\[1ex]\frac{1}{4\sqrt{2}}\left(-3-\sqrt{5}+\sqrt{30+6\sqrt{5}}\right)\,e^{-\frac{4i\pi}{5}}\\[1ex]\frac{1}{4\sqrt{2}}\left(3+\sqrt{5}+\sqrt{30+6\sqrt{5}}\right)\,e^{-\frac{i\pi}{5}}\end{pmatrix}\approx\begin{pmatrix} 1\\0.239\,e^{-\frac{4i\pi}{5}}\\2.090\,e^{-\frac{i\pi}{5}}\end{pmatrix}$}\\
\hline
$Y^{(4)}_{\bm{3}^\prime}\propto$ & \multicolumn{2}{|c|}{$\begin{pmatrix}1\\[1ex]\frac{1}{4\sqrt{2}}\left(-3+\sqrt{5}+\sqrt{30-6\sqrt{5}}\right)\,e^{\frac{2i\pi}{5}}\\[1ex]\frac{1}{4\sqrt{2}}\left(3-\sqrt{5}+\sqrt{30-6\sqrt{5}}\right)\,e^{\frac{3i\pi}{5}}\end{pmatrix}\approx\begin{pmatrix} 1\\0.585\,e^{\frac{2i\pi}{5}}\\0.855\,e^{\frac{3i\pi}{5}}\end{pmatrix}$}\\
\hline
$Y^{(6)}_{\bm{3} I}\propto$ & \multicolumn{2}{|c|}{$\begin{pmatrix}0\\0\\0 \end{pmatrix}$}\\
\hline
$Y^{(6)}_{\bm{3} II}\propto$ & \multicolumn{2}{|c|}{$\begin{pmatrix}1\\\frac{1}{\sqrt{2}}\left(3-\sqrt{5}\right)\,e^{-\frac{4i\pi}{5}}\\ \frac{1}{\sqrt{2}}\left(3-\sqrt{5}\right)\,e^{\frac{4i\pi}{5}}\end{pmatrix}\approx\begin{pmatrix} 1\\0.540\,e^{-\frac{4i\pi}{5}}\\0.540\,e^{\frac{4i\pi}{5}}\end{pmatrix}$}\\
\hline
$Y^{(6)}_{\bm{3}^\prime I}\propto$ & \multicolumn{2}{|c|}{$\begin{pmatrix}0\\0\\0 \end{pmatrix}$}\\
\hline
$Y^{(6)}_{\bm{3}^\prime II}\propto$ & \multicolumn{2}{|c|}{$\begin{pmatrix}1\\[1ex]\frac{1}{\sqrt{2}}\left(3+\sqrt{5}\right) e^{\frac{2i\pi}{5}}\\[2ex]\frac{1}{\sqrt{2}}\left(3+\sqrt{5}\right) e^{\frac{-2i\pi}{5}} \end{pmatrix}\approx\begin{pmatrix}1\\3.702\,e^{\frac{2i\pi}{5}}\\3.702\,e^{-\frac{2i\pi}{5}}  \end{pmatrix}$}\\
\hline \hline
\end{longtable}
\end{center}
\end{small}

\subsection{Modular Littlest seesaw models in $A_5$ and $A'_5$ }

In the following, we shall study the possible modular Littlest seesaw models which can be constructed in the $A_5$ or $A'_5$ modular group~\footnote{The Littlest seesaw model in the traditional $A_5$ flavor symmetry was investigated in~\cite{Ding:2017hdv}, the lepton mixing was predicted to be the GR1 pattern which preserves the first column of the golden ratio mixing matrix. The resulting solar mixing angle is too small to be compatible with current data.}. As shown in Appendix~\ref{app:N5}, the $A'_5$ modular group has only a unique singlet representation $\bm{1}$, and two triplet representations $\bm{3}$ and $\bm{3}'$. Hence the three generations of the left-handed lepton doublet fields $L$ can be assigned to either $\bm{3}$ or $\bm{3}'$, while both right-handed neutrinos $N^c_{\mathrm{atm}}$ and $N^c_{\mathrm{sol}}$ are invariant singlet under $A'_5$ and they are distinguished by their modular weights. As a result, both triplet modular forms $Y_{\text{atm}}$ and $Y_{\text{sol}}$ should be in the same representation of $A'_5$ as that of $L$. Because $A'_5$ can not be distinguished from $A_5$ in the representations $\bm{3}$, $\bm{3}'$ and $\bm{1}$, the same results in the following can also be achieved from $A_5$ modular symmetry as well. Furthermore, the right-handed neutrino mass terms couple with modular form singlet in the framework of modular flavor symmetry, and we see that the singlet modular form of level $N=5$ is only absent at weight two, as can be seen from table~\ref{tab:MFs-N5}. In order to forbid the cross term $N^c_{\text{atm}}N^c_{\text{sol}}$, the sum of the modular weights should be $k_{N^c_{\text{atm}}}+k_{N^c_{\text{sol}}}=2$, where $k_{N^c_{\text{atm}}}$  and $k_{N^c_{\text{sol}}}$ denote the modular weights of $N^c_{\text{atm}}$ and $N^c_{\text{sol}}$ respectively. Thus we could set $k_{N^c_{\text{atm}}}=0, k_{N^c_{\text{sol}}}=2$ or $k_{N^c_{\text{atm}}}=2, k_{N^c_{\text{sol}}}=0$. As a consequence, there are only four possible modular transformations of the lepton fields $L$, $N^c_{\text{atm}}$ and $N^c_{\text{sol}}$,
\begin{subequations}
\begin{eqnarray}
\label{eq:assign-A5-1}&&L\sim(\bm{3}, 2)\,,~~~N^{c}_{\text{atm}}\sim(\bm{1}, 0)\,,~~~N^{c}_{\text{sol}}\sim(\bm{1}, 2)\,,\\
\label{eq:assign-A5-2}&&L\sim(\bm{3}, 2)\,,~~~N^{c}_{\text{atm}}\sim(\bm{1}, 2)\,,~~~N^{c}_{\text{sol}}\sim(\bm{1}, 0)\,,\\
\label{eq:assign-A5-3}&&L\sim(\bm{3}', 2)\,,~~~N^{c}_{\text{atm}}\sim(\bm{1}, 0)\,,~~~N^{c}_{\text{sol}}\sim(\bm{1}, 2)\,,\\
\label{eq:assign-A5-4}&&L\sim(\bm{3}', 2)\,,~~~N^{c}_{\text{atm}}\sim(\bm{1}, 2)\,,~~~N^{c}_{\text{sol}}\sim(\bm{1}, 0)\,,
\end{eqnarray}
\end{subequations}
where the Higgs fields are assumed to be invariant under modular symmetry and the modular weights are zero without loss of generality. We have been concerned with modular forms of weights 2 and 4, since there are more than one linearly independent modular triplets at weight $k\geq6$ and the predictability would be reduced.

Since two pairs of residual symmetries which are related by group conjugation lead to the same predictions for lepton masses and mixing parameters, similar to the case of $S'_4$ we can only consider $\gamma_{\ell}=ST$ or $\gamma_{\ell}=T$. Given the representation matrices of $S$ and $T$ given in Eq.~\eqref{eq:Irr-matrix-N5}, we can determine the unitary transformation $U_{\ell}$ to be
\begin{eqnarray}
\nonumber \nonumber \gamma_{\ell}&=&ST ~:~U_{\ell}=\dfrac{1}{30}\begin{pmatrix}
 2\sqrt{75+30\sqrt{5}}\,e^{\frac{4\pi i}{5}}
 ~&~ 2\sqrt{75-15\sqrt{5}}\,e^{\frac{4\pi i}{5}}
 ~&~ 2\sqrt{75-15\sqrt{5}}\,e^{-\frac{\pi i}{5}}\\
2\sqrt{75-15\sqrt{5}}\,e^{-\frac{2\pi i}{5}}
~&~\left(15+\sqrt{75+30\sqrt{5}}\right)e^{\frac{3\pi i}{5}}
~&~~ \left(15-\sqrt{75+30\sqrt{5}}\right)e^{\frac{3\pi i}{5}}\\
 2\sqrt{75-15\sqrt{5}} ~&~ 15-\sqrt{75+30\sqrt{5}}
~&~ 15+\sqrt{75+30\sqrt{5}}\end{pmatrix}\\
&&\qquad\quad \quad~\approx  \left(
\begin{array}{ccc}
0.795\,e^{\frac{4\pi i}{5}}
 ~&~ 0.429\,e^{\frac{4\pi i}{5}}
 ~&~ 0.429\,e^{-\frac{\pi i}{5}}\\
0.429\,e^{-\frac{2\pi i}{5}}
~&~0.897\, e^{\frac{3\pi i}{5}}
~&~~ 0.103\,e^{\frac{3\pi i}{5}}\\
 0.429 ~&~ 0.103
~&~ 0.897
\end{array}
\right)\,, ~~\text{for}~~L\sim\bm{3}\,, \label{eq:Uell-ST-3}\\
\nonumber \gamma_{\ell}&=&ST ~:~U_{\ell}=\dfrac{1}{30}
\begin{pmatrix}
2\sqrt{75+15\sqrt{5}}\,e^{-\frac{2 \pi i}{5}}
~&~ 2\sqrt{75+15\sqrt{5}}\,e^{\frac{3\pi i}{5}}
~&~ 2\sqrt{75-30\sqrt{5}}\,e^{-\frac{2\pi i}{5}}\\
\left(15+\sqrt{75-30\sqrt{5}}\right)\,e^{\frac{\pi i}{5}}
~&~  \left(15-\sqrt{75-30\sqrt{5}}\right)\,e^{\frac{\pi i}{5}}
~&~~ 2\sqrt{75+15\sqrt{5}}\,e^{-\frac{4\pi i}{5}} \\
15-\sqrt{75-30\sqrt{5}}  ~&~ 15+\sqrt{75-30\sqrt{5}}
~&~ 2\sqrt{75+15\sqrt{5}}
\end{pmatrix}\\
&&\qquad\quad \quad~ \approx
\begin{pmatrix}
0.695\,e^{-\frac{2 \pi i}{5}}
~&~ 0.695\,e^{\frac{3\pi i}{5}}
~&~ 0.188\,e^{-\frac{2\pi i}{5}}\\
0.594\,e^{\frac{\pi i}{5}}
~&~  0.406\,e^{\frac{\pi i}{5}}
~&~~ 0.695\,e^{-\frac{4\pi i}{5}} \\
0.406  ~&~ 0.594
~&~ 0.695
\end{pmatrix}\,, ~~\text{for}~~L\sim\bm{3}'\,, \label{eq:Uell-ST-3prime}\\
\gamma_{\ell}&=&T~:~U_{\ell}=\left(
\begin{array}{ccc}
 1 ~&~ 0 ~&~ 0 \\
 0 ~&~ 1 ~&~ 0 \\
 0 ~&~ 0 ~&~ 1 \\
\end{array}\right)\,, ~~\text{for}~~L\sim\bm{3}~~\text{and}~~L\sim\bm{3}'\,, \label{eq:Uell-T-3-3prime}
\end{eqnarray}
up to phases and permutation of column vectors in our working basis. The VEVs $\langle\tau_{\mathrm{atm}}\rangle$ and $\langle\tau_{\mathrm{sol}}\rangle$ and be any modular fixed points in the upper half plane. Scanning all the possible alignments of the modular triplets $Y_{\text{atm}}\left(\langle\tau_{\mathrm{atm}}\rangle\right)$ and $Y_{\text{sol}}\left(\langle\tau_{\mathrm{sol}}\rangle\right)$, we find plenty of Littlest seesaw models can be achieved from the $A_5$ and $A'_5$  modular symmetries. The best fit values of $m_a$, $m_s$, $\eta$ and the corresponding predictions for neutrino masses and mixing parameters for each case are summarized in table~\ref{tab:BF-result-N5}. As explained in section~\ref{sec:modular-LSS}, the first column is the lepton mixing matrix is fixed by modular symmetry and it can be generally parametrized as

\begin{table}[t!]
\begin{center}
\resizebox{\textwidth}{!}{
\begin{tabular}{|c|c|c|c|c|c|c|c|c|c|c|c|c|}\hline\hline
Case
& $\chi^2_{\text{min}}$
& $m_a(\text{meV})$
& $m_s/m_a$
& $\eta/\pi$
& $\sin^2\theta_{12}$
& $\sin^2\theta_{13}$
& $\sin^2\theta_{23}$
& $\delta_{CP}/\pi$
& $\beta/\pi$
& $m_2(\text{meV})$
& $m_3(\text{meV})$
& $m_{\beta\beta}\text{(meV)}$\\
\hline
I
&$1.889$
& $26.749$
& $0.090$
& $1.223$
& $0.295$
& $0.02221$
& $0.486$
& $1.381$
& $1.220$
& $8.577$
& $50.323$
& $3.591$\\
\hline
II
& $2.314$
& $7.596$
& $0.609$
& $1.313$
& $0.305$
& $0.02233$
& $0.485$
& $1.019$
& $0.318$
& $8.589$
& $50.314$
& $3.213$\\
\hline
III
& $3.081$
& $11.223$
& $1.047$
& $1.331$
& $0.305$
& $0.02233$
& $0.512$
& $1.155$
& $1.895$
& $8.576$
& $50.324$
& $3.521$\\
\hline
IV
& $3.553$
& $38.658$
& $0.169$
& $1.348$
& $0.303$
& $0.02231$
& $0.456$
& $1.438$
& $1.047$
& $8.578$
& $50.323$
& $3.642$\\
\hline
V
& $9.320$
& $8.457$
& $0.749$
& $1.177$
& $0.295$
& $0.02238$
& $0.512$
& $1.616$
& $0.774$
& $8.603$
& $50.302$
& $3.606$\\
\hline
VI
& $13.837$
& $38.658$
& $0.169$
& $0.652$
& $0.303$
& $0.02232$
& $0.544$
& $1.562$
& $0.953$
& $8.572$
& $50.326$
& $3.641$\\
\hline
VII
& $14.425$
& $11.219$
& $1.050$
& $0.670$
& $0.305$
& $0.02231$
& $0.487$
& $1.844$
& $0.104$
& $8.596$
& $50.309$
& $3.521$\\
\hline
VIII
& $19.819$
& $37.369$
& $0.230$
& $0.328$
& $0.305$
& $0.02244$
& $0.575$
& $1.197$
& $1.460$
& $8.571$
& $50.327$
& $3.602$\\
\hline
IX
& $23.829$
& $10.899$
& $1.113$
& $1.686$
& $0.305$
& $0.02222$
& $0.514$
& $1.981$
& $1.683$
& $8.597$
& $50.308$
& $3.215$\\
\hline
X
& $8.443$
& $24.879$
& $0.212$
& $1.659$
& $0.269$
& $0.02230$
& $0.489$
& $1.214$
& $0.034$
& $8.577$
& $50.323$
& $2.639$\\
\hline
XI
& $10.640$
& $0.309$
& $11.252$
& $0.347$
& $0.329$
& $0.02211$
& $0.447$
& $1.507$
& $1.244$
& $8.606$
& $50.300$
& $3.625$\\
\hline
XII
& $18.454$
& $0.309$
& $11.248$
& $1.652$
& $0.329$
& $0.02207$
& $0.552$
& $1.493$
& $0.757$
& $8.610$
& $50.297$
& $3.624$\\
\hline
XIII
& $21.992$
& $24.765$
& $0.215$
& $0.343$
& $0.269$
& $0.02205$
& $0.509$
& $1.783$
& $1.968$
& $8.610$
& $50.297$
& $2.627$\\
\hline
XIV
& $24.340$
& $0.309$
& $8.255$
& $0.434$
& $0.334$
& $0.02154$
& $0.568$
& $1.347$
& $0.690$
& $8.627$
& $50.283$
& $2.632$\\
\hline\hline
\end{tabular}
}
\end{center}
\caption{\label{tab:BF-result-N5}Results of the $\chi^2$ analysis for the modular Littlest seesaw based on the finite modular group $A_5$ or $A'_5$. Note that the lightest neutrino mass is vanishing with $m_1=0$. }
\end{table}

\begin{eqnarray}
U_{PMNS}=\begin{pmatrix}
\cos\varphi ~&~ \times ~&~ \times \\
\cos\phi\sin\varphi ~&~ \times ~&~ \times \\
\sin\phi\sin\varphi ~&~ \times ~&~ \times
\end{pmatrix}\,,\label{eq:parameterization-fixed-colum}
\end{eqnarray}
where one can take $0\leq \varphi, \phi\leq\pi/2$ without loss of generality. The values of $\varphi$ and $\phi$ for each case are summarized in table~\ref{tab:fixed-column-parameters}. Comparing with the standard parametrization of lepton mixing matrix in Eq.~\eqref{eq:PMNS_def}, we can reach the following sum rules between the lepton mixing angles and Dirac CP phase $\delta_{CP}$,
\begin{eqnarray}
\nonumber &&\qquad\qquad\qquad \cos^2\theta_{12}\cos^2\theta_{13} = \cos^2\varphi\,,\\
\nonumber &&\cos\delta_{CP}=\dfrac{\csc2\theta_{23}\csc\theta_{13}}{8\sqrt{\cos^2\theta_{13}-\cos^2\varphi}}\Bigg\{\bigg[1+3\cos2\varphi-(3+\cos2\varphi)\cos2\theta_{13}\bigg]\sec\varphi\cos2\theta_{23}\\
\label{eq:cosdeltaCP-sum-rule}&&\qquad\qquad~~ +4\cos2\phi\sin\varphi\tan\varphi\cos^2\theta_{13}
\Bigg\}\,.
\end{eqnarray}
In the following, we shall present the assignments of matter fields, the moduli VEVs $\langle\tau_{\text{atm}}\rangle$ and $\langle\tau_{\text{sol}}\rangle$ as well as the corresponding alignments $Y_{\text{atm}}$ and $Y_{\text{sol}}$ for each possible modular Littlest seesaw models based on $A_5$ or $A'_5$ modular symmetry.

\begin{table}[hptb]
\begin{center}
\begin{tabular}{|c|c|c|c|c|c|c|c|}
\hline \hline
 & case I & case II & case III & case IV & case V & case VI & case VII\\
\hline
$\varphi/\pi$ & 0.188 & 0.192 & 0.192 & 0.191 & 0.188 & 0.191 & 0.192\\
\hline
$\phi/\pi$    & 0.271 & 0.316 & 0.316 & 0.251 & 0.229 & 0.249 & 0.184\\
\hline
 & case VIII & case IX & case X & case XI & case XII & case XIII & case XIV\\
\hline
$\varphi/\pi$ & 0.192 & 0.192 & 0.179 & 0.199 & 0.199 & 0.179 & 0.201\\
\hline
$\phi/\pi$    & 0.330 & 0.184 & 0.306 & 0.233 & 0.267 & 0.194 & 0.300\\
\hline \hline
\end{tabular}
\caption{\label{tab:fixed-column-parameters} The parameters $\varphi$ and $\phi$ of the fixed column shown in Eq.~\eqref{eq:parameterization-fixed-colum} for the modular Littlest seesaw of cases I$\sim$XIV in $A_5$ (or $A'_5$) modular symmetry. For the modular Littlest seesaw models based on $S'_4$ modular symmetry, we have $(\varphi/\pi, \phi/\pi)\simeq(0.196, 0.250)$, $(0.201, 0.243)$, and $(0.201, 0.257)$  for cases A and B, case C and case D respectively.  }
\end{center}
\end{table}

\subsubsection{$\gamma_{\ell}=ST$}

Since the charged lepton mass terms preserve the $Z^{ST}_3$ residual symmetry, the unitary transformation $U_{\ell}$ is of the form in Eq.~\eqref{eq:Uell-ST-3} or that in Eq.~\eqref{eq:Uell-ST-3prime} if the lepton doublet $L$ transforms as a triplet $\bm{3}$ or $\bm{3}'$ under the $A_5$ and $A'_5$ modular symmetry.

\begin{itemize}

\item{Case I}

The left-handed lepton fields $L$ are assumed to transform as a triplet $\bm{3}$ in this case, the right-handed neutrino fields $N^{c}_{\text{atm}}$ and $N^{c}_{\text{sol}}$ are invariant under the modular symmetry. The modular transformation of lepton fields $L$, $N^{c}_{\text{atm}}$, $N^{c}_{\text{sol}}$ is of the form in Eq.~\eqref{eq:assign-A5-1}. The Higgs fields are invariant under $A'_5$ with vanishing modular weight. One can assign the right-handed charged leptons to be modular singlets such as $e^{c}\sim(\bm{1}, 0)$, $\mu^{c}\sim(\bm{1}, 2)$, $\tau^{c}\sim(\bm{1}, 4)$. One can check that the charged lepton mass matrix would be diagonalized by the unitary transformation $U_{\ell}$ in Eq.~\eqref{eq:Uell-ST-3}, once the modulus $\tau_{ell}$ is stabilized at $\tau_{\ell}=\tau_{ST}$. Notice that the assignment of the modular transformation of the right-handed charged leptons is not unique~\cite{Ding:2019xna,Yao:2020zml}. Furthermore, we see that the modular invariance constrains both $Y_{\text{atm}}$ and $Y_{\text{sol}}$ to be triplet modular forms in the representation $\bm{3}$ and their modular weights are equal to $2$ and $4$ respectively. It is notable that the cross term $N^c_{\text{atm}}N^c_{\text{sol}}$ is forbidden by modular symmetry because there is no singlet modular form of weight two at level $N=5$. The VEVs of the $\tau_{\text{atm}}$ and $\tau_{\text{sol}}$ are assumed to be stabilized at the modular fixed points $\langle\tau_{\text{atm}}\rangle=TST^3\tau_S=\frac{7}{10}+\frac{i}{10}$ and
$\langle\tau_{\text{sol}}\rangle=T^3\tau_{ST}=\frac{5}{2}+i\frac{\sqrt{3}}{2}$ in this case. Thus the alignments of $Y_{\text{atm}}$ and $Y_{\text{sol}}$ in our working basis are found to be
\begin{eqnarray}
\nonumber Y_{\text{atm}} & = & Y^{(2)}_{\bm{3}}(\langle\tau_{\text{atm}}\rangle=TST^3\tau_S=\frac{7}{10}+\frac{i}{10}) \propto \begin{pmatrix}1\\\frac{1}{13\sqrt{2}}(-3+11i)\\\frac{1}{13\sqrt{2}}(-3+11i) \end{pmatrix}\approx\begin{pmatrix} 1\\-0.163+0.598 i\\-0.163+0.598i\end{pmatrix}\,, \nonumber\\
Y_{\text{sol}} & = & Y^{(4)}_{\bm{3}}(\langle\tau_{\text{sol}}\rangle=T^3\tau_{ST}=\frac{5}{2}+i\frac{\sqrt{3}}{2}) \propto \begin{pmatrix}1\\ \frac{1}{4\sqrt{2}}\left(-3-\sqrt{5}+\sqrt{30+6\sqrt{5}}\right)\\-\frac{1}{4\sqrt{2}}\left(3+\sqrt{5}+\sqrt{30+6\sqrt{5}}\right) \end{pmatrix} \approx\begin{pmatrix}1\\ 0.239\\ -2.090 \end{pmatrix}\,.~~~~~
\end{eqnarray}
The charged lepton permutation matrix is $P_{\ell}=P_{321}$, and the lepton mixing matrix turns out to be of the form
\begin{eqnarray}
U_{PMNS}\approx\begin{pmatrix}0.830 ~&~ \times ~&~ \times \\
0.367 ~&~ \times ~&~ \times \\
0.420 ~&~ \times ~&~ \times
\end{pmatrix}\,.
\end{eqnarray}
Accordingly the parameters $\varphi$ and $\phi$ are given by
\begin{eqnarray}
\varphi\approx0.188\pi\,,~~~\phi\approx 0.271\pi\,.
\end{eqnarray}
We display the contour plots of $\sin^2\theta_{12}$, $\sin^2\theta_{13}$, $\sin^{2}\theta_{23}$ and $\delta m^2/\Delta m^2$ in the plane $r\equiv m_s/m_a$ versus $\eta$ plane in figure~\ref{fig:contour-MLS-A5p}. We see that the experimental data can be well accommodated in a rather small region of parameter space which is indicated with red circle. The precisely measured reactor angle $\sin^2\theta_{13}$ and the mass ratio $\delta m^2/\Delta m^2$ impose strong constraint on the parameter values of $r$ and $\eta$.

\item{Case II }

The modular transformations of the lepton fields are the same as those in Eq.~\eqref{eq:assign-A5-1}. The atmospheric and solar moduli take the symmetry points $\langle\tau_{\text{atm}}\rangle=ST^3ST^2\tau_S=-\frac{13}{34}+\frac{i}{34}$, $\langle\tau_{\text{sol}}\rangle=ST\tau_S=-\frac{1}{2}+\frac{i}{2}$. Consequently the triplet modular forms $Y_{\text{atm}}$ and $Y_{\text{sol}}$ are aligned along the following directions,
\begin{eqnarray}
\nonumber Y_{\text{atm}} & = & Y^{(2)}_{\bm{3}}(\langle\tau_{\text{atm}}\rangle=ST^3ST^2\tau_S=-\frac{13}{34}+\frac{i}{34}) \propto \begin{pmatrix} 1\\\frac{1}{22\sqrt{2}}\left(11+11\sqrt{5}-\sqrt{1090-58\sqrt{5}}\right)\,e^{-\frac{4\pi i}{5}}\\
\frac{1}{22\sqrt{2}}\left(11+11\sqrt{5}+\sqrt{1090-58\sqrt{5}}\right)\,e^{\frac{4\pi i}{5}} \end{pmatrix} \\
\nonumber &&\hskip1.8in\qquad\qquad\qquad   \approx
\begin{pmatrix}
1\\
0.148\,e^{-\frac{4\pi i}{5}}\\
2.140\,e^{\frac{4\pi i}{5}}
\end{pmatrix}\,,\\
Y_{\text{sol}} & = & Y^{(4)}_{\bm{3}}(\langle\tau_{\text{sol}}\rangle=ST\tau_S=-\frac{1}{2}+\frac{i}{2})\propto  \begin{pmatrix}1\\
\frac{1+\sqrt{5}}{2\sqrt{2}}\,e^{\frac{4\pi i}{5}}\\ \frac{1+\sqrt{5}}{2\sqrt{2}}\,e^{-\frac{4\pi i}{5}} \end{pmatrix} \approx\begin{pmatrix}1\\
1.144\,e^{\frac{4\pi i}{5}}\\ 1.144\,e^{-\frac{4\pi i}{5}}\end{pmatrix}\,.
\end{eqnarray}
In order to be compatible with experimental data, the charged lepton permutation matrix should be $P_{\ell}=P_{231}$, and the lepton mixing matrix is determined to be of the following form,
\begin{eqnarray}
\label{eq:UPMNS-caseII}U_{PMNS}\approx\begin{pmatrix}0.824 ~&~ \times ~&~ \times \\
0.310 ~&~ \times ~&~ \times \\
0.474 ~&~ \times ~&~ \times
\end{pmatrix}\,,
\end{eqnarray}
which corresponds to
\begin{equation}
\varphi\approx 0.192\pi\,,~~~~\phi\approx 0.316\pi\,.
\end{equation}
The contour plots of $\sin^2\theta_{12}$, $\sin^2\theta_{13}$, $\sin^{2}\theta_{23}$ and $\delta m^2/\Delta m^2$ are shown in figure~\ref{fig:contour-MLS-A5p}. The agreement with experimental data can be achieved when $(\eta, r)$ is around the best fit point $(1.313\pi, 0.609)$.

\item{Case III }

The left-handed lepton doublet $L$ also transforms as a triplet $\bm{3}$ under $A_5$ and $A'_5$. The $\tau_{\text{atm}}$ and $\tau_{\text{sol}}$ are stabilized at $TST^3ST\tau_S=\frac{8}{13}+\frac{i}{13}$ and $\tau_S=i$ respectively. Consequently the alignments of the atmospheric and solar modular form triplets are given by
\begin{eqnarray}
\nonumber Y_{\text{atm}} & = & Y^{(2)}_{\bm{3}}(\langle\tau_{\text{atm}}\rangle=TST^3ST\tau_S=\frac{8}{13}+\frac{i}{13})
\propto\begin{pmatrix}1\\\frac{1}{22\sqrt{2}}\left(11-11\sqrt{5}+\sqrt{1090+58\sqrt{5}}\right)\,e^{\frac{2\pi i}{5}}\\\frac{1}{22\sqrt{2}}\left(-11+11\sqrt{5}+\sqrt{1090+58\sqrt{5}}\right)\,e^{\frac{3\pi i}{5}} \end{pmatrix}\\
\nonumber &&\hskip1.7in\qquad\qquad\qquad  \approx
\begin{pmatrix}
1\\
0.685\,e^{\frac{2\pi i}{5}}\\
1.560\,e^{\frac{3\pi i}{5}}
\end{pmatrix}\,,\\
Y_{\text{sol}} & = & Y^{(4)}_{\bm{3}}(\langle\tau_{\text{sol}}\rangle=\tau_S=i) \propto  \begin{pmatrix}1\\ \frac{1-\sqrt{5}}{2\sqrt{2}}\\ \frac{1-\sqrt{5}}{2\sqrt{2}} \end{pmatrix}\approx\begin{pmatrix}1\\ -0.437\\ -0.437 \end{pmatrix}\,.
\end{eqnarray}
The permutation matrix $P_{\ell}$ is taken to be $P_{\ell}=P_{231}$. The lepton mixing matrix is found to be of the same pattern as Eq.~\eqref{eq:UPMNS-caseII} with $\varphi\approx0.192\pi$ and $\phi \approx 0.316\pi$.

\item{Case IV }

The left-handed leptons $L$ are assigned to be a modular triplet $\bm{3'}$ in this case, the modular transformations of lepton fields take the pattern in Eq.~\eqref{eq:assign-A5-3}. Consequently both $Y_{\rm atm}$ and $Y_{\rm sol}$ are modular triplets in the representation $\bm{3'}$, they are aligned along the following directions,
\begin{eqnarray}
\nonumber Y_{\rm atm} & = & Y_{\bm{3}^{\prime}}^{(2)}\left(\langle\tau_{\rm atm}\rangle=\tau_S=i\right)\propto
\begin{pmatrix}
1\\
\frac{1}{2\sqrt{2}}\left(1-\sqrt{5}-2\sqrt{85-38\sqrt{5}}\right)\\
\frac{1}{2\sqrt{2}}\left(1-\sqrt{5}+2\sqrt{85-38\sqrt{5}}\right)
\end{pmatrix}\approx
\begin{pmatrix}
1\\
-0.558\\
-0.316
\end{pmatrix}\,,\\
\nonumber Y_{\rm sol} & = & Y_{\bm{3}^{\prime}}^{(4)}(\langle\tau_{\rm sol}\rangle=T^3S\tau_{ST}=\frac{7}{2}+i\frac{\sqrt{3}}{2})\propto
\begin{pmatrix}
1\\
\frac{1}{4\sqrt{2}}\left(-3+\sqrt{5}+\sqrt{30-6\sqrt{5}}\right)\,e^{\frac{4 \pi i}{5}}\\
\frac{1}{4\sqrt{2}}\left(3-\sqrt{5}+\sqrt{30-6\sqrt{5}}\right)\,e^{\frac{ \pi i}{5}}
\end{pmatrix}\\
&&\hskip1.9in\quad\qquad  \approx\begin{pmatrix}
1\\
0.585\,e^{\frac{4 \pi i}{5}}\\
0.855\,e^{\frac{ \pi i}{5}}
\end{pmatrix}\,.
\end{eqnarray}
We take the permutation matrix $P_{\ell}=P_{321}$, and the lepton mixing matrix is of the form
\begin{eqnarray}
U_{PMNS}=
\begin{pmatrix}
0.825~&~ \times ~&~ \times\\
0.398~&~ \times ~&~ \times \\
0.400 ~&~ \times ~&~ \times
\end{pmatrix}\,,
\end{eqnarray}
which implies $\varphi\approx 0.191\pi$, $\phi\approx 0.251\pi$.

\item{Case V }

The lepton fields transform as those in Eq.~\eqref{eq:assign-A5-1} under the action of modular symmetry. The complex moduli $\tau_{\text{atm}}$ and $\tau_{\text{sol}}$ are stabilized at $\langle\tau_{\text{atm}}\rangle=ST^2ST^3\tau_S=-\frac{17}{29}+\frac{i}{29}$ and $\langle\tau_{\text{sol}}\rangle=T^4ST^2\tau_{ST}=\frac{7}{2}+i\frac{\sqrt{3}}{6}$. The corresponding alignments of $Y_{\text{atm}}$ and $Y_{\text{sol}}$ are given by
\begin{eqnarray}
\nonumber Y_{\text{atm}} & = & Y^{(2)}_{\bm{3}}(\langle\tau_{\text{atm}}\rangle=ST^2ST^3\tau_S=-\frac{17}{29}+\frac{i}{29})\propto  \begin{pmatrix}1\\\frac{1}{22\sqrt{2}}\left(11+11\sqrt{5}-\sqrt{1090-58\sqrt{5}}\right)\,e^{\frac{4\pi i}{5}}\\\frac{1}{22\sqrt{2}}\left(11+11\sqrt{5}+\sqrt{1090-58\sqrt{5}}\right)\,e^{-\frac{4\pi i}{5}} \end{pmatrix}\,, \\
\nonumber &&\hskip2.0in\qquad\qquad\quad   \approx\begin{pmatrix}1\\ 0.148\,e^{\frac{4\pi i}{5}}\\2.140\,e^{-\frac{4\pi i}{5}} \end{pmatrix}\,, \\
\nonumber Y_{\text{sol}} & = & Y^{(4)}_{\bm{3}}(\langle\tau_{\text{sol}}\rangle=T^4ST^2\tau_{ST}=\frac{7}{2}+i\frac{\sqrt{3}}{6})\propto\begin{pmatrix}1\\\frac{1}{4\sqrt{2}}\left(-3+\sqrt{5}+\sqrt{30-6\sqrt{5}}\right)\,e^{\frac{2 \pi i}{5}}\\\frac{1}{4\sqrt{2}}\left(3-\sqrt{5}+\sqrt{30-6\sqrt{5}}\right)\,e^{\frac{3 \pi i}{5}} \end{pmatrix} \\
&&\hskip2.0in\quad\quad\qquad  \approx\begin{pmatrix}
1\\
0.585\,e^{\frac{2 \pi i}{5}}\\
0.855\,e^{\frac{3 \pi i}{5}} \end{pmatrix}\,.
\end{eqnarray}
The charged lepton permutation matrix should be $P_{\ell}=P_{231}$ in order to match with the experimental data. The lepton mixing matrix takes the following form
\begin{eqnarray}
U_{PMNS}=\begin{pmatrix}0.830 ~&~ \times ~&~ \times \\
0.420  ~&~ \times ~&~ \times \\
0.367  ~&~ \times ~&~ \times
\end{pmatrix}\,.
\end{eqnarray}
Consequently the parameters $\varphi$ and $\phi$ are $\varphi\approx 0.188\pi$, $
\phi\approx0.229\pi$.

\item{Case VI }

The modular transformations of the lepton fields $L$ are given in Eq.~\eqref{eq:assign-A5-3}. Consequently the modular forms $Y_{\rm atm}$ and $Y_{\rm sol}$ transform as $\bm{3'}$, and they are aligned along the following directions,
\begin{small}
\begin{eqnarray}
\nonumber Y_{\rm atm} & = & Y^{(2)}_{\bm{3}^{\prime}}\left(\langle\tau_{\rm atm}\rangle=T^{4}\tau_S=4+i\right)\propto
\begin{pmatrix}
1\\
\frac{1}{2\sqrt{2}}\left(-1+\sqrt{5}+2\sqrt{85-38\sqrt{5}}\right)\,e^{\frac{\pi i}{5}}\\
\frac{1}{2\sqrt{2}}\left(-1+\sqrt{5}-2\sqrt{85-38\sqrt{5}}\right)\,e^{-\frac{\pi i}{5}}
\end{pmatrix}\approx
\begin{pmatrix}
1\\
0.558\, \,e^{\frac{\pi i}{5}}\\
0.316 \,e^{-\frac{\pi i}{5}}
\end{pmatrix}\,,\\
Y_{\rm sol} & = & Y^{(4)}_{\bm{3}^{\prime}}(\langle\tau_{\rm atm}\rangle=T\tau_{ST}=\frac{1}{2}+i\frac{\sqrt{3}}{2}) \propto
\begin{pmatrix}
1\\
\frac{1}{4\sqrt{2}}\left(-3+\sqrt{5}+\sqrt{30-6\sqrt{5}}\right)\,e^{\frac{2\pi i}{5}}\\
\frac{1}{4\sqrt{2}}\left(3-\sqrt{5}+\sqrt{30-6\sqrt{5}}\right)\,e^{\frac{3\pi i}{5}}
\end{pmatrix}\approx
\begin{pmatrix}
1\\
0.585\,\,e^{\frac{2\pi i}{5}}\\
0.855\,e^{\frac{3\pi i}{5}}
\end{pmatrix}\,.~~~~~
\end{eqnarray}
\end{small}
The charged lepton permutation matrix is $P_{\ell}=P_{231}$, and the lepton mixing matrix is determined to be
\begin{eqnarray}
U_{PMNS} =
\begin{pmatrix}
0.825~&~\times ~&~\times \\
0.400~&~\times ~&~\times\\
0.398~&~\times ~&~\times
\end{pmatrix}\,,
\end{eqnarray}
which leads to $\varphi\approx 0.191\pi$, $\phi\approx 0.249\pi$.

\item{Case VII }

The modular transformations of the lepton fields are the same as these in Eq.~\eqref{eq:assign-A5-1}. The atmospheric and solar modular forms $Y_{\text{atm}}$ and $Y_{\text{sol}}$ take the following forms
\begin{eqnarray}
\nonumber Y_{\text{atm}} & = & Y^{(2)}_{\bm{3}}(\langle\tau_{\text{atm}}\rangle=T^4ST\tau_S=\frac{7}{2}+\frac{i}{2})\propto \begin{pmatrix} 1\\\frac{1}{22\sqrt{2}}\left(-11+11\sqrt{5}+\sqrt{1090+58\sqrt{5}}\right)\,e^{-\frac{3\pi i}{5}}\\
\frac{1}{22\sqrt{2}}\left(11-11\sqrt{5}+\sqrt{1090+58\sqrt{5}}\right)\,e^{-\frac{2\pi i}{5}} \end{pmatrix} \\
\nonumber &&\hskip1.5in\qquad\qquad\quad\approx
\begin{pmatrix}
1\\
1.560\,e^{-\frac{3\pi i}{5}}\\
0.685\,e^{-\frac{2\pi i}{5}}
\end{pmatrix} \,,\\
Y_{\text{sol}} & = & Y^{(4)}_{\bm{3}}(\langle\tau_{\text{sol}}\rangle=\tau_{S}=i) \propto \begin{pmatrix}1\\\frac{1-\sqrt{5}}{2\sqrt{2}}\\\frac{1-\sqrt{5}}{2\sqrt{2}}\end{pmatrix} \approx\begin{pmatrix}1
\\ -0.437\\ -0.437 \end{pmatrix}\,.
\end{eqnarray}
The charged lepton permutation matrix is $P_{\ell}=P_{312}$, and the lepton mixing matrix is of the following pattern
\begin{eqnarray}
\label{eq:PMNS-caseVII}U_{PMNS}=\begin{pmatrix}0.824 ~&~ \times ~&~ \times \\
0.474~&~ \times ~&~ \times\\
0.310 ~&~ \times ~&~ \times
\end{pmatrix}\,,
\end{eqnarray}
which implies $\varphi\approx 0.192\pi$ and $\phi\approx 0.184\pi$. Notice that the same modular Littlest seesaw model can be obtained from the alignments of case II by choosing the permutation matrix $P_{\ell}=P_{321}$.

\item{Case VIII }

The modular transformations of the lepton fields are as those in Eq.~\eqref{eq:assign-A5-3}. The atmospheric and solar moduli are stabilized at $\langle\tau_{\rm atm}\rangle=\tau_S=i$ and $\langle\tau_{\rm sol}\rangle=T^3ST^2ST^4\tau_{ST}=\frac{63}{26}+i\frac{\sqrt{3}}{78}$ respectively. The VEVs of the modular forms $Y_{\rm atm}$ and $Y_{\rm sol}$ in the representation $\bm{3'}$ are determined to be
\begin{eqnarray}
\nonumber Y_{\rm atm} & = &  Y^{(2)}_{\bm{3}^{\prime}}\left(\langle\tau_{\rm atm}\rangle=\tau_S=i\right) \propto
\begin{pmatrix}
1\\
\frac{1}{2\sqrt{2}}\left(1-\sqrt{5}-2\sqrt{85-38\sqrt{5}}\right)\\
\frac{1}{2\sqrt{2}}\left(1-\sqrt{5}+2\sqrt{85-38\sqrt{5}}\right)
\end{pmatrix} \approx
\begin{pmatrix}
1\\
-0.558\\
-0.316
\end{pmatrix}\,,\\
\nonumber Y_{\rm sol} & = & Y^{(4)}_{\bm{3}^{\prime}}(\langle\tau_{\rm sol}\rangle=T^3ST^2ST^4\tau_{ST}=\frac{63}{26}+i\frac{\sqrt{3}}{78}) \propto
\begin{pmatrix}
1\\
\frac{1}{4\sqrt{2}}\left(3-\sqrt{5}+\sqrt{30-6\sqrt{5}}\right)\,e^{-\frac{3 \pi i}{5}}\\
\frac{1}{4\sqrt{2}}\left(-3+\sqrt{5}+\sqrt{30-6\sqrt{5}}\right)\,e^{-\frac{2 \pi i}{5}}
\end{pmatrix}\\
&&\hskip2.0in\qquad\qquad\qquad \approx
\begin{pmatrix}
1\\
0.855\,e^{-\frac{3 \pi i}{5}}\\
0.585\,e^{-\frac{2 \pi i}{5}}
\end{pmatrix}\,.
\end{eqnarray}
The current neutrino oscillation data can be accommodated for the permutation $P_{\ell}=P_{321}$, and the fixed column of the lepton mixing matrix is given by
\begin{eqnarray}
U_{PMNS}=
\begin{pmatrix}
0.824~&~\times ~&~\times \\
0.288~&~\times ~&~\times \\
0.487~&~\times ~&~\times
\end{pmatrix}\,,
\end{eqnarray}
which corresponds to $\varphi\approx 0.192\pi$, $\phi\approx0.330\pi$.

\item{Case IX }

The modular transformations of lepton fields are given in Eq.~\eqref{eq:assign-A5-1}. The modular symmetry is broken down by the moduli VEVs $\langle\tau_{\text{atm}}\rangle=TST^3ST\tau_S=\frac{8}{13}+\frac{i}{13}$ and $\langle\tau_{\text{sol}}\rangle=\tau_{S}=i$ in this case. The modular triplets $Y_{\text{atm}}$ and $Y_{\text{sol}}$ are found to be
\begin{eqnarray}
\nonumber Y_{\text{atm}} & = & Y^{(2)}_{\bm{3}}(\langle\tau_{\text{atm}}\rangle=TST^3ST\tau_S=\frac{8}{13}+\frac{i}{13})
\propto \begin{pmatrix}1\\\frac{1}{22\sqrt{2}}\left(11-11\sqrt{5}+\sqrt{1090+58\sqrt{5}}\right)\,e^{\frac{2\pi i}{5}}\\
\frac{1}{22\sqrt{2}}\left(-11+11\sqrt{5}+\sqrt{1090+58\sqrt{5}}\right)\,e^{\frac{3\pi i}{5}} \end{pmatrix}\\
\nonumber &&\hskip2.0in\qquad\quad\quad   \approx
\begin{pmatrix}
1\\
0.685\,e^{\frac{2\pi i}{5}}\\
1.560\,e^{\frac{3\pi i}{5}}
\end{pmatrix}\,,\\
Y_{\text{sol}} & = & Y^{(4)}_{\bm{3}}(\langle\tau_{\text{sol}}\rangle=\tau_{S}=i)\propto \begin{pmatrix}1\\ \frac{1-\sqrt{5}}{2\sqrt{2}}\\\frac{1-\sqrt{5}}{2\sqrt{2}}\end{pmatrix}
\approx\begin{pmatrix}1
\\ -0.437\\-0.437 \end{pmatrix}\,.
\end{eqnarray}
The experimental data can be well accommodated of $P_{\ell}=P_{321}$, and the first column of the lepton mixing matrix is found to be of the same form as that of case VII shown in Eq.~\eqref{eq:PMNS-caseVII}. We note that the same modular Littlest seesaw model can be reached from the alignments of case III for the charged lepton permutation $P_{\ell}=P_{321}$.

\end{itemize}

\subsubsection{$\gamma_{\ell}=T$}

In this section, we shall study the modular Littlest seesaw models in which the $A_5$ or $A'_5$ modular symmetry is broken down to the $Z^T_5$ subgroup in the charged lepton sector with $\langle\tau_{\ell}\rangle=i\infty$. As a consequence, the charged lepton diagonalization matrix $U_{\ell}$ is a unit matrix up to phases and permutations column vectors.

\begin{figure}[t!]
\begin{center}
\includegraphics[width=0.98\textwidth]{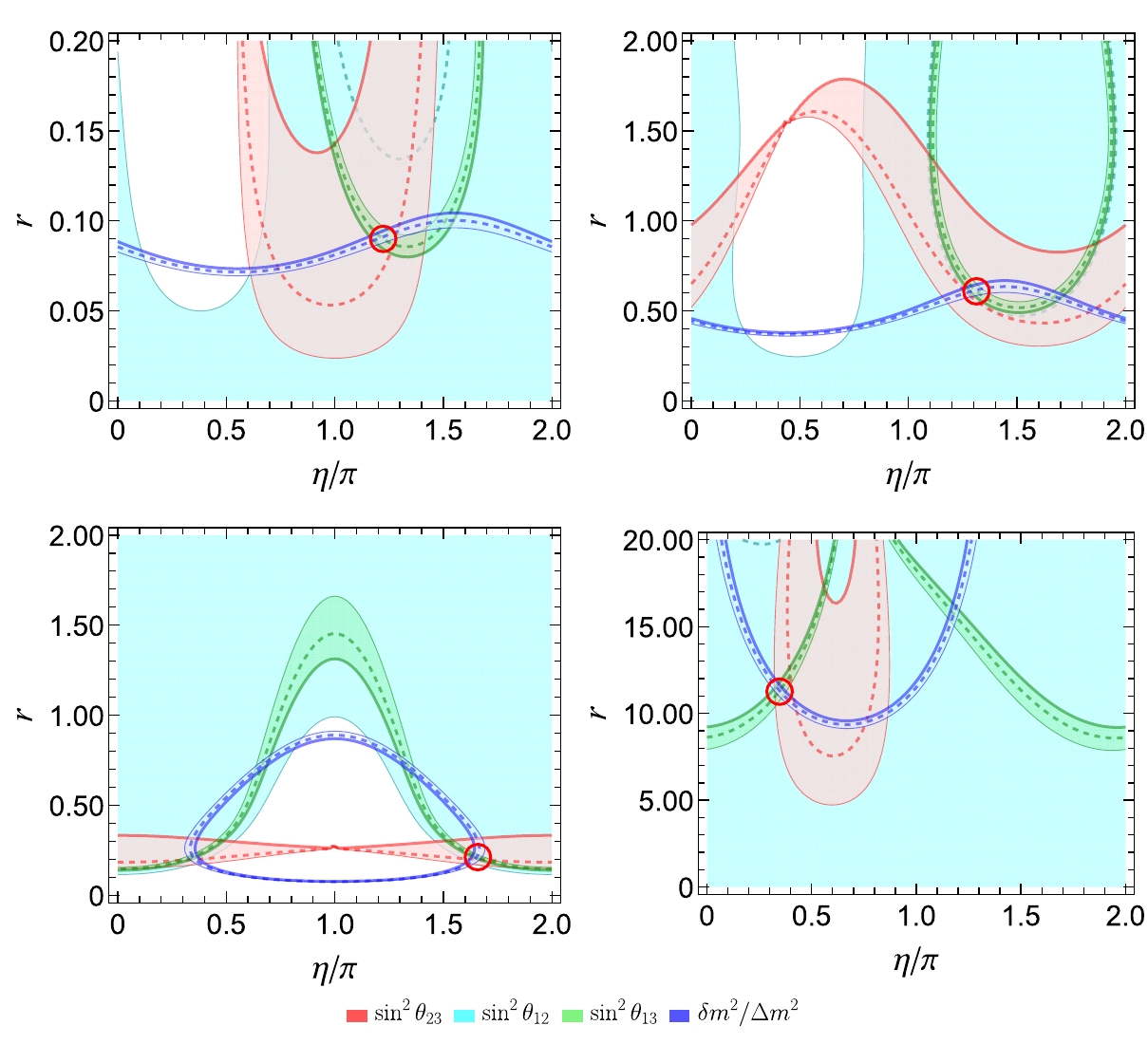}
\end{center}
\caption{\label{fig:contour-MLS-A5p} The contour plots of $\sin^2\theta_{12}$, $\sin^2\theta_{13}$, $\sin^{2}\theta_{23}$ and $\delta m^2/\Delta m^2$ in the $\eta/\pi-r$ plane for the four possible modular Littlest seesaw models in $A^\prime_5$ modular symmetry. The panels in the upper-left, upper-right, lower-left, lower-right are for case I, case II, case X and case XI respectively. The cyan, red, green and blue areas denote the $3\sigma$ regions of $\sin^{2}\theta_{23}$, $\sin^{2}\theta_{13}$ and $m_{2}^{2}/m_{3}^{2}$ respectively. The solid lines denote the $3\sigma$ upper bounds, the thin lines denote the $3\sigma$ lower bounds
and the dashed lines refer to their best fit values, as adopted from latest neutrino global fit~\cite{Capozzi:2025wyn}. The red circle indicates the phenomenologically viable region of parameter space.}
\end{figure}

\begin{itemize}

\item{Case X  }

The lepton fields transform in the manner shown in Eq.~\eqref{eq:assign-A5-3} under the $A_5$ or $A'_5$ modular symmetry. The VEVs of the atmospheric and solar moduli are $\langle\tau_{\rm atm}\rangle=\tau_S=i$ and $\langle\tau_{\rm sol}\rangle=T^2ST^4\tau_{ST}=\frac{45}{26}+i\frac{\sqrt{3}}{26}$ respectively. Accordingly the alignments of the modular forms $Y_{\rm atm}$ and $Y_{\rm sol}$ are given by
\begin{small}
\begin{eqnarray}
\nonumber Y_{\rm atm} & = & Y^{(2)}_{\bm{3}^{\prime}}(\langle\tau_{\rm atm}\rangle=\tau_S=i)\propto \begin{pmatrix}
1\\
\frac{1}{2\sqrt{2}}\left(1-\sqrt{5}-2\sqrt{85-38\sqrt{5}}\right)\\
\frac{1}{2\sqrt{2}}\left(1-\sqrt{5}+2\sqrt{85-38\sqrt{5}}\right)
\end{pmatrix}\approx\begin{pmatrix}
1\\
-0.558\\
-0.316
\end{pmatrix} \,,\\
Y_{\rm sol} & = & Y^{(4)}_{\bm{3}^{\prime}}(\langle\tau_{\rm sol}\rangle=T^2ST^4\tau_{ST}=\frac{45}{26}+i\frac{\sqrt{3}}{26})\propto
\begin{pmatrix}
1\\
-\frac{1}{4\sqrt{2}}\left(3+\sqrt{5}+\sqrt{30+6\sqrt{5}}\right)\\
-\frac{1}{4\sqrt{2}}\left(3+\sqrt{5}-\sqrt{30+6\sqrt{5}}\right)
\end{pmatrix}  \approx
\begin{pmatrix}
1\\
-2.090\\
0.239
\end{pmatrix}\,,~~~~~~
\end{eqnarray}
\end{small}
Good agreement with current neutrino oscillation data can be achieved for the charged lepton permutation $P_{\ell}=P_{312}$. The lepton mixing matrix turns out to be of the following form
\begin{eqnarray}
U_{PMNS}=
\begin{pmatrix}
0.845~&~\times ~&~\times\\
0.306~&~\times ~&~\times\\
0.438~&~\times ~&~\times
\end{pmatrix}\,.
\end{eqnarray}
Hence the parameters $\varphi$ and $\phi$ in Eq.~\eqref{eq:parameterization-fixed-colum} are $\varphi\approx 0.179\pi$, $\phi\approx0.306\pi$. The contour plots of $\sin^2\theta_{12}$, $\sin^2\theta_{13}$, $\sin^{2}\theta_{23}$ and $\delta m^2/\Delta m^2$ are shown in figure~\ref{fig:contour-MLS-A5p}. The values of $\eta$ and $r$ are subject to strong constraint from $\sin^2\theta_{13}$ and $\delta m^2/\Delta m^2$, and they should be in the region around the point $(1.659\pi, 0.212)$ in order to match with the experimental data.

\item{ Case XI }

The modular weights and transformations of lepton fields under $A_5$ or $A'_5$ modular symmetry is listed in Eq.~\eqref{eq:assign-A5-3}. The values of the modular forms $Y_{\rm atm}$ and $Y_{\rm sol}$ are of the following form:
\begin{eqnarray}
\nonumber Y_{\rm atm} & = & Y^{(2)}_{\bm{3}^{\prime}}(\langle\tau_{\rm atm}\rangle=TST^4\tau_S=\frac{13}{17}+\frac{i}{17})\propto
\begin{pmatrix}
1\\
\frac{1}{2\sqrt{2}}\left(-1-\sqrt{5}+2\sqrt{85+38\sqrt{5}}\right)
\,e^{\frac{\pi i}{5}}\\
\frac{1}{2\sqrt{2}}\left(1+\sqrt{5}+2\sqrt{85+38\sqrt{5}}\right)\,e^{\frac{4\pi i}{5}}
\end{pmatrix}\\
\nonumber &&\hskip2.0in\quad\quad\quad \approx
\begin{pmatrix}
1\\
8.075\,e^{\frac{\pi i}{5}}\\
10.363\,e^{\frac{4\pi i}{5}}
\end{pmatrix}\,,\\
Y_{\rm sol} & = & Y^{(4)}_{\bm{3}^{\prime}}(\langle\tau_{\rm sol}\rangle=ST^3\tau_{ST}=-\frac{5}{14}+i\frac{\sqrt{3}}{14})\propto
\begin{pmatrix}
1\\
\frac{1}{4\sqrt{2}}\left(-3-\sqrt{5}+\sqrt{30+6\sqrt{5}}\right)\\
-\frac{1}{4\sqrt{2}}\left(3+\sqrt{5}+\sqrt{30+6\sqrt{5}}\right)
\end{pmatrix}\approx
\begin{pmatrix}
1\\
0.239\\
-2.090
\end{pmatrix}\,.~~~~~
\end{eqnarray}
The permutation matrix is taken to be $P_{\ell}=P_{132}$, and the lepton mixing matrix is of the form
\begin{eqnarray}
\label{eq:UPMNS-caseXI}U_{PMNS}=
\begin{pmatrix}
0.810 ~&~\times ~&~\times\\
0.436 ~&~\times ~&~\times\\
0.392 ~&~\times ~&~\times
\end{pmatrix}\,,
\end{eqnarray}
which leads to $\varphi\approx 0.199\pi$ and $\phi\approx0.233\pi$. We display the contour plots of $\sin^2\theta_{12}$, $\sin^2\theta_{13}$, $\sin^{2}\theta_{23}$ and $\delta m^2/\Delta m^2$ in figure~\ref{fig:contour-MLS-A5p}.

\item{Case XII }

The left-handed leptons $L$ is assigned to a triplet $\bm{3'}$ of $A_5$ or $A'_5$, the modular weights and representations of $N^{c}_{\rm atm}$ and $N^{c}_{\rm sol}$ are given in Eq.~\eqref{eq:assign-A5-3}. The moduli
$\tau_{\rm atm}$ and $\tau_{\rm sol}$ are stabilized at $\langle\tau_{\rm atm}\rangle=ST\tau_S=-\frac{1}{2}+\frac{i}{2}$ and $\langle\tau_{\rm sol}\rangle=TST^3\tau_{ST}=\frac{9}{14}+i\frac{\sqrt{3}}{14}$ respectively, The modular triplets $Y_{\rm atm} $ and $Y_{\rm sol}$ are aligned along the following directions  \begin{eqnarray}
\nonumber Y_{\rm atm} & = & Y^{(2)}_{\bm{3}^{\prime}}(\langle\tau_{\rm atm}\rangle=ST\tau_S=-\frac{1}{2}+\frac{i}{2})\propto
\begin{pmatrix}
1\\
\frac{1}{2\sqrt{2}}\left(-1-\sqrt{5}+2\sqrt{85+38\sqrt{5}}\right)
\,e^{\frac{3\pi i}{5}}\\
\frac{1}{2\sqrt{2}}\left(1+\sqrt{5}+2\sqrt{85+38\sqrt{5}}\right)\,e^{\frac{2\pi i}{5}}
\end{pmatrix}\\
\nonumber &&\hskip1.5in\qquad\qquad\quad   \approx
\begin{pmatrix}
1\\
8.075\,e^{\frac{3\pi i}{5}}\\
10.363\,e^{\frac{2\pi i}{5}}
\end{pmatrix}\,,\\
\nonumber Y_{\rm sol} & = & Y^{(4)}_{\bm{3}^{\prime}}(\langle\tau_{\rm sol}\rangle=TST^3\tau_{ST}=\frac{9}{14}+i\frac{\sqrt{3}}{14}) \propto
\begin{pmatrix}
1\\
\frac{1}{4\sqrt{2}}\left(-3-\sqrt{5}+\sqrt{30+6\sqrt{5}}\right)\,e^{\frac{4 \pi i}{5}}\\
\frac{1}{4\sqrt{2}}\left(3+\sqrt{5}+\sqrt{30+6\sqrt{5}}\right)\,e^{\frac{\pi i}{5}}
\end{pmatrix}\\
&&\hskip2.0in\quad\quad\qquad   \approx
\begin{pmatrix}
1\\
0.239\,e^{\frac{4 \pi i}{5}}\\
2.090\,e^{\frac{\pi i}{5}}
\end{pmatrix}\,.
\end{eqnarray}
The experimental data can be accommodated for $P_{\ell}=\mathbb{1}_3$, and the lepton mixing matrix takes the following form,
\begin{eqnarray}
\label{eq:UPMNS-caseXII}U_{PMNS}=
\begin{pmatrix}
0.810 ~&~\times ~&~\times\\
0.392 ~&~\times ~&~\times \\
0.436 ~&~\times ~&~\times
\end{pmatrix}\,,
\end{eqnarray}
which gives to $\varphi\approx 0.199\pi$ and $\phi\approx0.267\pi$. In the limit of sequential dominance where the atmospheric neutrino $N^{c}_{\text{atm}}$ gives the dominant contribution to the light neutrino mass over $N^{c}_{\text{sol}}$, one column of the lepton mixing matrix is proportional to $P_{132}Y^{*}_{\text{atm}}\propto(0.0759, 0.787e^{-2\pi i/5}, 0.613e^{-3\pi i/5})^T$, where $P_{132}$ arises from the CG coefficient in the working basis. As a result, it gives $\theta_{13}\approx4.353^\circ$ and $\theta_{23}\approx52.075^{\circ}$ at leading order. The contribution of $N^{c}_{\text{sol}}$ would generate the solar mixing angle $\theta_{12}$ and introduces small corrections to the leading order values of $\theta_{13}$ and $\theta_{23}$.

\item{Case XIII  }

The modular transformation of the lepton fields is of the pattern in Eq.~\eqref{eq:assign-A5-3}. The VEVs of the modular forms $Y_{\rm atm} $ and $Y_{\rm sol}$ are given by
\begin{eqnarray}
\nonumber Y_{\rm atm} & = & Y^{(2)}_{\bm{3}^{\prime}}(\langle\tau_{\rm atm}\rangle=\tau_S=i)\propto
\begin{pmatrix}
1\\
\frac{1}{2\sqrt{2}}\left(1-\sqrt{5}-2\sqrt{85-38\sqrt{5}}\right)
\\
\frac{1}{2\sqrt{2}}\left(1-\sqrt{5}+2\sqrt{85-38\sqrt{5}}\right)
\end{pmatrix}\approx
\begin{pmatrix}
1\\
-0.558\\
-0.316
\end{pmatrix}\,,\\
Y_{\rm sol} & = & Y^{(4)}_{\bm{3}^{\prime}}(\langle\tau_{\rm sol}\rangle=T^3ST^2\tau_{ST}=\frac{5}{2}+i\frac{\sqrt{3}}{6}) \propto
\begin{pmatrix}
1\\
-\frac{1}{4\sqrt{2}}\left(3+\sqrt{5}+\sqrt{30+6\sqrt{5}}\right)\\
\frac{1}{4\sqrt{2}}\left(-3-\sqrt{5}+\sqrt{30+6\sqrt{5}}\right)
\end{pmatrix}\approx
\begin{pmatrix}
1\\
-2.090\\
0.239
\end{pmatrix}\,.~~~~~
\end{eqnarray}
The charged lepton permutation matrix is taken to be $P_{\ell}=P_{213}$, and the lepton mixing matrix is determined to be of the following form
\begin{eqnarray}
U_{PMNS}=
\begin{pmatrix}
0.845 ~&~\times ~&~\times\\
0.438~&~\times ~&~\times\\
0.306~&~\times ~&~\times
\end{pmatrix}\,,
\end{eqnarray}
which implies $\varphi \approx0.179\pi$, $\phi\approx0.194\pi$. We notice that this modular Littlest seesaw model can also be obtained from the alignment of case X by choosing the permutation $P_{\ell}=P_{213}$.

\item{Case XIV }

In the last case, the assignment of the lepton fields under modular symmetry is given in Eq.~\eqref{eq:assign-A5-3}. The stabilizded values of the moduli are $\langle\tau_{\rm atm}\rangle=ST^4\tau_S=-\frac{4}{17}+\frac{i}{17}$ and $\langle\tau_{\rm sol}\rangle=ST^3\tau_{ST}=-\frac{5}{14}+i\frac{\sqrt{3}}{14}$. Thus the alignments of the modular forms $Y_{\rm atm}$ and $Y_{\rm sol}$ are determined as
\begin{eqnarray}
\nonumber Y_{\rm atm} & = & Y^{(2)}_{\bm{3}^{\prime}}(\tau_{\rm atm}=ST^4\tau_S=-\frac{4}{17}+\frac{i}{17})\propto
\begin{pmatrix}
1\\
\frac{1}{2\sqrt{2}}\left(-1-\sqrt{5}+2\sqrt{85+38\sqrt{5}}\right)\,e^{-\frac{3 \pi i}{5}}
\\
\frac{1}{2\sqrt{2}}\left(1+\sqrt{5}+2\sqrt{85+38\sqrt{5}}\right)\,e^{-\frac{2 \pi i}{5}}
\end{pmatrix}\\
\nonumber &&\hskip2.0in\quad\quad\quad   \approx
\begin{pmatrix}
1\\
8.075\,e^{-\frac{3 \pi i}{5}}\\
10.363\,e^{-\frac{2 \pi i}{5}}
\end{pmatrix}\,,\\
Y_{\rm sol} & = & Y^{(4)}_{\bm{3}^{\prime}}(\tau_{\rm sol}=ST^3\tau_{ST}=-\frac{5}{14}+i\frac{\sqrt{3}}{14})\propto
\begin{pmatrix}
1\\
\frac{1}{4\sqrt{2}}\left(-3-\sqrt{5}+\sqrt{30+6\sqrt{5}}\right)\\
-\frac{1}{4\sqrt{2}}\left(3+\sqrt{5}+\sqrt{30+6\sqrt{5}}\right)
\end{pmatrix}\approx
\begin{pmatrix}
1\\
0.239\\
-2.090
\end{pmatrix}\,.~~~~
\end{eqnarray}
In order to match with the neutrino oscillation data,
$P_{\ell}=\mathbb{1}_3$ is adopted, Then the lepton mixing matrix is of the following pattern
\begin{eqnarray}
U_{PMNS}=
\begin{pmatrix}
0.807 ~&~\times ~&~\times\\
0.347 ~&~\times ~&~\times\\
0.477 ~&~\times ~&~\times
\end{pmatrix}\,,
\end{eqnarray}
which leads to $\varphi\approx0.201\pi$, $\phi\approx0.300\pi$.

\end{itemize}

\begin{figure}[hptb]
\begin{center}
\includegraphics[width=0.95\textwidth]{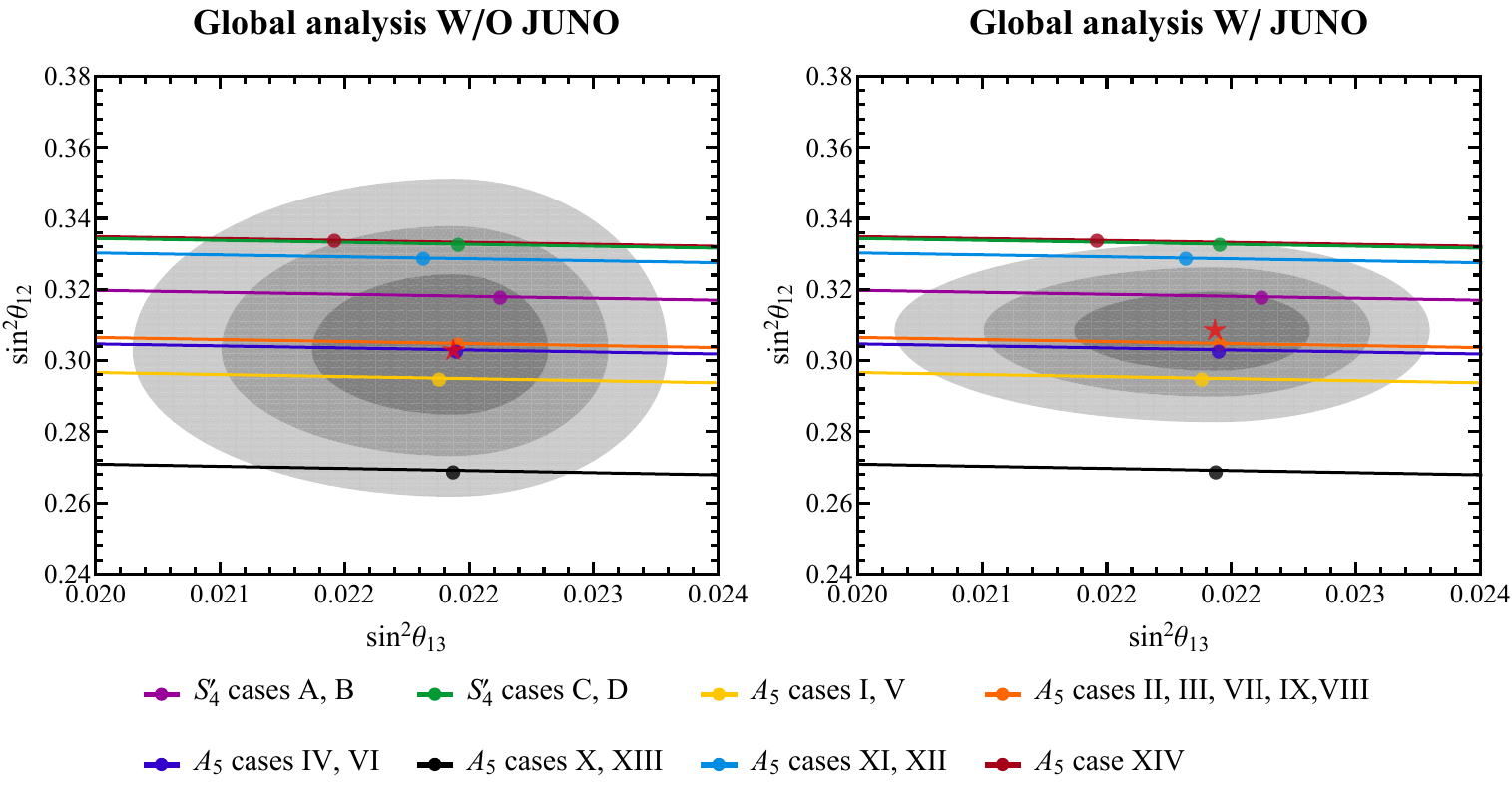}
\caption{\label{fig:theta12-13-corre-Bari}Correlation between $\sin^2\theta_{12}$ and $\sin^2\theta_{13}$ in the modular Littlest seesaw models based on $S'_4$ and $A_5$ modular symmetries, and the small dots indicate the best fit values of the models. The gray regions denote the $1\sigma$, $2\sigma$ and $3\sigma$ regions of $\sin^2\theta_{12}$ and $\sin^2\theta_{13}$ for the normal mass hierarchy~\cite{Capozzi:2025wyn}, and the red pentagram stands for the best-fit point of the neutrino global analysis~\cite{Capozzi:2025wyn}. The left and right panels are for the global analysis without and with JUNO respectively. }
\end{center}
\end{figure}

\begin{figure}[hptb]
\begin{center}
\includegraphics[width=0.95\textwidth]{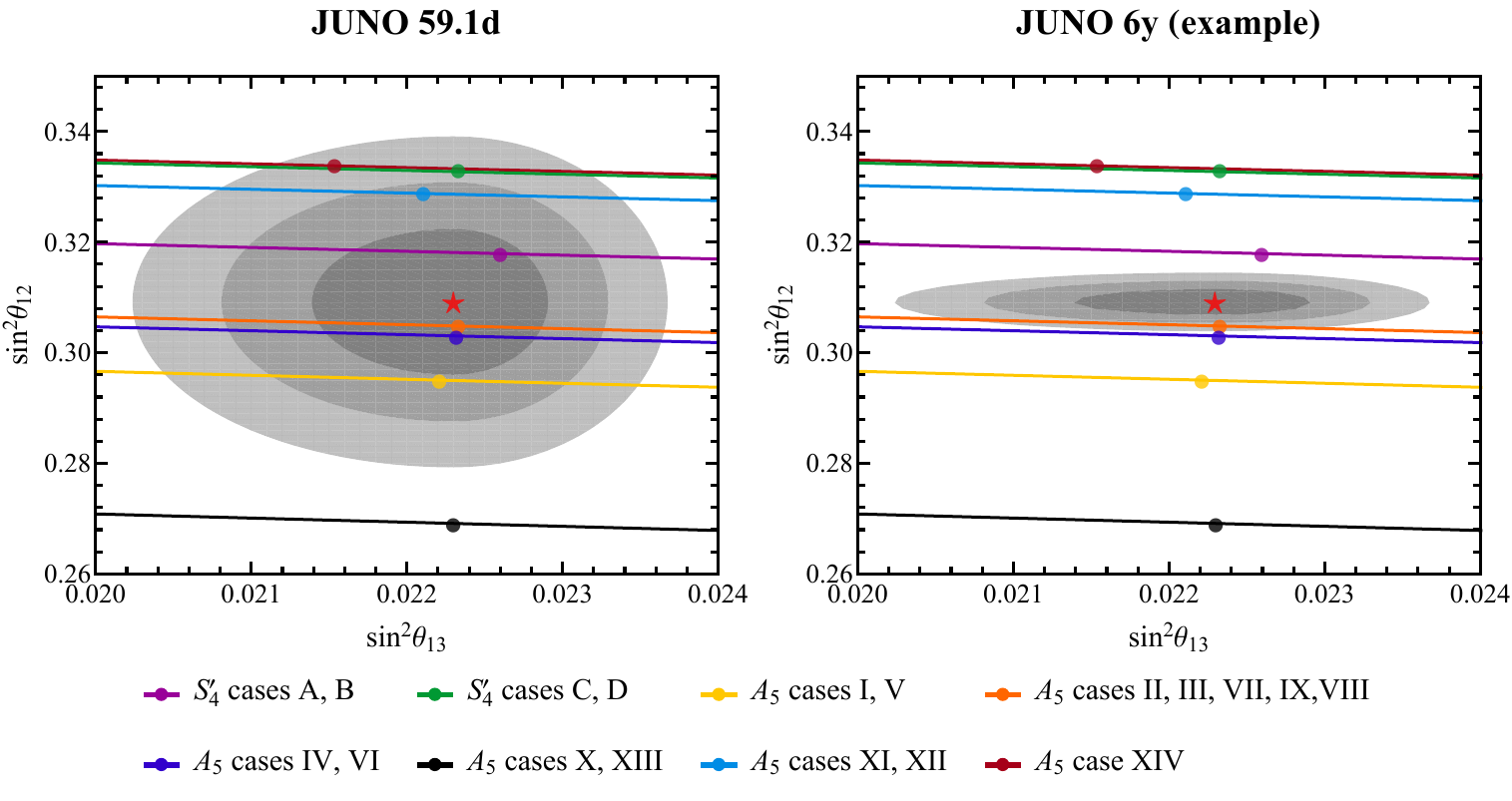}
\caption{\label{fig:theta12-13-corre-JUNO}Correlation between $\sin^2\theta_{12}$ and $\sin^2\theta_{13}$ in the modular Littlest seesaw models based on $S'_4$ and $A_5$ modular symmetries, and the small dots indicate the best fit values of the models. The left and right panels are for JUNO 59.1 days and JUNO 6 years of data taking respectively. The central value and uncertainty of $\sin^2\theta_{12}$ are taken from the latest JUNO measurement~\cite{JUNO:2025gmd}, its $1\sigma$ uncertainty is expected to be reduced to $0.5\%$ after six years of data collection~\cite{JUNO:2022mxj}. The gray regions denote the $1\sigma$, $2\sigma$ and $3\sigma$ regions of $\sin^2\theta_{12}$ and $\sin^2\theta_{13}$ for the normal mass hierarchy~\cite{JUNO:2025gmd,Capozzi:2025wyn}, and the red pentagram stands for their central values~\cite{JUNO:2025gmd,Capozzi:2025wyn}.  }
\end{center}
\end{figure}

\section{\label{sec:pheno}Phenomenological implications of JUNO and future long baseline neutrino oscillation experiments}

As shown in sections~\ref{sec:MLS-N4} and \ref{sec:MLS-N5}, the modular Littlest seesaw models are very predictive, the light neutrino mass matrix depends on only three free parameters $m_a$, $m_s$ and $\eta$ to explain all the light neutrino masses, mixing angles and CP violation phases. The experimental data can only be accommodated in a small region in the $\eta-r$ plane with $r\equiv m_s/m_a$, as can be seen from figures~\ref{fig:mixing-pars-S4pLSS} and~\ref{fig:contour-MLS-A5p}. The best fit values of $m_a$, $m_s/m_a$, $\eta$ and the corresponding predictions for neutrino masses and mixing parameters are listed in tables~\ref{tab:BF-results-S4p} and \ref{tab:BF-result-N5}.  It is notable that the first column of the lepton mixing matrix is fixed by modular symmetry. In the $S'_4$ modular symmetry, the lepton mixing matrix is the TM1 pattern shown in Eq.~\eqref{eq:PMNS-caseA} for case A and case B, and it takes the form in Eq.~\eqref{eq:PMNS-caseC} and Eq.~\eqref{eq:PMNS-caseD} for cases C and D respectively. The fixed first column can be parameterized as $(\cos\varphi, \cos\phi\sin\varphi, \sin\phi\sin\varphi)^{T}$, where the values of $\phi$ and $\varphi$ are listed in table~\ref{tab:fixed-column-parameters} for the modular Littlest seesaw models based on $A_5$. Notice that $(\varphi/\pi, \phi/\pi)\simeq(0.196, 0.250)$, $(0.201, 0.243)$, and $(0.201, 0.257)$  for cases A and B, case C and case D respectively. As a consequence, the solar angle $\theta_{12}$ and reactor angle $\theta_{13}$ are related as $\cos^2\theta_{12}\cos^2\theta_{13}=\cos^2\varphi$ shown in Eq.~\eqref{eq:cosdeltaCP-sum-rule}. Given the precisely measured value of $\theta_{13}$ by reactor neutrino oscillation experiments, we can get sharp prediction for the solar mixing angle $\sin^2\theta_{12}$.

Recently the Jiangmen Underground Neutrino Observatory (JUNO), using the first 59.1 days of data, reported the high-precision determination of the solar oscillation parameters $\sin^2\theta_{12}$ and  $\Delta m^2_{21}$ for NO neutrino mass spectrum~\cite{JUNO:2025gmd}. The best-fit values and the $1\sigma$ uncertainties are given as
\begin{equation}\label{eq:s12sq_JUNO_59day}
\sin^2\theta_{12}=0.3092\pm0.0087,~~~ \Delta m^2_{21}=(7.50\pm0.12)\times 10^{-5} \text{eV}^2\,.
\end{equation}
It is expected that the JUNO can further improve the precision of $\Delta m^2_{21}$ and $\sin^2\theta_{12}$ to a world-leading precision of $0.3\%$ and $0.5\%$ respectively after six years of data collection~\cite{JUNO:2022mxj}. We plot the correlation between $\theta_{12}$ and $\theta_{13}$ in figures~\ref{fig:theta12-13-corre-Bari} and~\ref{fig:theta12-13-corre-JUNO}, in comparison with the latest neutrino global analysis by Italian Bari group and JUNO measurement. The best fit value of $\sin^2\theta_{12}$ and $\sin^2\theta_{13}$ predicted by the $S'_4$ and $A_5$ modular Littlest seesaw models are shown in small dots. We see that case X and case XIII of $A_5$ are excluded by the JUNO's first result, because their prediction for solar angle $\sin^2\theta_{12}\approx0.269$ is too small and it is outside the $3\sigma$ range preferred by JUNO. It is notable that the uncertainty of $\sin^2\theta_{12}$ would be reduced significantly after six years running of JUNO. If the current best fit value of $\sin^2\theta_{12}$ would not change significantly, only the cases II, III, VII, IX, VIII are marginally compatible with future prospective JUNO precision data, and all the other modular Littlest seesaw models are expected to be ruled out, as can be seen from figure~\ref{fig:theta12-13-corre-JUNO}. Hence JUNO can provide a crucial tests of the modular Littlest seesaw models in near future.

\begin{figure}[t!]
\begin{center}
\includegraphics[width=0.98\textwidth]{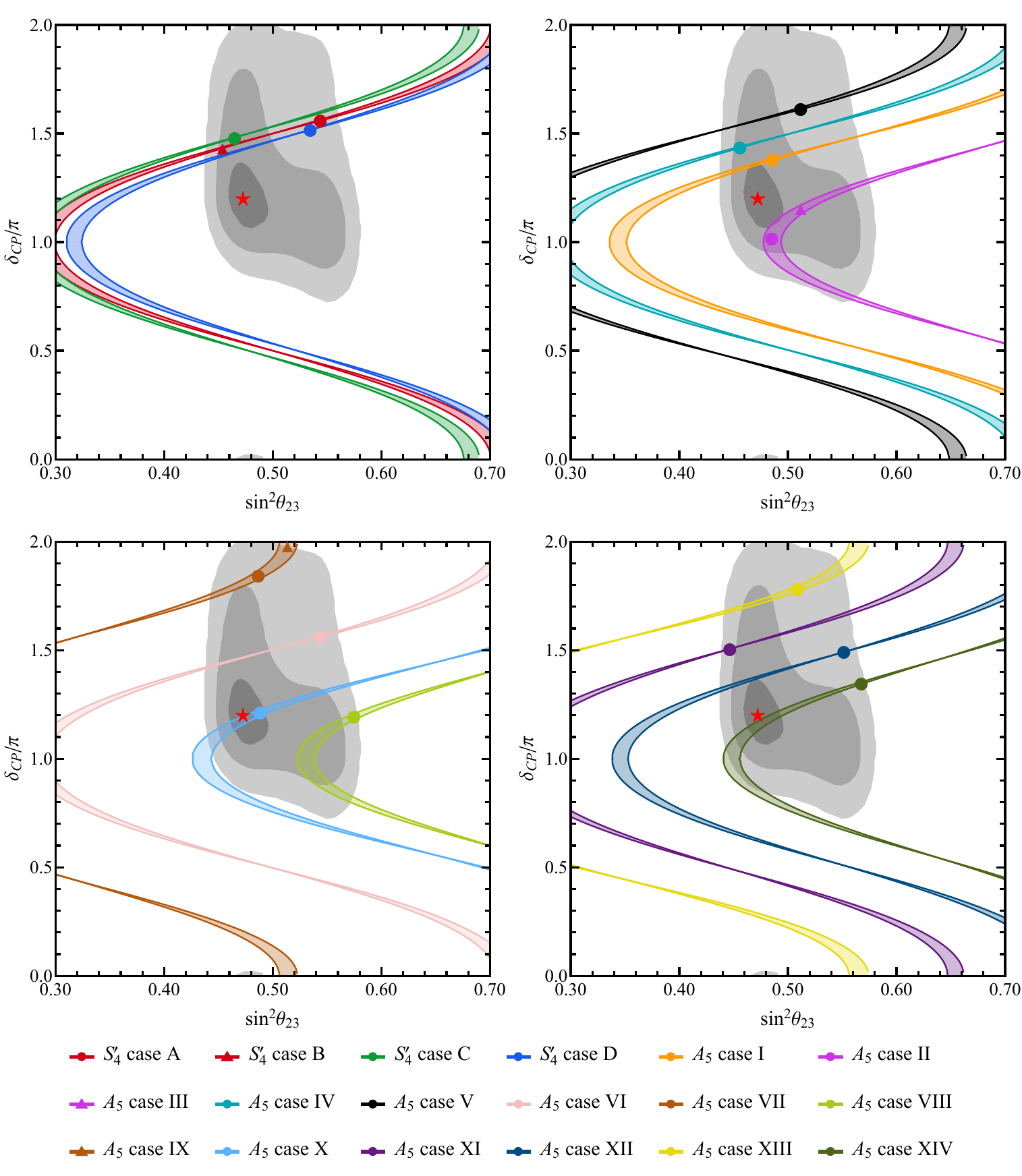}
\caption{\label{fig:deltaCP-theta23-corre-Bari2}Correlation between $\delta_{CP}$ and $\sin^2\theta_{23}$ in the modular Littlest seesaw models based on $S'_4$ and $A_5$ modular symmetries, and the small dots and triangles indicate the best fit values of the models. The top-left panel is for the modular Littlest seesaw models based on $S'_4$ modular symmetry, and the other three panels are for $A_5$ modular group. The gray regions denote the $1\sigma$, $2\sigma$ and $3\sigma$ regions of $\delta_{CP}$ and $\sin^2\theta_{23}$ for the normal mass hierarchy~\cite{Capozzi:2025wyn}, and the red pentagram stands for the best-fit point of the neutrino global analysis~\cite{Capozzi:2025wyn}.}
\end{center}
\end{figure}

From the first column of lepton mixing matrix fixed by modular symmetry, one can express $\cos\delta_{CP}$ in terms of the mixing angles $\theta_{13}$ and $\theta_{23}$, as shown in Eq.~\eqref{eq:cosdeltaCP-sum-rule}. The variation of $\delta_{CP}$ with respect to $\sin^2\theta_{23}$ for every modular Littlest seesaw model is displayed in figure~\ref{fig:deltaCP-theta23-corre-Bari2}, where the width of each curve arises from varying $\theta_{13}$ over its $3\sigma$ allowed range~\cite{Capozzi:2025wyn}. The small dots and triangles indicate the best fit values of $\delta_{CP}$ and $\sin^2\theta_{23}$ predicted by the model. We note that the lepton mixing is the TM1 pattern in both case A and case B, consequently the same sum rule for $\cos\delta_{CP}$ in Eq.~\eqref{eq:cos-deltaCP-sumR-AB} is reached. Nevertheless the best fit values of $\delta_{CP}$ and $\sin^2\theta_{23}$ in these two cases are different, they are represented by red dot and red triangle respectively in the upper-left panel of figure~\ref{fig:deltaCP-theta23-corre-Bari2}. Analogously the lepton mixing matrices of case II and case III for $A_5$ modular symmetry have a common first column, thus they give rise to the same sum rule of $\cos\delta_{CP}$ represented by the purple band in the upper-right panel of figure~\ref{fig:deltaCP-theta23-corre-Bari2}. Similarly case VII and case IX of $A_5$ also predict the same sum rule and an identical first column of the mixing matrix. From figures~\ref{fig:mixing-pars-S4pLSS} and~\ref{fig:contour-MLS-A5p}, we can see that the modular Littlest seesaw models are very predictive so that $\sin^2\theta_{23}$ as well as $\delta_{CP}$ can only vary in small region around the best fit point.

The forthcoming long baseline neutrino oscillation experiments DUNE~\cite{DUNE:2020ypp} and T2HK~\cite{Hyper-KamiokandeProto-:2015xww,Hyper-Kamiokande:2018ofw} aim at first precision measurement of $\delta_{CP}$ and determining the neutrino mass ordering and the octant of the atmospheric mixing angle $\theta_{23}$, by the precise measurement of both the appearance oscillation mode $\nu_{\mu}\rightarrow \nu_e$ and disappearance mode $\nu_{\mu}\rightarrow \nu_{\mu}$ as well as their CP conjugates. The projected sensitivities on
$\sin^2\theta_{23}$ and $\delta_{CP}$ strongly depend on their true values.  For $\theta_{23}=40^{\circ}$ or $50^{\circ}$, the precision on $\theta_{23}$ is about $0.2^{\circ}$ ($0.13^{\circ}$) for DUNE (T2HK). The precision gets worse at $\theta_{23}=45^{\circ}$ and it increases to approximately $2^{\circ}$ ($0.95^{\circ}$) for DUNE (T2HK). Regarding the CP violation, the combination of DUNE and T2HK is expected to measure $\delta_{CP}$ with $1\sigma$ uncertainty between $4.5^{\circ}$ and $11^{\circ}$ for all values of $\delta_{CP}$ after 10 years of running in parallel~\cite{Ballett:2016daj}. For the cases II, III, VII, VIII, IX based on $A_5$ modular symmetry compatible with JUNO six year precision data, they can be further discriminated by measurements of $\sin^2\theta_{23}$ and $\delta_{CP}$ at DUNE and T2HK,
as can be seen from table~\ref{tab:BF-result-N5}. For example, case II and case VII predict $(\sin^2\theta_{23}, \delta_{CP})\approx(0.485, 1.019\pi)$ and $(0.487, 1.844\pi)$ respectively, the values of the Dirac CP phase $\delta_{CP}$ are significantly different although the values of $\theta_{23}$ are of quite similar size.

A conspicuous feature of modular Littlest seesaw is that the neutrino mass spectrum is normal ordering and the lightest neutrino is massless. The uncertainty of $\Delta m^2_{31}$ by T2HK~\cite{Hyper-Kamiokande:2018ofw} and JUNO~\cite{JUNO:2022mxj} is expected to reach $0.6\%$ and $0.2\%$ respectively. Thus the determination of neutrino mass ordering in future neutrino facilities will yield a general test of modular Littlest seesaw models. Moreover, the effective Majorana neutrino mass is predicted in the range $2.3\;\text{meV}<m_{\beta\beta}<3.7\;\text{meV}$, as shown in tables~\ref{tab:BF-results-S4p} and \ref{tab:BF-result-N5}. The current most stringent limit on the effective neutrino mass is $m_{\beta\beta} < (28- 122)~\text{meV}$ at 90\% C.L. from KamLAND-Zen collaboration~\cite{KamLAND-Zen:2024eml}. The next generation neutrinoless double decay experiments aim to explore the full inverted ordering region of parameter space, they hope to reach a sensitivity of order $10\; \text{meV}$. For instance, the projected sensitivities of  LEGEND-1000~\cite{LEGEND:2021bnm} and nEXO~\cite{nEXO:2021ujk} are $(9 - 21)~\text{meV}$
and $(4.7-20.3)~\text{meV}$ respectively for ten years of running time.
Hence the modular Littlest seesaw models predicts the neutrinoless double decay to be beyond foreseeable experimental capabilities.

\section{\label{sec:conclusion}Conclusion}

Modular symmetry is an appealing framework for addressing the flavor puzzle of SM. In this framework, the Yukawa couplings are promoted to modular forms of the complex modulus $\tau$ and they transforms non-trivially under the modular group. Given the modular weights and transformation properties of matter fields under the modular group, the Yukawa couplings would be strongly constrained by modular invariance and the number of free parameters would be significantly reduced. Hence modular symmetry allows to construct highly predictive flavor models and the introduction of flavon fields is not mandatory.

Modular Littlest seesaw implements the idea of Littlest seesaw in the context of modular symmetry~\cite{Ding:2019gof}, and it is based on two right-handed neutrinos and three modular fixed points. The three generations of left-handed leptons are assigned to a triplet of the finite modular group $\Gamma_N$ or $\Gamma'_N$, the two right-handed neutrinos $N^{c}_\text{atm}$ and $N^{c}_\text{sol}$ are modular singlets. The corresponding neutrino Yukawa couplings $Y_{\text{atm}}$ and $Y_{\text{sol}}$ are triplet modular forms with complex modulus at the fixed points $\langle\tau_{\text{atm}}\rangle$ and $\langle\tau_{\text{sol}}\rangle$ respectively. The VEV of the complex modulus in the charged lepton Yukawa couplings are assumed to be stabilized at the fixed point $\langle\tau_{\ell}\rangle$. The modular Littlest seesaw model is very predictive and the resulting light neutrino mass matrix only depend on three real parameters $m_a$, $r=m_s/m_a$ and $\eta$. The lightest neutrino is massless with $m_1=0$ in the modular Littlest seesaw and the first column of the lepton mixing matrix is $Y_{\text{atm}}\times Y_{\text{sol}}/|Y_{\text{atm}}\times Y_{\text{sol}}|$, where the Clebsch-Gordan contraction coefficients are neglected.

In the present work, we have extended the previous modular Littlest seesaw analysis from $S_4$ modular group to the double cover group $S_4'$ as well as to $A_5$. An exhaustive scan over all possible modular Littlest Seesaw constructions is performed, and a large number of viable fixed points and modular form alignments are identified by confronting the models with the latest global fits to neutrino oscillation data~\cite{Capozzi:2025wyn}. As a result, we find four (two additional) viable models based on $S'_4$ and fourteen based on $A_5$, which we denote as cases A–D and I–XIV, respectively. The double cover $A'_5$ does not yield any new phenomenologically distinct scenarios, since the singlet and triplet representations of $A'_5$ coincide with those of $A_5$. In these new constructions, the two columns of the neutrino Yukawa coupling matrix take more general forms than in the conventional Littlest Seesaw, leading to novel mixing patterns beyond TM1. As a consequence, new sum rule relations among the lepton mixing angles and CP violation phases are obtained, as shown in Eq.~\eqref{eq:cosdeltaCP-sum-rule}. These sum rules offer powerful and sensitive probes of the modular Littlest Seesaw models.

The phenomenological implications of the modular Littlest seesaw at future neutrino facilities such as JUNO, DUNE and T2HK, are investigated. The JUNO's first results already exclude the case X and case XIII based on $A_5$, since their predicted value of the solar mixing angle $\sin^2\theta_{12}\approx0.269$ lies outside the JUNO-preferred
$3\sigma$ range. With a projected $1\sigma$ precision of about $0.5\%$ on $\sin^2\theta_{12}$ after six years of data taking, JUNO will provide a stringent test of
all the remaining models, as illustrated in figure~\ref{fig:theta12-13-corre-JUNO}.
They can be further discriminated by precise measurements of $\theta_{23}$ and $\delta_{CP}$ at DUNE and T2HK. The neutrino mass spectrum is normal ordering with $m_1=0$ in modular Littlest seesaw models, consequently the determination of neutrino mass ordering by JUNO, DUNE and T2HK offers another decisive test of the modular Littlest seesaw. Furthermore, the effective Majorana neutrino mass $m_{\beta\beta}$ is predicted to below the sensitivity of future ton scale experiments searching for neutrinoless double decay. Consequently, the paradigm of modular Littlest Seesaw models would be strongly disfavoured if a signal of neutrinoless double beta decay were observed in future experiments.

In addition to the $S'_4$, $A_5$ and $A'_5$ modular symmetries, we have also explored possible modular Littlest Seesaw models based on other finite modular groups, namely
$A_4\times Z_2$, $GL(2, 3)$ and $2O$~\cite{Liu:2021gwa}. We find that the groups
$GL(2, 3)$ and $2O$ reproduce the same phenomenologically viable models as the
$S_4$ modular group, corresponding to cases A and B, while no phenomenologically viable model emerges from $A_4\times Z_2$. Building on the phenomenological success of explaining neutrino masses and mixing, it is appealing to implement these modular Littlest seesaw models in the context of Grand Unified Theories (GUTs) so that quark masses and CKM mixing parameters can also be accommodated, and predictivity of the leptonic sector is kept. This is left for future work.

\section*{Acknowledgements}

We are grateful to Professors Eligio Lisi and Antonio Marrone for useful correspondence on the correlation between $\theta_{12}$ and $\theta_{13}$ in neutrino data global fitting. We acknowledge Professor Cai-Chang Li for sharing the program extracting the sum rules. GJD and EHS are supported by the National Natural Science Foundation of China under Grant Nos.~12375104, 12547106 and Guizhou Provincial Major Scientific and Technological Program XKBF (2025)010. JNL is supported by the National Natural Science Foundation of China with Grant No.~12505133. SFK acknowledges the STFC Consolidated Grant ST/X000583/1 and his work was funded by a Leverhulme Trust Emeritus Fellowship Grant.

\begin{appendix}

\section{\label{app:N4-group-MF} Finite modular group $\Gamma'_4\cong S_4^\prime$ and modular forms of level $N=4$ }

The homogeneous finite modular group $\Gamma^\prime_4\cong S^\prime_4$ at level $N=4$ has 48 group elements which can be generated by $S$ and $T$ satisfying the following multiplication rules~\cite{Novichkov:2020eep,Liu:2020akv,Liu:2020msy,Ding:2022nzn}:
\begin{equation}
 S^4=T^4=(ST)^3=1,~~~ S^2T=TS^2 \,.
\end{equation}
The quotient group $S^\prime_4$ over $Z^{S^2}_2$ is isomorphic to the group $S_4$, where $Z^{S^2}_2$ denotes the cyclic group $Z_2$ generated by $S^2$. The group $S^\prime_4$ has four singlet representations $\bm{1},\,\bm{1}^\prime,\,\widehat{\bm{1}},\,\widehat{\bm{1}}^\prime$, two doublet representations $\bm{2},\,\widehat{\bm{2}}$, and four triplet representations $\bm{3},\,\bm{3}^\prime,\,\widehat{\bm{3}},\widehat{\bm{3}}^\prime$. In these irreducible representations,  the generators $S$ and $T$ are represented by the following matrices,
\begin{eqnarray}
\nonumber\bm{1}: & S=1, & T=1, \\
\nonumber\bm{1}^{\prime}: & S=-1, & T=-1, \\
\nonumber\widehat{\bm{1}}: & S=i, & T=-i, \\
\nonumber\widehat{\bm{1}}^{\prime}: & S=-i, & T=i, \\
\nonumber\bm{2}: & S=\frac{1}{2}\begin{pmatrix}-1 ~&~ \sqrt{3} \\ \sqrt{3} ~&~ 1\end{pmatrix}, & T=\begin{pmatrix}1 ~&~ 0 \\ 0 ~&~ -1\end{pmatrix}, \\
\nonumber\widehat{\bm{2}}: & S=\frac{i}{2}\begin{pmatrix}-1 & \sqrt{3} \\ \sqrt{3} & 1\end{pmatrix}, & T=-i\begin{pmatrix}1 ~&~ 0 \\ 0 ~&~ -1\end{pmatrix}, \\
\nonumber\bm{3}: & S=\frac{1}{2}\begin{pmatrix}0 ~&~ \sqrt{2} ~&~ \sqrt{2} \\ \sqrt{2} ~&~ -1 ~&~ 1 \\ \sqrt{2} ~&~ 1 ~&~ -1\end{pmatrix}, & T=\begin{pmatrix}1 ~&~ 0 ~&~ 0 \\ 0 ~&~ i ~&~ 0 \\ 0 ~&~ 0 ~&~ -i\end{pmatrix}, \\
\nonumber\bm{3}^{\prime}: & S=-\frac{1}{2}\begin{pmatrix}0 & \sqrt{2} & \sqrt{2} \\ \sqrt{2} ~&~ -1 ~&~ 1 \\ \sqrt{2} ~&~ 1 ~&~ -1\end{pmatrix}, & T=-\begin{pmatrix}1 ~&~ 0 ~&~ 0 \\ 0 ~&~ i ~&~ 0 \\ 0 ~&~ 0 ~&~ -i\end{pmatrix}, \\
\nonumber\widehat{\bm{3}}: & S=\frac{i}{2}\begin{pmatrix}0 ~&~ \sqrt{2} ~&~ \sqrt{2} \\ \sqrt{2} ~&~ -1 ~&~ 1 \\ \sqrt{2} ~&~ 1 ~&~ -1\end{pmatrix}, &  T=-i\begin{pmatrix}1 ~&~ 0 ~&~ 0 \\ 0 ~&~ i ~&~ 0 \\ 0 ~&~ 0 ~&~ -i\end{pmatrix}, \\
\widehat{\bm{3}}^{\prime}: & S=-\frac{i}{2}\begin{pmatrix}0 ~&~ \sqrt{2} ~&~ \sqrt{2} \\ \sqrt{2} ~&~ -1 ~&~ 1 \\ \sqrt{2} ~&~ 1 ~&~ -1\end{pmatrix}, & ~~ T=i\begin{pmatrix}1 ~&~ 0 ~&~ 0 \\ 0 ~&~ i ~&~ 0 \\ 0 ~&~ 0 ~&~ -i\end{pmatrix}\,.
\end{eqnarray}
Notice that the above convention for the triplet representations are different from that of~\cite{Novichkov:2020eep}. We report the tensor products between different irreducible representations of $S'_4$ and the relevant Clebsch–Gordan (CG) coefficients in table~\ref{tab:CG-S4p}.

\begin{center}
\setlength\LTcapwidth{\textwidth}
\begin{longtable}{|c|c|c|c|c|c|}
\caption{\label{tab:CG-S4p} The tensor products and CG coefficients of $S_4^\prime$ group.}
\\ \toprule \hline

\endfirsthead
\multicolumn{6}{c}{{\bfseries \tablename\ \thetable{} -- continued from previous page}}\\

\hline
\endhead

\multicolumn{6}{c}{{\bfseries \tablename\ \thetable{} -- continues on next page}}\\
\endfoot

\hline
\endlastfoot

\hline
\multicolumn{3}{|c|}{$\bm{1}\otimes \bm{1}=\bm{1}^\prime\otimes \bm{1}^\prime=\widehat{\bm{1}}\otimes\widehat{\bm{1}}^\prime=\bm{1}$} & \multicolumn{3}{c|}{$\bm{1}\otimes\widehat{\bm{1}}=\bm{1}^\prime\otimes\widehat{\bm{1}}^\prime=\widehat{\bm{1}}$} \\
\hline
\multicolumn{3}{|c|}{$\bm{1}\sim \alpha\beta$} & \multicolumn{3}{c|}{$\widehat{\bm{1}}\sim \alpha\beta$} \\
\hline
\multicolumn{3}{|c|}{$\bm{1}\otimes\bm{1}^\prime=\widehat{\bm{1}}\otimes\widehat{\bm{1}}=\widehat{\bm{1}}^\prime\otimes\widehat{\bm{1}}^\prime=\bm{1}^\prime$} & \multicolumn{3}{c|}{$\bm{1}\otimes\widehat{\bm{1}}^\prime=\bm{1}^\prime\otimes\widehat{\bm{1}}=\widehat{\bm{1}}^\prime$}\\
\hline
\multicolumn{3}{|c|}{$\bm{1}^\prime\sim\alpha\beta$} & \multicolumn{3}{c|}{$\widehat{\bm{1}}^\prime\sim\alpha\beta$}\\
\hline
\multicolumn{3}{|c|}{$\bm{1}\otimes\bm{2}=\widehat{\bm{1}}^\prime\otimes\widehat{\bm{2}}=\bm{2}$} & \multicolumn{3}{c|}{$\bm{1}\otimes \widehat{\bm{2}}=\widehat{\bm{1}}\otimes\bm{2}=\widehat{\bm{2}}$} \\
\hline
\multicolumn{3}{|c|}{$\bm{2}\sim\alpha\begin{pmatrix}\beta_1\\\beta_2 \end{pmatrix}$} & \multicolumn{3}{c|}{$\widehat{\bm{2}}\sim\alpha\begin{pmatrix}\beta_1\\\beta_2 \end{pmatrix}$}\\
\hline
\multicolumn{3}{|c|}{$\bm{1}^\prime\otimes\bm{2}=\widehat{\bm{1}}\otimes\widehat{\bm{2}}=\bm{2}$} & \multicolumn{3}{c|}{$\bm{1}^\prime\otimes \widehat{\bm{2}}=\widehat{\bm{1}}^\prime\otimes\bm{2}=\widehat{\bm{2}}$} \\
\hline
\multicolumn{3}{|c|}{$\bm{2}\sim\alpha\begin{pmatrix}\beta_2\\-\beta_1 \end{pmatrix}$} & \multicolumn{3}{c|}{$\widehat{\bm{2}}\sim\alpha\begin{pmatrix}\beta_2\\ -\beta_1 \end{pmatrix}$}\\
\hline
\multicolumn{3}{|c|}{$\bm{1}\otimes\bm{3}=\bm{1}^\prime\otimes\bm{3}^\prime=\widehat{\bm{1}}\otimes\widehat{\bm{3}}^\prime=\widehat{\bm{1}}^\prime\otimes\widehat{\bm{3}}=\bm{3}$} & \multicolumn{3}{c|}{$\bm{1}\otimes\bm{3}^\prime=\bm{1}^\prime\otimes\bm{3}=\widehat{\bm{1}}\otimes\widehat{\bm{3}}=\widehat{\bm{1}}^\prime\otimes\widehat{\bm{3}}^\prime=\bm{3}^\prime$} \\
\hline
\multicolumn{3}{|c|}{$\bm{3}\sim\alpha\begin{pmatrix}\beta_1\\\beta_2\\\beta_3 \end{pmatrix}$} & \multicolumn{3}{c|}{$\bm{3}^\prime\sim\alpha\begin{pmatrix}\beta_1\\\beta_2\\\beta_3 \end{pmatrix}$}\\
\hline
\multicolumn{3}{|c|}{$\bm{1}\otimes\widehat{\bm{3}}=\bm{1}^\prime\otimes\widehat{\bm{3}}^\prime=\widehat{\bm{1}}\otimes\bm{3}=\widehat{\bm{1}}^\prime\otimes\bm{3}^\prime=\widehat{\bm{3}}$} & \multicolumn{3}{c|}{$\bm{1}\otimes\widehat{\bm{3}}^\prime=\bm{1}^\prime\otimes\widehat{\bm{3}}=\widehat{\bm{1}}\otimes\bm{3}^\prime=\widehat{\bm{1}}^\prime\otimes\bm{3}=\widehat{\bm{3}}^\prime$} \\
\hline
\multicolumn{3}{|c|}{$\widehat{\bm{3}}\sim\alpha\begin{pmatrix}\beta_1\\\beta_2\\\beta_3 \end{pmatrix}$} & \multicolumn{3}{c|}{$\widehat{\bm{3}}^{\prime}\sim\alpha\begin{pmatrix}\beta_1\\\beta_2\\\beta_3 \end{pmatrix}$}\\
\hline
\multicolumn{2}{|c|}{$\bm{2}\otimes\bm{2}=\bm{1}\oplus\bm{1}^\prime\oplus\bm{2}$} & \multicolumn{2}{c|}{$\bm{2}\otimes\widehat{\bm{2}}=\widehat{\bm{1}}\oplus\widehat{\bm{1}}^\prime\oplus\widehat{\bm{2}}$} & \multicolumn{2}{c|}{$\widehat{\bm{2}}\otimes\widehat{\bm{2}}=\bm{1}\oplus\bm{1}^\prime\oplus\bm{2}$}\\
\hline
\multicolumn{2}{|c|}{$
\begin{array}{lcl}
\bm{1} & \sim & \alpha_1\beta_1+\alpha_2\beta_2\\
\bm{1}^\prime & \sim & \alpha_1\beta_2-\alpha_2\beta_1\\
\bm{2} &\sim & \begin{pmatrix} -\alpha_1\beta_1+\alpha_2\beta_2\\ \alpha_1\beta_2+\alpha_2\beta_1 \end{pmatrix}
\end{array}
$} & \multicolumn{2}{c|}{$
\begin{array}{lcl}
\widehat{\bm{1}} & \sim & \alpha_1\beta_1+\alpha_2\beta_2\\
\widehat{\bm{1}}^\prime & \sim & \alpha_1\beta_2-\alpha_2\beta_1\\
\widehat{\bm{2}} &\sim & \begin{pmatrix} -\alpha_1\beta_1+\alpha_2\beta_2\\ \alpha_1\beta_2+\alpha_2\beta_1 \end{pmatrix}
\end{array}
$}& \multicolumn{2}{c|}{$
\begin{array}{lcl}
\bm{1}& \sim & \alpha_1\beta_2-\alpha_2\beta_1\\
\bm{1}^\prime & \sim & \alpha_1\beta_1+\alpha_2\beta_2\\
\bm{2} &\sim & \begin{pmatrix} -\alpha_1\beta_2-\alpha_2\beta_1\\ -\alpha_1\beta_1+\alpha_2\beta_2 \end{pmatrix}
\end{array}$}\\
\hline
\multicolumn{3}{|c|}{$\bm{2}\otimes\bm{3}=\widehat{\bm{2}}\otimes\widehat{\bm{3}}^\prime=\bm{3}\oplus\bm{3}^\prime$} & \multicolumn{3}{c|}{$\bm{2}\otimes\bm{3}^\prime=\widehat{\bm{2}}\otimes\widehat{\bm{3}}=\bm{3}\oplus\bm{3}^\prime$}\\
\hline
\multicolumn{3}{|c|}{$
\begin{array}{lcl}
\bm{3} & \sim & \begin{pmatrix} 2\alpha_1\beta_1\\ -\alpha_1\beta_2+\sqrt{3}\,\alpha_2\beta_3 \\ -\alpha_1\beta_3+\sqrt{3}\,\alpha_2\beta_2\end{pmatrix}\\
\bm{3}^\prime & \sim & \begin{pmatrix}-2\alpha_2\beta_1\\ \sqrt{3}\,\alpha_1\beta_3+\alpha_2\beta_2\\ \sqrt{3}\,\alpha_1\beta_2+\alpha_2\beta_3 \end{pmatrix}
\end{array}
$} & \multicolumn{3}{c|}{$
\begin{array}{lcl}
\bm{3} & \sim &\begin{pmatrix}-2\alpha_2\beta_1\\ \sqrt{3}\,\alpha_1\beta_3+\alpha_2\beta_2\\ \sqrt{3}\,\alpha_1\beta_2+\alpha_2\beta_3 \end{pmatrix}\\
\bm{3}^\prime & \sim &  \begin{pmatrix} 2\alpha_1\beta_1\\ -\alpha_1\beta_2+\sqrt{3}\,\alpha_2\beta_3 \\ -\alpha_1\beta_3+\sqrt{3}\,\alpha_2\beta_2\end{pmatrix}
\end{array}
$}\\
\hline
\multicolumn{3}{|c|}{$\bm{2}\otimes\widehat{\bm{3}}=\widehat{\bm{2}}\otimes\bm{3}=\widehat{\bm{3}}\oplus\widehat{\bm{3}}^\prime$} & \multicolumn{3}{c|}{$\bm{2}\otimes\widehat{\bm{3}}^\prime=\widehat{\bm{2}}\otimes\bm{3}^\prime=\widehat{\bm{3}}\oplus\widehat{\bm{3}}^\prime$}\\
\hline
\multicolumn{3}{|c|}{$
\begin{array}{lcl}
\widehat{\bm{3}} & \sim & \begin{pmatrix} 2\alpha_1\beta_1\\ -\alpha_1\beta_2+\sqrt{3}\,\alpha_2\beta_3 \\ -\alpha_1\beta_3+\sqrt{3}\,\alpha_2\beta_2\end{pmatrix}\\
\widehat{\bm{3}}^\prime & \sim & \begin{pmatrix}-2\alpha_2\beta_1\\ \sqrt{3}\,\alpha_1\beta_3+\alpha_2\beta_2\\ \sqrt{3}\,\alpha_1\beta_2+\alpha_2\beta_3 \end{pmatrix}
\end{array}
$} & \multicolumn{3}{c|}{$
\begin{array}{lcl}
\widehat{\bm{3}} & \sim &\begin{pmatrix}-2\alpha_2\beta_1\\ \sqrt{3}\,\alpha_1\beta_3+\alpha_2\beta_2\\ \sqrt{3}\,\alpha_1\beta_2+\alpha_2\beta_3 \end{pmatrix}\\
\widehat{\bm{3}}^\prime & \sim &  \begin{pmatrix} 2\alpha_1\beta_1\\ -\alpha_1\beta_2+\sqrt{3}\,\alpha_2\beta_3 \\ -\alpha_1\beta_3+\sqrt{3}\,\alpha_2\beta_2\end{pmatrix}
\end{array}
$}\\
\hline
\multicolumn{3}{|c|}{$\bm{3}\otimes\bm{3}=\bm{3}^\prime\otimes\bm{3}^\prime=\widehat{\bm{3}}\otimes\widehat{\bm{3}}^\prime=\bm{1}\oplus\bm{2}\oplus\bm{3}\oplus\bm{3}^\prime$} & \multicolumn{3}{c|}{$\bm{3}\otimes\widehat{\bm{3}}=\bm{3}^\prime\otimes\widehat{\bm{3}}^\prime=\widehat{\bm{1}}\oplus\widehat{\bm{2}}\oplus\widehat{\bm{3}}\oplus\widehat{\bm{3}}^\prime$}\\
\hline
\multicolumn{3}{|c|}{$
\begin{array}{lcl}
\bm{1} & \sim & \alpha_1\beta_1+\alpha_2\beta_3+\alpha_3\beta_2\\
\bm{2} & \sim & \begin{pmatrix}2\alpha_1\beta_1-\alpha_2\beta_3-\alpha_3\beta_2\\
\sqrt{3}\,\alpha_2\beta_2+\sqrt{3}\,\alpha_3\beta_3 \end{pmatrix}\\
\bm{3} & \sim & \begin{pmatrix}\alpha_2\beta_3-\alpha_3\beta_2\\
\alpha_1\beta_2-\alpha_2\beta_1\\
-\alpha_1\beta_3+\alpha_3\beta_1
\end{pmatrix}\\
\bm{3}^\prime & \sim & \begin{pmatrix}
\alpha_2\beta_2-\alpha_3\beta_3\\
-\alpha_1\beta_3-\alpha_3\beta_1\\
\alpha_1\beta_2+\alpha_2\beta_1
\end{pmatrix}
\end{array}
$} & \multicolumn{3}{c|}{$
\begin{array}{lcl}
\widehat{\bm{1}}& \sim & \alpha_1\beta_1+\alpha_2\beta_3+\alpha_3\beta_2\\
\widehat{\bm{2}}& \sim & \begin{pmatrix}2\alpha_1\beta_1-\alpha_2\beta_3-\alpha_3\beta_2\\
\sqrt{3}\,\alpha_2\beta_2+\sqrt{3}\,\alpha_3\beta_3 \end{pmatrix}\\
\widehat{\bm{3}}& \sim & \begin{pmatrix}\alpha_2\beta_3-\alpha_3\beta_2\\
\alpha_1\beta_2-\alpha_2\beta_1\\
-\alpha_1\beta_3+\alpha_3\beta_1
\end{pmatrix}\\
\widehat{\bm{3}}^\prime& \sim & \begin{pmatrix}
\alpha_2\beta_2-\alpha_3\beta_3\\
-\alpha_1\beta_3-\alpha_3\beta_1\\
\alpha_1\beta_2+\alpha_2\beta_1
\end{pmatrix}
\end{array}
$}\\
\hline
\multicolumn{3}{|c|}{$\bm{3}\otimes\bm{3}^\prime=\widehat{\bm{3}}\otimes\widehat{\bm{3}}=\widehat{\bm{3}}^\prime\otimes\widehat{\bm{3}}^\prime=\bm{1}^\prime\oplus\bm{2}\oplus\bm{3}\oplus\bm{3}^\prime$} & \multicolumn{3}{c|}{$\bm{3}\otimes\widehat{\bm{3}}^\prime=\bm{3}^\prime\otimes\widehat{\bm{3}}=\widehat{\bm{1}}^\prime\oplus\widehat{\bm{2}}\oplus\widehat{\bm{3}}\oplus\widehat{\bm{3}}^\prime$}\\
\hline
\multicolumn{3}{|c|}{$
\begin{array}{lcl}
\bm{1}^\prime & \sim & \alpha_1\beta_1+\alpha_2\beta_3+\alpha_3\beta_2\\
\bm{2} & \sim & \begin{pmatrix}
\sqrt{3}\,\alpha_2\beta_2+\sqrt{3}\,\alpha_3\beta_3\\
-2\alpha_1\beta_1+\alpha_2\beta_3+\alpha_3\beta_2
 \end{pmatrix}\\
\bm{3} & \sim & \begin{pmatrix}
\alpha_2\beta_2-\alpha_3\beta_3\\
-\alpha_1\beta_3-\alpha_3\beta_1\\
\alpha_1\beta_2+\alpha_2\beta_1
\end{pmatrix}\\
\bm{3}^\prime & \sim & \begin{pmatrix}\alpha_2\beta_3-\alpha_3\beta_2\\
\alpha_1\beta_2-\alpha_2\beta_1\\
-\alpha_1\beta_3+\alpha_3\beta_1
\end{pmatrix}
\end{array}
$} & \multicolumn{3}{c|}{$
\begin{array}{lcl}
\widehat{\bm{1}}^\prime & \sim & \alpha_1\beta_1+\alpha_2\beta_3+\alpha_3\beta_2\\
\widehat{\bm{2}} & \sim & \begin{pmatrix}
\sqrt{3}\,\alpha_2\beta_2+\sqrt{3}\,\alpha_3\beta_3\\
-2\alpha_1\beta_1+\alpha_2\beta_3+\alpha_3\beta_2
 \end{pmatrix}\\
\widehat{\bm{3}} & \sim & \begin{pmatrix}
\alpha_2\beta_2-\alpha_3\beta_3\\
-\alpha_1\beta_3-\alpha_3\beta_1\\
\alpha_1\beta_2+\alpha_2\beta_1
\end{pmatrix}\\
\widehat{\bm{3}}^\prime & \sim & \begin{pmatrix}\alpha_2\beta_3-\alpha_3\beta_2\\
\alpha_1\beta_2-\alpha_2\beta_1\\
-\alpha_1\beta_3+\alpha_3\beta_1
\end{pmatrix}
\end{array}
$}\\
\hline \bottomrule
\end{longtable}
\end{center}
As shown in Eq.~\eqref{eq:MF-w13hp-N4}, there are three linearly independent weight $k=1$ modular forms at level $N=4$, and they can be organized into a triplet $Y^{(1)}_{\widehat{\bm{3}}^\prime}(\tau)=(Y_1(\tau), Y_2(\tau), Y_3(\tau))^T$ transforming as $\widehat{\bm{3}}^\prime$ under the action of $S'_4$. The higher weight modular forms can be generated by the tensor product of $Y^{(1)}_{\widehat{\bm{3}}^\prime}$. At the weight $k=2$, we have two independent modular multiplets,
\begin{eqnarray}
\nonumber Y^{(2)}_{\bm{2}} &= \frac{1}{\sqrt{3}}\left(Y^{(1)}_{\widehat{\bm{3}}^\prime}Y^{(1)}_{\widehat{\bm{3}}^\prime}\right)_{\bm{2}}  &= \left(\begin{array}{c}
Y^2_2+Y^2_3\\
2(Y_2Y_3-Y^2_1)/\sqrt{3}
\end{array}\right)=\left(\begin{array}{c}
\vartheta^4_1+\vartheta^4_2\\
-2\sqrt{3}\vartheta^2_1\vartheta^2_2
\end{array}\right)\,,\\
Y^{(2)}_{\bm{3}} &=-\left(Y^{(1)}_{\widehat{\bm{3}}^\prime}Y^{(1)}_{\widehat{\bm{3}}^\prime}\right)_{\bm{3}}  &= \left(\begin{array}{c}
Y^2_3-Y^2_2\\
2Y_1Y_3\\
-2Y_1Y_2\end{array}\right)=\left(\begin{array}{c}
\vartheta^4_1-\vartheta^4_2 \\
2\sqrt{2}\vartheta^3_1\vartheta_2\\
2\sqrt{2}\vartheta_1\vartheta^3_2
\end{array}\right)\,.
\end{eqnarray}
From the expressions of $Y_{1,2,3}(\tau)$ in terms of $\vartheta_1(\tau)$ and $\vartheta_2(\tau)$ in Eq.~\eqref{eq:MF-w13hp-N4}, we can see $\left(Y^{(1)}_{\widehat{\bm{3}}^\prime}Y^{(1)}_{\widehat{\bm{3}}^\prime}\right)_{\bm{1}^\prime}=0$ so that the modular forms $Y_1(\tau)$, $Y_2(\tau)$, $Y_3(\tau)$ satisfy the following constraint
\begin{eqnarray}
Y_1^2+2Y_2Y_3=0\,.
\end{eqnarray}
For weight $k=3$, we have three independent modular multiplets in the representations $\widehat{\bm{1}}^\prime$, $\widehat{\bm{3}}$, $\widehat{\bm{3}}^\prime$ as follows,
\begin{eqnarray}
\nonumber Y^{(3)}_{\widehat{\bm{1}}^\prime} &= \frac{1}{3\sqrt{2}}\left(Y^{(1)}_{\widehat{\bm{3}}^\prime}Y^{(2)}_{\bm{3}}\right)_{\widehat{\bm{1}}^\prime}  &=\frac{1}{\sqrt{2}}Y_1\left(-Y^2_2+Y^2_3\right)=\vartheta_1\vartheta_2\left(\vartheta_1^4-\vartheta_2^4\right)\,, \\
\nonumber Y^{(3)}_{\widehat{\bm{3}}} &= -\left(Y^{(1)}_{\widehat{\bm{3}}^\prime}Y^{(2)}_{\bm{3}}\right)_{\widehat{\bm{3}}}  &=
\left(\begin{array}{c} -4 Y_1Y_2Y_3 \\ -2Y_1^2Y_2-Y_2^2Y_3+Y_3^3 \\ Y_2^3-2Y_1^2Y_3-Y_2Y_3^2 \end{array}\right)
=\left(\begin{array}{c} 4\sqrt{2}\,\vartheta_1^3\vartheta_2^3 \\ \vartheta_1^6+3\vartheta_1^2\vartheta_2^4 \\ -\vartheta_2^2\left(3\vartheta_1^4+\vartheta_2^4\right) \end{array}\right) \,,\\
Y^{(3)}_{\widehat{\bm{3}}^\prime} &= \left(Y^{(1)}_{\widehat{\bm{3}}^\prime}Y^{(2)}_{\bm{3}}\right)_{\widehat{\bm{3}}^\prime}  &=\left(\begin{array}{c} 2Y_1\left(Y_2^2+Y_3^2\right) \\ -Y_2^3-2Y_1^2Y_3+Y_2Y_3^2 \\ -2Y_1^2Y_2+Y_2^2Y_3-Y_3^3 \end{array}\right)
=\left(\begin{array}{c} 2\sqrt{2}\,\vartheta_1\vartheta_2\left(\vartheta_1^4+\vartheta_2^4\right) \\ -5\vartheta_1^4\vartheta_2^2+\vartheta_2^6 \\ -\vartheta_1^6+5\vartheta_1^2\vartheta_2^4
\end{array}\right) \,.
\end{eqnarray}
The weight $k=4$ modular forms of level $N=4$ can be arranged into a singlet $\bm{1}$, a doublet $\bm{2}$ and two triplets $\bm{3}$ and $\bm{3}^\prime$,
\begin{eqnarray}
\nonumber Y^{(4)}_{\bm{1}}&=&\left(Y^{(2)}_{\bm{3}}Y^{(2)}_{\bm{3}}\right)_{\bm{1}}
=-8Y_1^2Y_2Y_3+\left(Y_2^2-Y_3^2\right)^2
=\vartheta_1^8 + 14\vartheta_1^4\vartheta_2^4 + \vartheta_2^8\,,\\
\nonumber Y^{(4)}_{\bm{2}}&=&\frac{1}{2}\left(Y^{(2)}_{\bm{3}}Y^{(2)}_{\bm{3}}\right)_{\bm{2}}
=\left(\begin{array}{c} Y_2^4+4Y_1^2Y_2Y_3-2Y_2^2Y_3^2+Y_3^4 \\ 2\sqrt{3}Y_1^2\left(Y_2^2+Y_3^2\right)\end{array}\right)
=\left(\begin{array}{c} \vartheta_1^8 - 10\vartheta_1^4\vartheta_2^4 + \vartheta_2^8 \\ 4\sqrt{3}\,\vartheta_1^2\vartheta_2^2(\vartheta_1^4+\vartheta_2^4) \end{array}\right)\,,\\
\nonumber Y^{(4)}_{\bm{3}}&=&-\frac{1}{2}\left(Y^{(2)}_{\bm{2}}Y^{(2)}_{\bm{3}}\right)_{\bm{3}}
=\left(\begin{array}{c}Y_2^4-Y_3^4 \\ -2Y_1^3Y_2+3Y_1Y_2^2Y_3+Y_1Y_3^3 \\ -Y_1Y_2^3+2Y_1^3Y_3-3Y_1Y_2Y_3^2\end{array}\right)
=\left(\begin{array}{c} -\vartheta_1^8+\vartheta_2^8 \\ \sqrt{2}\,\vartheta_1^3\vartheta_2\left(\vartheta_1^4+7\vartheta_2^4\right) \\ \sqrt{2}\,\vartheta_1\vartheta_2^3\left(7\vartheta_1^4+\vartheta_2^4\right) \end{array}\right)\,,\\
Y^{(4)}_{\bm{3}^\prime}&=&\frac{1}{2\sqrt{6}}\left(Y^{(2)}_{\bm{2}}Y^{(2)}_{\bm{3}}\right)_{\bm{3}^\prime}
=\dfrac{1}{3\sqrt{2}}\left(\begin{array}{c} 2\left(-Y_2^2+Y_3^2\right)\left(Y_1^2-Y_2Y_3\right) \\ -3Y_1Y_2^3-2Y_1^3Y_3-Y_1Y_2Y_3^2 \\ 2Y_1^3Y_2+Y_1Y_2^2Y_3+3Y_1Y_3^2\end{array}\right)
=\vartheta_1\vartheta_2\left(\vartheta_1^4-\vartheta_2^4\right)
\left(\begin{array}{c}
\sqrt{2}\vartheta_1\vartheta_2 \\
-\vartheta_2^2 \\
\vartheta_1^2
\end{array}\right)\,.
\end{eqnarray}
The linearly independent weight 5 modular forms can be constructed from the tensor products of weight-1 and weight-4 modular forms as follows,
\begin{small}
\begin{eqnarray}
\nonumber
Y^{(5)}_{\widehat{\bm{2}}}
& = & \frac{1}{3\sqrt{2}}\left(Y^{(1)}_{\widehat{\bm{3}}^{\prime}}Y^{(4)}_{\bm{3}}\right)_{\widehat{\bm{2}}}
=\dfrac{1}{\sqrt{6}}\left(Y_2^2-Y_3^2\right)
\left(\begin{array}{c}
-2Y_1^3+2Y_1Y_2Y_3\\
-\sqrt{3}Y_1\left(Y_2^2+Y_3^2\right)
\end{array}\right)
=\vartheta_1\vartheta_2\left(\vartheta_1^4-\vartheta_2^4\right)
\left(\begin{array}{c}
2\sqrt{3}\,\vartheta_1^2\vartheta_2^2\\
\vartheta_1^4+\vartheta_2^4
\end{array}\right)\,,\\
\nonumber
Y^{(5)}_{\widehat{\bm{3}}}
& = & \left(Y^{(1)}_{\widehat{\bm{3}}^{\prime}}Y^{(4)}_{\bm{3}}\right)_{\widehat{\bm{3}}}
=\left(\begin{array}{c}
-2Y_1\left(Y_1^2-2Y_2Y_3\right)\left(Y_2^2+Y_3^2\right)\\
-2Y_1^4Y_3-Y_2^4Y_3+Y_3^5+Y_1^2\left(Y_2^3+3Y_2Y_3^2\right)\\
-2Y_1^4Y_2+Y_2^5+Y_2Y_3^4+Y_1^2\left(3Y_2^2Y_3+Y_3^3\right)
\end{array}\right)
=\left(\begin{array}{c}
-8\sqrt{2}\,\vartheta_1^3\vartheta_2^3\left(\vartheta_1^4+\vartheta_2^4\right) \\
\vartheta_1^2\left(\vartheta_1^8-14\vartheta_1^4\vartheta_2^4-3\vartheta_2^8\right) \\
\vartheta_2^2\left(3\vartheta_1^8+14\vartheta_1^4\vartheta_2^4-\vartheta_2^8\right)
\end{array}\right)\,,\\
\nonumber
Y^{(5)}_{\widehat{\bm{3}}^\prime I}
& = & \left(Y^{(1)}_{\widehat{\bm{3}}^{\prime}}Y^{(4)}_{\bm{2}}\right)_{\widehat{\bm{3}}^\prime}
=\left(\begin{array}{c}
2Y_1\left[4Y_1^2Y_2Y_3+\left(Y_2^2-Y_3^2\right)^2\right]\\
-Y_2^5+2Y_1^2Y_2^2Y_3+2Y_2^3Y_3^2+6Y_1^2Y_3^2-Y_2Y_3^4\\
-Y_3\left(Y_2^2-Y_3^2\right)^2+2Y_1^2Y_2\left(3Y_2^2+Y_3^2\right)
\end{array}\right)
=\left(\begin{array}{c}
2\sqrt{2}\,\vartheta_1\vartheta_2\left(\vartheta_1^8-10\vartheta_1^4\vartheta_2^4+\vartheta_2^8\right) \\
\vartheta_2^2\left(13\vartheta_1^8+2\vartheta_1^4\vartheta_2^4+\vartheta_2^8\right) \\
-\vartheta_1^2\left(\vartheta_1^8+2\vartheta_1^4\vartheta_2^4+13\vartheta_2^8\right)
\end{array}\right)\,,\\
Y^{(5)}_{\widehat{\bm{3}}^\prime II}
& = & \left(Y^{(1)}_{\widehat{\bm{3}}^{\prime}}Y^{(4)}_{\bm{1}}\right)_{\widehat{\bm{3}}^\prime}
=\left[-8Y_1^2Y_2Y_3+\left(Y_2^2-Y_3^2\right)^2\right]
\left(\begin{array}{c}
Y_1 \\
Y_2 \\
Y_3
\end{array}\right)
=\left(\vartheta_1^8+14\vartheta_1^4\vartheta_2^4+\vartheta_2^8\right)
\left(\begin{array}{c}
\sqrt{2}\,\vartheta_1\vartheta_2 \\
-\vartheta_2^2 \\
\vartheta_1^2
\end{array}\right)\,.
\end{eqnarray}
\end{small}
where $Y^{(5)}_{\widehat{\bm{3}}^\prime\,I}$ and $Y^{(5)}_{\widehat{\bm{3}}^\prime\,II}$ denote the two independent modular forms in the representation $\widehat{\bm{3}}^\prime$ of $S'_4$. Finally, the independent weight 6 modular form multiplets of level 4 are given by
\begin{eqnarray}
\nonumber
Y^{(6)}_{\bm{1}}
& = &\left(Y^{(2)}_{\bm{2}}Y^{(4)}_{\bm{2}}\right)_{\bm{1}}
=\left(Y_2^2+Y_3^2\right)\left[-4Y_1^4+8Y_1^2Y_2Y_3+\left(Y_2^2-Y_3^2\right)^2\right]
=\vartheta_1^{12}-33\vartheta_1^{8}\vartheta_2^{4}-33\vartheta_1^{4}\vartheta_2^{8}+\vartheta_2^{12}\,,\\
\nonumber
Y^{(6)}_{\bm{1}^\prime}
& = & \frac{1}{6\sqrt{3}}\left(Y^{(2)}_{\bm{2}}Y^{(4)}_{\bm{2}}\right)_{\bm{1}^\prime}
=\dfrac{1}{9}\left[4Y_1^4Y_2Y_3-Y_2Y_3\left(Y_2^2-Y_3^2\right)^2+4Y_1^2\left(Y_2^4+Y_3^4\right)\right]
=\vartheta_1^2\vartheta_2^2\left(\vartheta_1^4-\vartheta_2^4\right)^2\,,\\
\nonumber
Y^{(6)}_{\bm{2}}
& = & -\left(Y^{(2)}_{\bm{2}}Y^{(4)}_{\bm{2}}\right)_{\bm{2}}
=\dfrac{1}{\sqrt{3}}
\left(\begin{array}{c}
\sqrt{3}\left(Y_2^2+Y_3^2\right)\left[4Y_1^4+\left(Y_2^2-Y_3^2\right)^2\right]\\
8Y_1^4Y_2Y_3-2Y_2Y_3\left(Y_2^2-Y_3^2\right)^2-4Y_1^2\left(Y_2^4+6Y_2^2Y_3^2+Y_3^4\right)
\end{array}\right)\\
\nonumber & = & \left(\vartheta_1^8+14\vartheta_1^4\vartheta_2^4+\vartheta^8_2\right)
\left(\begin{array}{c}
\vartheta_1^4+\vartheta_2^4\\
-2\sqrt{3}\,\vartheta_1^2\vartheta_2^2
\end{array}\right)\,,\\
\nonumber
Y^{(6)}_{\bm{3} I}
& = & \dfrac{1}{2}\left(Y^{(2)}_{\bm{3}}Y^{(4)}_{\bm{2}}\right)_{\bm{3}}
=\left(\begin{array}{c}
-\left(Y_2^2-Y_3^2\right)\left[4Y_1^2Y_2Y_3+\left(Y_2^2-Y_3^2\right)^2\right]\\
-Y_1Y_3\left(Y_2^2-Y_3^2\right)^2-2Y_1^3\left(3Y_2^3+5Y_2Y_3^2\right)\\
Y_1\left(Y_2^5+10Y_1^2Y_2^2Y_3-2Y_2^3Y_3^2+6Y_1^2Y_3^3+Y_2Y_3^4\right)
\end{array}\right)\\
\nonumber & = & \left(\begin{array}{c}
\left(\vartheta_1^4-\vartheta_2^4\right)\left(\vartheta_1^8-10\vartheta_1^4\vartheta_2^4+\vartheta_2^8\right)\\
\sqrt{2}\vartheta_1^3\vartheta_2\left(-\vartheta_1^8+22\vartheta_1^4\vartheta_2^4+11\vartheta_2^8\right)\\
\sqrt{2}\vartheta_1\vartheta_2^3\left(11\vartheta_1^8+22\vartheta_1^4\vartheta_2^4-\vartheta_2^8\right)
\end{array}\right)
\,,\\
\nonumber
Y^{(6)}_{\bm{3} II}
& = & \left(Y^{(2)}_{\bm{3}}Y^{(4)}_{\bm{1}}\right)_{\bm{3}}
=\left[-8Y_1^2Y_2Y_3+\left(Y_2^2-Y_3^2\right)^2\right]
\left(\begin{array}{c}
Y^2_3-Y^2_2\\
2Y_1Y_3\\
-2Y_1Y_2\end{array}\right)
= \left(\vartheta_1^8+14\vartheta_1^4\vartheta_2^4+\vartheta^8_2\right)
\left(\begin{array}{c}
\vartheta_1^4-\vartheta_2^4\\
2\sqrt{2}\vartheta_1^3\vartheta_2\\
2\sqrt{2}\vartheta_1\vartheta_2^3
\end{array}\right)\,,\\
Y^{(6)}_{\bm{3}^\prime}
& = & -\frac{1}{2\sqrt{6}}\left(Y^{(2)}_{\bm{3}}Y^{(4)}_{\bm{2}}\right)_{\bm{3}^\prime}
=\dfrac{1}{\sqrt{2}}Y_1\left(Y_2^2-Y_3^2\right)
\left(\begin{array}{c}
-2Y_1\left(Y_2^2+Y_3^2\right)\\
Y_2^3+2Y_1^2Y_3-Y_2Y_3^2\\
2Y_1^2Y_2-Y_2^2Y_3+Y_3^3
\end{array}\right)\nonumber\\
& = & \vartheta_1\vartheta_2\left(\vartheta_1^4-\vartheta_2^4\right)
\left(\begin{array}{c}
2\sqrt{2}\,\vartheta_1\vartheta_2\left(\vartheta_1^4+\vartheta_2^{4}\right)\\
-5\vartheta_1^4\vartheta_2^2+\vartheta_2^6\\
-\vartheta_1^6+5\vartheta_1^2\vartheta_2^4
\end{array}\right)\,.
\end{eqnarray}
Given any modulus $\tau$, one can determine the value of $Y^{(k)}_{\bm{r}}(\tau)$ from the expressions of $\vartheta_1(\tau)$ and $\vartheta_2(\tau)$ in Eq.~\eqref{eq:vartheta12}. In particular, the modular multiplet $Y^{(k)}_{\bm{r}}(\tau_f)$ at the modular fixed point $\tau=\tau_f$ are aligned along specific direction, as shown in Eq.~\eqref{eq-modular-multiplet-0} and Eq.~\eqref{eq:Ytauf}. We are concerned with the triplet modular form multiplets which are relevant to the modular Littlest seesaw models. We list the alignment of the triplet modular multiplets up to weight 6 at level $N=4$ in table~\ref{tab:tb-fp-S4p}.

{\fontsize{8pt}{4pt}
\begin{center}
\setlength\LTcapwidth{\textwidth}
\begin{longtable}{|c|c|c|c|c|c|c|c|c|c|}
\caption{\label{tab:tb-fp-S4p}
The alignments of modular triplets of level $N=4$ at the modular fixed points, where we consider modular weights up to 6.}
\endfirsthead
\multicolumn{10}{c}{{\bfseries \tablename\ \thetable{} -- continued from previous page}}
\endhead
\endlastfoot
\hline
\multicolumn{10}{|c|}{The alignments of triplet modular forms $Y_{\bm{3},\,\bm{3}^\prime,\,\widehat{\bm{3}},\,\widehat{\bm{3}}^\prime}(\gamma\tau_{S})$ up to weight 6}\\
\hline
$\gamma$
& $\gamma\tau_S$
&  \multicolumn{2}{c|}{$\begin{array}{l}Y^{(1)}_{\widehat{\bm{3}}^\prime}(\gamma\tau_S)\end{array}$}
&  \multicolumn{2}{c|}{$\begin{array}{l}Y^{(2)}_{\bm{3}}(\gamma\tau_S)\\Y^{(6)}_{\bm{3} II}(\gamma\tau_S)\end{array}$ }
&  \multicolumn{2}{c|}{$\begin{array}{l}Y^{(3)}_{\widehat{\bm{3}}}(\gamma\tau_S)\\Y^{(5)}_{\widehat{\bm{3}}^\prime I}(\gamma\tau_S)\end{array}$ }
& $\begin{array}{l}Y^{(3)}_{\widehat{\bm{3}}^\prime}(\gamma\tau_S)\\Y^{(4)}_{\bm{3}}(\gamma\tau_S)\\Y^{(5)}_{\widehat{\bm{3}}}(\gamma\tau_S)\\Y^{(6)}_{\bm{3}^\prime}(\gamma\tau_S)\end{array}$
& $\begin{array}{l}Y^{(4)}_{\bm{3}^\prime}(\gamma\tau_S)\\Y^{(5)}_{\widehat{\bm{3}}^\prime II}(\gamma\tau_S)\\Y^{(6)}_{\bm{3} I}(\gamma\tau_S)\end{array}$\\
\hline
$\{1,S,S^2,S^3\}$
& $i$
& \multicolumn{2}{c|}{ $\begin{pmatrix}1\\1-\frac{1}{\sqrt{2}}\\-1-\frac{1}{\sqrt{2}}\end{pmatrix}$ }
&  \multicolumn{2}{c|}{$\begin{pmatrix}1\\-\frac{1}{2}-\frac{1}{\sqrt{2}}\\\frac{1}{2}-\frac{1}{\sqrt{2}}\end{pmatrix}$ }
&  \multicolumn{2}{c|}{$\begin{pmatrix}1\\-2-\frac{1}{\sqrt{2}}\\2-\frac{1}{\sqrt{2}}\end{pmatrix}$ }
& $\begin{pmatrix}1\\\frac{1}{\sqrt{2}}\\\frac{1}{\sqrt{2}}\end{pmatrix}$
& $\begin{pmatrix}1\\1-\frac{1}{\sqrt{2}}\\-1-\frac{1}{\sqrt{2}}\end{pmatrix}$ \\
\hline
$\begin{array}{l}\{T,TS,\\TS^2,TS^3\}\end{array}$
& $1+i$
&  \multicolumn{2}{c|}{$\begin{pmatrix}1\\i\left(1-\frac{1}{\sqrt{2}}\right)\\i\left(1+\frac{1}{\sqrt{2}}\right)\end{pmatrix}$ }
&  \multicolumn{2}{c|}{$\begin{pmatrix}1\\i\left(-\frac{1}{2}-\frac{1}{\sqrt{2}}\right)\\i\left(-\frac{1}{2}+\frac{1}{\sqrt{2}}\right)\end{pmatrix}$ }
&  \multicolumn{2}{c|}{$\begin{pmatrix}1\\i\left(-2-\frac{1}{\sqrt{2}}\right)\\i\left(-2+\frac{1}{\sqrt{2}}\right)\end{pmatrix}$ }
& $\begin{pmatrix}1\\\frac{i}{\sqrt{2}}\\-\frac{i}{\sqrt{2}}\end{pmatrix}$
& $\begin{pmatrix}1\\i\left(1-\frac{1}{\sqrt{2}}\right)\\i\left(1+\frac{1}{\sqrt{2}}\right)\end{pmatrix}$ \\
\hline
$\begin{array}{l}\{T^2,T^2S,\\T^2S^2,T^2S^3\}\end{array}$
& $2+i$
& \multicolumn{2}{c|}{$\begin{pmatrix}1\\-1+\frac{1}{\sqrt{2}}\\1+\frac{1}{\sqrt{2}}\end{pmatrix}$ }
& \multicolumn{2}{c|}{$\begin{pmatrix}1\\\frac{1}{2}+\frac{1}{\sqrt{2}}\\-\frac{1}{2}+\frac{1}{\sqrt{2}}\end{pmatrix}$ }
& \multicolumn{2}{c|}{$\begin{pmatrix}1\\2+\frac{1}{\sqrt{2}}\\-2+\frac{1}{\sqrt{2}}\end{pmatrix}$ }
& $\begin{pmatrix}1\\-\frac{1}{\sqrt{2}}\\-\frac{1}{\sqrt{2}}\end{pmatrix}$
& $\begin{pmatrix}1\\-1+\frac{1}{\sqrt{2}}\\1+\frac{1}{\sqrt{2}}\end{pmatrix}$ \\
\hline
$\begin{array}{l}\{T^3, T^3S,\\T^3S^2,(ST)^2\}\end{array}$
& $3+i$
& \multicolumn{2}{c|}{$\begin{pmatrix}1\\i\left(-1+\frac{1}{\sqrt{2}}\right)\\i\left(-1-\frac{1}{\sqrt{2}}\right)\end{pmatrix}$ }
& \multicolumn{2}{c|}{$\begin{pmatrix}1\\
i\left(\frac{1}{2}+\frac{1}{\sqrt{2}}\right)\\
i\left(\frac{1}{2}-\frac{1}{\sqrt{2}}\right)\end{pmatrix}$}
& \multicolumn{2}{c|}{$\begin{pmatrix}1\\
i\left(2+\frac{1}{\sqrt{2}}\right)\\
i\left(2-\frac{1}{\sqrt{2}}\right)\end{pmatrix}$}
& $\begin{pmatrix}1\\-\frac{i}{\sqrt{2}}\\\frac{i}{\sqrt{2}}\end{pmatrix}$
& $\begin{pmatrix}1\\
i\left(-1+\frac{1}{\sqrt{2}}\right)\\
i\left(-1-\frac{1}{\sqrt{2}}\right)\end{pmatrix}$ \\
\hline
$\begin{array}{l}\{(TS)^2, TS^3T,\\ ST^3, TST\}\end{array}$
& $\frac{1}{2}+\frac{i}{2}$
& \multicolumn{2}{c|}{$\begin{pmatrix}1\\\frac{1}{2}+\frac{i}{2}\\-\frac{1}{2}+\frac{i}{2}\end{pmatrix}$ }
& \multicolumn{2}{c|}{$\begin{pmatrix}1\\-1-i\\1-i\end{pmatrix}$ }
& \multicolumn{2}{c|}{$\begin{pmatrix}1\\-\frac{1}{4}-\frac{i}{4}\\\frac{1}{4}-\frac{i}{4}\end{pmatrix}$ }
& $\begin{pmatrix}0\\1\\-i\end{pmatrix}$
& $\begin{pmatrix}1\\\frac{1}{2}+\frac{i}{2}\\-\frac{1}{2}+\frac{i}{2}\end{pmatrix}$ \\
\hline
$\begin{array}{l}\{ST,STS,\\ S^3T, STS^3\}\end{array}$
& $-\frac{1}{2}+\frac{i}{2}$
& \multicolumn{2}{c|}{$\begin{pmatrix}1\\\frac{1}{2}-\frac{i}{2}\\-\frac{1}{2}-\frac{i}{2}\end{pmatrix}$ }
& \multicolumn{2}{c|}{$\begin{pmatrix}1\\-1+i\\1+i\end{pmatrix}$ }
& \multicolumn{2}{c|}{$\begin{pmatrix}1\\-\frac{1}{4}+\frac{i}{4}\\\frac{1}{4}+\frac{i}{4}\end{pmatrix}$}
& $\begin{pmatrix}0\\1\\i\end{pmatrix}$
& $\begin{pmatrix}1\\\frac{1}{2}-\frac{i}{2}\\-\frac{1}{2}-\frac{i}{2}\end{pmatrix}$ \\
\hline
$\begin{array}{l}\{ST^2,ST^2S,\\ S^3T^2,ST^2S^3\}\end{array}$
& $-\frac{2}{5}+\frac{i}{5}$
& \multicolumn{2}{c|}{$\begin{pmatrix}1\\1+\frac{1}{\sqrt{2}}\\-1+\frac{1}{\sqrt{2}}\end{pmatrix}$ }
& \multicolumn{2}{c|}{$\begin{pmatrix}1\\-\frac{1}{2}+\frac{1}{\sqrt{2}}\\\frac{1}{2}+\frac{1}{\sqrt{2}}\end{pmatrix}$ }
& \multicolumn{2}{c|}{$\begin{pmatrix}1\\-2+\frac{1}{\sqrt{2}}\\2+\frac{1}{\sqrt{2}}\end{pmatrix}$ }
& $\begin{pmatrix}1\\-\frac{1}{\sqrt{2}}\\-\frac{1}{\sqrt{2}}\end{pmatrix}$
& $\begin{pmatrix}1\\1+\frac{1}{\sqrt{2}}\\-1+\frac{1}{\sqrt{2}}\end{pmatrix}$ \\
\hline
$\begin{array}{l}\{T^3S^3T,\\ T^2ST^3,\\T^3ST,\\ T^2S^3T^3\}\end{array}$
& $\frac{5}{2}+\frac{i}{2}$
& \multicolumn{2}{c|}{$\begin{pmatrix}1\\-\frac{1}{2}-\frac{i}{2}\\\frac{1}{2}-\frac{i}{2}\end{pmatrix}$ }
& \multicolumn{2}{c|}{$\begin{pmatrix}1\\1+i\\-1+i\end{pmatrix}$ }
& \multicolumn{2}{c|}{$\begin{pmatrix}1\\\frac{1}{4}+\frac{i}{4}\\-\frac{1}{4}+\frac{i}{4}\end{pmatrix}$ }
& $\begin{pmatrix}0\\1\\-i\end{pmatrix}$
& $\begin{pmatrix}1\\-\frac{1}{2}-\frac{i}{2}\\\frac{1}{2}-\frac{i}{2}\end{pmatrix}$ \\
\hline
$\begin{array}{l}\{T^2ST,\\ TS^3T^3,\\T^2S^3T,\\ TST^3\}\end{array}$
& $\frac{3}{2}+\frac{i}{2}$
& \multicolumn{2}{c|}{$\begin{pmatrix}1\\-\frac{1}{2}+\frac{i}{2}\\\frac{1}{2}+\frac{i}{2}\end{pmatrix}$ }
& \multicolumn{2}{c|}{$\begin{pmatrix}1\\1-i\\-1-i\end{pmatrix}$ }
& \multicolumn{2}{c|}{$\begin{pmatrix}1\\\frac{1}{4}-\frac{i}{4}\\-\frac{1}{4}-\frac{i}{4}\end{pmatrix}$ }
& $\begin{pmatrix}0\\1\\i\end{pmatrix}$
& $\begin{pmatrix}1\\-\frac{1}{2}+\frac{i}{2}\\\frac{1}{2}+\frac{i}{2}\end{pmatrix}$ \\
\hline
$\begin{array}{l}\{TS^3T^2,\\TST^2S^3,\\ TST^2,\\ TST^2S\}\end{array}$
& $\frac{3}{5}+\frac{i}{5}$
& \multicolumn{2}{c|}{$\begin{pmatrix}1\\
i\left(1+\frac{1}{\sqrt{2}}\right)\\
i\left(1-\frac{1}{\sqrt{2}}\right)\end{pmatrix}$ }
& \multicolumn{2}{c|}{$\begin{pmatrix}1\\
i\left(-\frac{1}{2}+\frac{1}{\sqrt{2}}\right)\\
i\left(-\frac{1}{2}-\frac{1}{\sqrt{2}}\right)\end{pmatrix}$ }
& \multicolumn{2}{c|}{$\begin{pmatrix}1\\
i\left(-2+\dfrac{1}{\sqrt{2}}\right)\\
i\left(-2-\dfrac{1}{\sqrt{2}}\right)\end{pmatrix}$ }
& $\begin{pmatrix}1\\-\frac{i}{\sqrt{2}}\\\frac{i}{\sqrt{2}}\end{pmatrix}$
& $\begin{pmatrix}1\\
i\left(1+\frac{1}{\sqrt{2}}\right)\\
i\left(1-\frac{1}{\sqrt{2}}\right)\end{pmatrix}$ \\
\hline
$\begin{array}{l}\{ ST^2ST,\\ T^3ST^2,\\ST^2S^3T,\\ T^3S^3T^2\}\end{array}$
& $-\frac{3}{5}+\frac{i}{5}$
&\multicolumn{2}{c|}{ $\begin{pmatrix}1\\i\left(-1-\frac{1}{\sqrt{2}}\right)\\i\left(-1+\frac{1}{\sqrt{2}}\right)\end{pmatrix}$ }
&\multicolumn{2}{c|}{ $\begin{pmatrix}1\\
i\left(\frac{1}{2}-\frac{1}{\sqrt{2}}\right)\\
i\left(\frac{1}{2}+\frac{1}{\sqrt{2}}\right)\end{pmatrix}$ }
&\multicolumn{2}{c|}{ $\begin{pmatrix}1\\
i\left(2-\frac{1}{\sqrt{2}}\right)\\
i\left(2+\frac{1}{\sqrt{2}}\right)\end{pmatrix}$ }
& $\begin{pmatrix}1\\\frac{i}{\sqrt{2}}\\-\frac{i}{\sqrt{2}}\end{pmatrix}$
& $\begin{pmatrix}1\\
i\left(-1-\frac{1}{\sqrt{2}}\right)\\
i\left(-1+\frac{1}{\sqrt{2}}\right)\end{pmatrix}$ \\
\hline
$\begin{array}{l}\{T^2ST^2,\\(ST^2)^2,\\ T^2S^3T^2,\\ ST^2S^3T^2\}\end{array}$
& $\frac{8}{5}+\frac{i}{5}$
& \multicolumn{2}{c|}{$\begin{pmatrix}1\\-1-\frac{1}{\sqrt{2}}\\1-\frac{1}{\sqrt{2}}\end{pmatrix}$ }
& \multicolumn{2}{c|}{$\begin{pmatrix}1\\\frac{1}{2}-\frac{1}{\sqrt{2}}\\-\frac{1}{2}-\frac{1}{\sqrt{2}}\end{pmatrix}$ }
& \multicolumn{2}{c|}{$\begin{pmatrix}1\\2-\frac{1}{\sqrt{2}}\\-2-\frac{1}{\sqrt{2}}\end{pmatrix}$ }
& $\begin{pmatrix}1\\\frac{1}{\sqrt{2}}\\\frac{1}{\sqrt{2}}\end{pmatrix}$
& $\begin{pmatrix}1\\-1-\frac{1}{\sqrt{2}}\\1-\frac{1}{\sqrt{2}}\end{pmatrix}$ \\
\hline
\multicolumn{10}{|c|}{The alignments of triplet modular forms $Y_{\bm{3},\,\bm{3}^\prime,\,\widehat{\bm{3}},\,\widehat{\bm{3}}^\prime}(\gamma\tau_{ST})$ up to weight 6}\\
\hline
$\gamma$
& $\gamma\tau_{ST}$
& \multicolumn{3}{c|}{$\begin{array}{l}Y^{(1)}_{\widehat{\bm{3}}^\prime}(\gamma\tau_{ST})\\Y^{(4)}_{\bm{3}}(\gamma\tau_{ST})\\Y^{(4)}_{\bm{3}^\prime}(\gamma\tau_{ST})\end{array}$}
& \multicolumn{3}{c|}{$\begin{array}{l}Y^{(2)}_{\bm{3}}(\gamma\tau_{ST})\\Y^{(5)}_{\widehat{\bm{3}}}(\gamma\tau_{ST})\\Y^{(5)}_{\widehat{\bm{3}}^\prime I}(\gamma\tau_{ST})\end{array}$}
& $\begin{array}{l}Y^{(3)}_{\widehat{\bm{3}}}(\gamma\tau_{ST})\\Y^{(3)}_{\widehat{\bm{3}}^\prime}(\gamma\tau_{ST})\\Y^{(6)}_{\bm{3} I}(\gamma\tau_{ST})\\Y^{(6)}_{\bm{3}^\prime}(\gamma\tau_{ST})\end{array}$
& $\begin{array}{l}Y^{(5)}_{\widehat{\bm{3}}^\prime II}(\gamma\tau_{ST})\\Y^{(6)}_{\bm{3} II}(\gamma\tau_{ST})\end{array}$\\
\hline
$\begin{array}{l}\{1, ST, \\(ST)^2\}\end{array}$
& \multirow{2}{*}[-2ex]{$-\frac{1}{2}+i\frac{\sqrt{3}}{2}$}
& \multicolumn{3}{c|}{\multirow{2}{*}{$\begin{pmatrix}1\\\frac{1}{2} \left(\sqrt{3}-1\right) \xi ^7\\\frac{1}{2} \left(\sqrt{3}+1\right) \xi ^5\end{pmatrix}$ }}
& \multicolumn{3}{c|}{\multirow{2}{*}{$\begin{pmatrix}1\\\frac{1}{2} \left(\sqrt{3}+1\right) \xi ^3\\\frac{1}{2} \left(\sqrt{3}-1\right) \xi \end{pmatrix}$}}
& \multirow{2}{*}{$\begin{pmatrix}1\\\xi ^7\\\xi \end{pmatrix}$}
& \multirow{2}{*}{$\begin{pmatrix}0\\0\\0\end{pmatrix}$} \\
\cline{1-1}
$\begin{array}{l}\{S^2, S^3T,\\ T^3S\}\end{array}$ & & \multicolumn{3}{c|}{} & \multicolumn{3}{c|}{} & &\\
\hline
$\begin{array}{l}\{S,TS^2,\\ TS^3T\}\end{array}$
& \multirow{2}{*}[-2ex]{$\frac{1}{2}+i\frac{\sqrt{3}}{2}$}
& \multicolumn{3}{c|}{ \multirow{2}{*}{$\begin{pmatrix}1\\\frac{1}{2} \left(\sqrt{3}-1\right) \xi \\\frac{1}{2} \left(\sqrt{3}+1\right) \xi ^3\end{pmatrix}$}}
& \multicolumn{3}{c|}{ \multirow{2}{*}{$\begin{pmatrix}1\\\frac{1}{2} \left(\sqrt{3}+1\right) \xi ^5\\\frac{1}{2} \left(\sqrt{3}-1\right) \xi ^7\end{pmatrix}$ }}
&  \multirow{2}{*}{$\begin{pmatrix}1\\\xi \\\xi ^7\end{pmatrix}$}
&  \multirow{2}{*}{$\begin{pmatrix}0\\0\\0\end{pmatrix}$ }\\
\cline{1-1}
$\begin{array}{l}\{S^3,T,\\ TST\}\end{array}$ & & \multicolumn{3}{c|}{} & \multicolumn{3}{c|}{} & &\\
\hline
$\begin{array}{l}\{T^2, T^2ST,\\TS^3\}\end{array}$
& \multirow{2}{*}[-2ex]{$\frac{1}{2} \left(3+i \sqrt{3}\right)$}
& \multicolumn{3}{c|}{\multirow{2}{*}{$\begin{pmatrix}1\\\frac{1}{2} \left(\sqrt{3}-1\right) \xi ^3\\\frac{1}{2} \left(\sqrt{3}+1\right) \xi \end{pmatrix}$ }}
& \multicolumn{3}{c|}{\multirow{2}{*}{$\begin{pmatrix}1\\\frac{1}{2} \left(\sqrt{3}+1\right) \xi ^7\\\frac{1}{2} \left(\sqrt{3}-1\right) \xi^5 \end{pmatrix}$ }}
& \multirow{2}{*}{$\begin{pmatrix}1\\\xi ^3\\\xi ^5\end{pmatrix}$}
& \multirow{2}{*}{$\begin{pmatrix}0\\0\\0\end{pmatrix}$} \\
\cline{1-1}
$\begin{array}{l}\{TS,T^2S^2,\\T^2S^3T\}\end{array}$ & & \multicolumn{3}{c|}{} & \multicolumn{3}{c|}{} & &\\
\hline
$\begin{array}{l}\{T^3,T^3ST,\\T^2S^3\}\end{array}$
& \multirow{2}{*}[-2ex]{$\frac{1}{2} \left(5+i \sqrt{3}\right)$}
& \multicolumn{3}{c|}{\multirow{2}{*}{$\begin{pmatrix}1\\\frac{1}{2} \left(\sqrt{3}-1\right) \xi ^5\\\frac{1}{2} \left(\sqrt{3}+1\right) \xi ^7\end{pmatrix}$}}
&\multicolumn{3}{c|}{ \multirow{2}{*}{$\begin{pmatrix}1\\\frac{1}{2} \left(\sqrt{3}+1\right) \xi \\\frac{1}{2} \left(\sqrt{3}-1\right) \xi ^3\end{pmatrix}$ }}
& \multirow{2}{*}{$\begin{pmatrix}1\\\xi ^5\\\xi ^3\end{pmatrix}$}
& \multirow{2}{*}{$\begin{pmatrix}0\\0\\0\end{pmatrix}$} \\
\cline{1-1}
$\begin{array}{l}\{T^2S,T^3S^2,\\ T^3S^3T\}\end{array}$ & & \multicolumn{3}{c|}{} & \multicolumn{3}{c|}{} & &\\
\hline
$\begin{array}{l}\{(TS)^2,\\TS^3T^2,\\ST^2S\}\end{array}$
& \multirow{2}{*}[-2ex]{$\frac{1}{6}\left(3+i\sqrt{3}\right)$}
& \multicolumn{3}{c|}{\multirow{2}{*}{$\begin{pmatrix}1\\\frac{1}{2} \left(\sqrt{3}+1\right) \xi \\\frac{1}{2} \left(\sqrt{3}-1\right) \xi ^3\end{pmatrix}$}}
&\multicolumn{3}{c|}{ \multirow{2}{*}{$\begin{pmatrix}1\\\frac{1}{2} \left(\sqrt{3}-1\right) \xi ^5\\\frac{1}{2} \left(\sqrt{3}+1\right) \xi ^7\end{pmatrix}$ }}
& \multirow{2}{*}{$\begin{pmatrix}1\\\xi ^5\\\xi ^3\end{pmatrix}$}
& \multirow{2}{*}{$\begin{pmatrix}0\\0\\0\end{pmatrix}$} \\
\cline{1-1}
$\begin{array}{l}\{ ST^3, TST^2,\\ST^2S^3\}\end{array}$ & & \multicolumn{3}{c|}{} & \multicolumn{3}{c|}{} & &\\
\hline
$\begin{array}{l}\{ST^2, ST^2ST,\\ STS^3\}\end{array}$
& \multirow{2}{*}[-2ex]{$\dfrac{1}{6}(-3+i\sqrt{3})$}
& \multicolumn{3}{c|}{\multirow{2}{*}{$\begin{pmatrix}1\\\frac{1}{2} \left(\sqrt{3}+1\right) \xi ^7\\\frac{1}{2} \left(\sqrt{3}-1\right) \xi ^5\end{pmatrix}$ }}
&\multicolumn{3}{c|}{ \multirow{2}{*}{$\begin{pmatrix}1\\\frac{1}{2} \left(\sqrt{3}-1\right) \xi ^3\\\frac{1}{2} \left(\sqrt{3}+1\right) \xi \end{pmatrix}$ }}
& \multirow{2}{*}{$\begin{pmatrix}1\\\xi ^3\\\xi ^5\end{pmatrix}$}
& \multirow{2}{*}{$\begin{pmatrix}0\\0\\0\end{pmatrix}$} \\
\cline{1-1}
$\begin{array}{l}\{STS, S^3T^2,\\ST^2S^3T\}\end{array}$ & & \multicolumn{3}{c|}{} & \multicolumn{3}{c|}{} & &\\
\hline
$\begin{array}{l}\{ T^2ST^3,\\ T^3ST^2,\\ ST^2S^3T^2\}\end{array}$
& \multirow{2}{*}[-2ex]{$\frac{1}{6}  \left(15 +i\sqrt{3}\right)$}
& \multicolumn{3}{c|}{\multirow{2}{*}[-1ex]{$\begin{pmatrix}1\\\frac{1}{2} \left(\sqrt{3}+1\right) \xi ^5\\\frac{1}{2} \left(\sqrt{3}-1\right) \xi ^7\end{pmatrix}$ }}
& \multicolumn{3}{c|}{\multirow{2}{*}[-1ex]{$\begin{pmatrix}1\\\frac{1}{2} \left(\sqrt{3}-1\right) \xi \\\frac{1}{2} \left(\sqrt{3}+1\right) \xi ^3\end{pmatrix}$} }
& \multirow{2}{*}[-1ex]{$\begin{pmatrix}1\\\xi \\\xi ^7\end{pmatrix}$}
& \multirow{2}{*}[-1ex]{$\begin{pmatrix}0\\0\\0\end{pmatrix}$} \\
\cline{1-1}
$\begin{array}{l}\{ T^3S^3T^2,\\(ST^2)^2,\\ T^2S^3T^3\}\end{array}$ & & \multicolumn{3}{c|}{} & \multicolumn{3}{c|}{} & &\\
\hline
$\begin{array}{l}\{ TS^3T^3,\\ T^2S^3T^2,\\ TST^2S\}\end{array}$
& \multirow{2}{*}[-3ex]{$\frac{1}{6} \left(9+i \sqrt{3}\right)$}
& \multicolumn{3}{c|}{\multirow{2}{*}[-1ex]{$\begin{pmatrix}1\\\frac{1}{2} \left(\sqrt{3}+1\right) \xi ^3\\\frac{1}{2} \left(\sqrt{3}-1\right) \xi \end{pmatrix}$ }}
&\multicolumn{3}{c|}{\multirow{2}{*}[-1ex]{ $\begin{pmatrix}1\\\frac{1}{2} \left(\sqrt{3}-1\right) \xi^7 \\\frac{1}{2} \left(\sqrt{3}+1\right) \xi ^5\end{pmatrix}$}}
& \multirow{2}{*}[-1ex]{$\begin{pmatrix}1\\\xi ^7\\\xi \end{pmatrix}$}
& \multirow{2}{*}[-1ex]{$\begin{pmatrix}0\\0\\0\end{pmatrix}$} \\
\cline{1-1}
$\begin{array}{l}\{TST^2S^3,\\ TST^3,\\T^2ST^2\}\end{array}$ & & \multicolumn{3}{c|}{} & \multicolumn{3}{c|}{} & &\\
\hline
\multicolumn{10}{|c|}{The alignments of triplet modular forms $Y_{\bm{3},\,\bm{3}^\prime,\,\widehat{\bm{3}},\,\widehat{\bm{3}}^\prime}(\gamma\tau_{TS})$ of weight 1 up to 6}\\
\hline
$\gamma$
& $\gamma\tau_{TS}$
& \multicolumn{3}{c|}{$\begin{array}{l}Y^{(1)}_{\widehat{\bm{3}}^\prime}(\gamma\tau_{TS})\\Y^{(4)}_{\bm{3}}(\gamma\tau_{TS})\\Y^{(4)}_{\bm{3}^\prime}(\gamma\tau_{TS})\end{array}$}
& \multicolumn{3}{c|}{$\begin{array}{l}Y^{(2)}_{\bm{3}}(\gamma\tau_{TS})\\Y^{(5)}_{\widehat{\bm{3}}}(\gamma\tau_{TS})\\Y^{(5)}_{\widehat{\bm{3}}^\prime I}(\gamma\tau_{TS})\end{array}$}
& $\begin{array}{l}Y^{(3)}_{\widehat{\bm{3}}}(\gamma\tau_{TS})\\Y^{(3)}_{\widehat{\bm{3}}^\prime}(\gamma\tau_{TS})\\Y^{(6)}_{\bm{3} I}(\gamma\tau_{TS})\\Y^{(6)}_{\bm{3}^\prime}(\gamma\tau_{TS})\end{array}$
& $\begin{array}{l}Y^{(5)}_{\widehat{\bm{3}}^\prime II}(\gamma\tau_{TS})\\Y^{(6)}_{\bm{3} II}(\gamma\tau_{TS})\end{array}$\\
\hline
$\begin{array}{l}\{1, TS, \\(TS)^2\}\end{array}$ &\multirow{2}{*}[-2ex]{$\frac{1}{2} \left(1+i \sqrt{3}\right)$}
& \multicolumn{3}{c|}{\multirow{2}{*}{$\begin{pmatrix}1\\\frac{1}{2} \left(\sqrt{3}-1\right) \xi \\\frac{1}{2} \left(\sqrt{3}+1\right) \xi ^3\end{pmatrix}$ }}
& \multicolumn{3}{c|}{\multirow{2}{*}{$\begin{pmatrix}1\\\frac{1}{2} \left(\sqrt{3}+1\right) \xi ^5\\\frac{1}{2} \left(\sqrt{3}-1\right) \xi ^7\end{pmatrix}$ }}
& \multirow{2}{*}{$\begin{pmatrix}1\\\xi \\\xi ^7\end{pmatrix}$}
& \multirow{2}{*}{$\begin{pmatrix}0\\0\\0\end{pmatrix}$} \\
\cline{1-1}
$\begin{array}{l}\{S^2,TS^3,\\ ST^3\}\end{array}$& & \multicolumn{3}{c|}{} & \multicolumn{3}{c|}{} & &\\
\hline
$\begin{array}{l}\{S,STS,T^3\}\end{array}$ & \multirow{2}{*}[-2ex]{$\frac{1}{2}\left(-1+i\sqrt{3}\right)$}
& \multicolumn{3}{c|}{\multirow{2}{*}{$\begin{pmatrix}1\\\frac{1}{2} \left(\sqrt{3}-1\right) \xi ^7\\\frac{1}{2} \left(\sqrt{3}+1\right) \xi ^5\end{pmatrix}$ }}
& \multicolumn{3}{c|}{\multirow{2}{*}{$\begin{pmatrix}1\\\frac{1}{2} \left(\sqrt{3}+1\right) \xi ^3\\\frac{1}{2} \left(\sqrt{3}-1\right) \xi \end{pmatrix}$ }}
& \multirow{2}{*}{$\begin{pmatrix}1\\\xi ^7\\\xi \end{pmatrix}$}
& \multirow{2}{*}{$\begin{pmatrix}0\\0\\0\end{pmatrix}$} \\
\cline{1-1}
$\begin{array}{l}\{S^3, STS^3,\\T^3S^2\}\end{array}$ & & \multicolumn{3}{c|}{} & \multicolumn{3}{c|}{} & &\\
\hline
$\begin{array}{l}\{T,T^2S,\\ TS^3T^3\}\end{array}$ & \multirow{2}{*}[-2ex]{$\frac{1}{2} \left(3+i \sqrt{3}\right)$}
& \multicolumn{3}{c|}{\multirow{2}{*}{$\begin{pmatrix}1\\\frac{1}{2} \left(\sqrt{3}-1\right) \xi ^3\\\frac{1}{2} \left(\sqrt{3}+1\right) \xi \end{pmatrix}$ }}
& \multicolumn{3}{c|}{\multirow{2}{*}{$\begin{pmatrix}1\\\frac{1}{2} \left(\sqrt{3}+1\right) \xi ^7\\\frac{1}{2} \left(\sqrt{3}-1\right) \xi^5 \end{pmatrix}$ }}
& \multirow{2}{*}{$\begin{pmatrix}1\\\xi ^3\\\xi ^5\end{pmatrix}$}
& \multirow{2}{*}{$\begin{pmatrix}0\\0\\0\end{pmatrix}$} \\
\cline{1-1}
$\begin{array}{l}\{TS^2,T^2S^3,\\ TST^3\}\end{array}$ & & \multicolumn{3}{c|}{} & \multicolumn{3}{c|}{} & &\\
\hline
$\begin{array}{l}\{T^2, T^3S,\\ T^2S^3T^3\}\end{array}$ & \multirow{2}{*}[-2ex]{$\frac{1}{2} \left(5+i \sqrt{3}\right)$}
& \multicolumn{3}{c|}{\multirow{2}{*}[-1ex]{$\begin{pmatrix}1\\\frac{1}{2} \left(\sqrt{3}-1\right) \xi ^5\\\frac{1}{2} \left(\sqrt{3}+1\right) \xi ^7\end{pmatrix}$ }}
& \multicolumn{3}{c|}{\multirow{2}{*}[-1ex]{$\begin{pmatrix}1\\\frac{1}{2} \left(\sqrt{3}+1\right) \xi \\\frac{1}{2} \left(\sqrt{3}-1\right) \xi ^3\end{pmatrix}$ }}
& \multirow{2}{*}[-1ex]{$\begin{pmatrix}1\\\xi ^5\\\xi ^3\end{pmatrix}$}
& \multirow{2}{*}[-1ex]{$\begin{pmatrix}0\\0\\0\end{pmatrix}$} \\
\cline{1-1}
$\begin{array}{l}\{(ST)^2,\\ T^2ST^3,\\T^2S^2\}\end{array}$ & & \multicolumn{3}{c|}{} & \multicolumn{3}{c|}{} & &\\
\hline
$\begin{array}{l}\{ST,ST^2S,\\ T^3ST^2\}\end{array}$ & \multirow{2}{*}[-2ex]{$\frac{1}{6}(-3+i\sqrt{3})$}
& \multicolumn{3}{c|}{\multirow{2}{*}[-1ex]{$\begin{pmatrix}1\\\frac{1}{2} \left(\sqrt{3}+1\right) \xi ^7\\\frac{1}{2} \left(\sqrt{3}-1\right) \xi ^5\end{pmatrix}$ }}
& \multicolumn{3}{c|}{\multirow{2}{*}[-1ex]{$\begin{pmatrix}1\\\frac{1}{2} \left(\sqrt{3}-1\right) \xi^3 \\\frac{1}{2} \left(\sqrt{3}+1\right) \xi \end{pmatrix}$}}
& \multirow{2}{*}[-1ex]{$\begin{pmatrix}1\\\xi ^3\\\xi ^5\end{pmatrix}$}
& \multirow{2}{*}[-1ex]{$\begin{pmatrix}0\\0\\0\end{pmatrix}$} \\
\cline{1-1}
$\begin{array}{l}\{S^3T,\\ST^2S^3,\\T^3S^3T^2\}\end{array}$ & & \multicolumn{3}{c|}{} & \multicolumn{3}{c|}{} & &\\
\hline
$\begin{array}{l}\{ST^2, TST,\\ TST^2S\}\end{array}$ & \multirow{2}{*}[-2ex]{$\frac{1}{6} \left(3+i\sqrt{3}\right)$}
& \multicolumn{3}{c|}{\multirow{2}{*}[-1ex]{$\begin{pmatrix}1\\\frac{1}{2} \left(\sqrt{3}+1\right) \xi \\\frac{1}{2} \left(\sqrt{3}-1\right) \xi ^3\end{pmatrix}$ }}
& \multicolumn{3}{c|}{\multirow{2}{*}[-1ex]{$\begin{pmatrix}1\\\frac{1}{2} \left(\sqrt{3}-1\right) \xi ^5\\\frac{1}{2} \left(\sqrt{3}+1\right) \xi^7 \end{pmatrix}$ }}
& \multirow{2}{*}[-1ex]{$\begin{pmatrix}1\\\xi ^5\\\xi ^3\end{pmatrix}$}
& \multirow{2}{*}[-1ex]{$\begin{pmatrix}0\\0\\0\end{pmatrix}$} \\
\cline{1-1}
$\begin{array}{l}\{ TS^3T,\\TST^2S^3,\\ S^3T^2\}\end{array}$ & & \multicolumn{3}{c|}{} & \multicolumn{3}{c|}{} & &\\
\hline
$\begin{array}{l}\{ T^3S^3T,\\ ST^2ST,\\ T^2S^3T^2\}\end{array}$ & \multirow{2}{*}[-2ex]{$\frac{1}{6} \left(15+i\sqrt{3}\right)$}
& \multicolumn{3}{c|}{\multirow{2}{*}[-1ex]{$\begin{pmatrix}1\\\frac{1}{2} \left(\sqrt{3}+1\right) \xi ^5\\\frac{1}{2} \left(\sqrt{3}-1\right) \xi ^7\end{pmatrix}$ }}
& \multicolumn{3}{c|}{\multirow{2}{*}[-1ex]{$\begin{pmatrix}1\\\frac{1}{2} \left(\sqrt{3}-1\right) \xi \\\frac{1}{2} \left(\sqrt{3}+1\right) \xi ^3\end{pmatrix}$ }}
& \multirow{2}{*}[-1ex]{$\begin{pmatrix}1\\\xi \\\xi ^7\end{pmatrix}$}
& \multirow{2}{*}[-1ex]{$\begin{pmatrix}0\\0\\0\end{pmatrix}$} \\
\cline{1-1}
$\begin{array}{l}\{T^3ST,\\ST^2S^3T,\\T^2ST^2\}\end{array}$ & & \multicolumn{3}{c|}{} & \multicolumn{3}{c|}{} & &\\
\hline
$\begin{array}{l}\{ T^2ST,\\(ST^2)^2,\\ TST^2\}\end{array}$ & \multirow{2}{*}[-2ex]{$\frac{1}{6} \left(9+i \sqrt{3}\right)$}
& \multicolumn{3}{c|}{\multirow{2}{*}[-1ex]{$\begin{pmatrix}1\\\frac{1}{2} \left(\sqrt{3}+1\right) \xi ^3\\\frac{1}{2} \left(\sqrt{3}-1\right) \xi \end{pmatrix}$ }}
& \multicolumn{3}{c|}{\multirow{2}{*}[-1ex]{$\begin{pmatrix}1\\\frac{1}{2} \left(\sqrt{3}-1\right) \xi ^7\\\frac{1}{2} \left(\sqrt{3}+1\right) \xi ^5\end{pmatrix}$ }}
& \multirow{2}{*}[-1ex]{$\begin{pmatrix}1\\\xi ^7\\\xi \end{pmatrix}$}
& \multirow{2}{*}[-1ex]{$\begin{pmatrix}0\\0\\0\end{pmatrix}$} \\
\cline{1-1}
$\begin{array}{l}\{TS^3T^2,\\T^2S^3T,\\ ST^2S^3T^2\}\end{array}$ & & \multicolumn{3}{c|}{} & \multicolumn{3}{c|}{} & &\\
\hline
\multicolumn{10}{|c|}{The alignments of triplet modular forms $Y_{\bm{3},\,\bm{3}^\prime,\,\widehat{\bm{3}},\,\widehat{\bm{3}}^\prime}(\gamma\tau_{T})$ up to weight 6}\\
\hline
$\gamma$
& $\gamma\tau_{T}$
& \multicolumn{3}{c|}{$\begin{array}{l}Y^{(1)}_{\widehat{\bm{3}}^\prime}(\gamma\tau_{T})\\Y^{(3)}_{\widehat{\bm{3}}^\prime}(\gamma\tau_{T})\\Y^{(5)}_{\widehat{\bm{3}}^\prime I}(\gamma\tau_{T})\\Y^{(5)}_{\widehat{\bm{3}}^\prime II}(\gamma\tau_{T})\end{array}$}
& \multicolumn{3}{c|}{$\begin{array}{l}Y^{(2)}_{\bm{3}}(\gamma\tau_{TS})\\Y^{(4)}_{\bm{3}}(\gamma\tau_{T})\\Y^{(6)}_{\bm{3} I}(\gamma\tau_{T})\\Y^{(6)}_{\bm{3} II}(\gamma\tau_{T})\end{array}$}
& $\begin{array}{l}Y^{(3)}_{\widehat{\bm{3}}}(\gamma\tau_{T})\\Y^{(5)}_{\widehat{\bm{3}}}(\gamma\tau_{T})\end{array}$
& $\begin{array}{l}Y^{(4)}_{\bm{3}^\prime}(\gamma\tau_{T})\\Y^{(6)}_{\bm{3}^\prime}(\gamma\tau_{T})\end{array}$\\
\hline
$\begin{array}{l}\{1, T, T^2, T^3\}\end{array}$
& \multirow{2}{*}[-2ex]{$i\infty$}
& \multicolumn{3}{c|}{\multirow{2}{*}{$\begin{pmatrix}0\\0\\1\end{pmatrix}$ }}
& \multicolumn{3}{c|}{\multirow{2}{*}{$\begin{pmatrix}1\\0\\0\end{pmatrix}$ }}
& \multirow{2}{*}{$\begin{pmatrix}0\\1\\0\end{pmatrix}$}
& \multirow{2}{*}{$\begin{pmatrix}0\\0\\0\end{pmatrix}$} \\
\cline{1-1}
$\begin{array}{l}\{S^2,\,TS^2,\\T^2S^2,\,T^3S^2\}\end{array}$ & &\multicolumn{3}{c|}{} &\multicolumn{3}{c|}{} & & \\
\hline
$\begin{array}{l}\{S,ST\\,ST^2, ST^3\}\end{array}$
& \multirow{2}{*}[-2ex]{$0$}
& \multicolumn{3}{c|}{\multirow{2}{*}{$\begin{pmatrix}1\\\frac{1}{\sqrt{2}}\\-\frac{1}{\sqrt{2}}\end{pmatrix}$ }}
& \multicolumn{3}{c|}{\multirow{2}{*}{$\begin{pmatrix}0\\1\\1\end{pmatrix}$}}
& \multirow{2}{*}{$\begin{pmatrix}1\\-\frac{1}{\sqrt{2}}\\\frac{1}{\sqrt{2}}\end{pmatrix}$}
& \multirow{2}{*}{$\begin{pmatrix}0\\0\\0\end{pmatrix}$} \\
\cline{1-1}
$\begin{array}{l}\{S^3, S^3T,\\ S^3T^2,(TS)^2\}\end{array}$ & &\multicolumn{3}{c|}{} & \multicolumn{3}{c|}{} & &\\
\hline
$\begin{array}{l}\{(ST)^2,\\ T^3S^3T,\\ T^3S^3T^2,\\ STS\}\end{array}$
& \multirow{2}{*}[-3ex]{$-1$}
& \multicolumn{3}{c|}{\multirow{2}{*}{$\begin{pmatrix}1\\-\frac{i}{\sqrt{2}}\\-\frac{i}{\sqrt{2}}\end{pmatrix}$ }}
& \multicolumn{3}{c|}{\multirow{2}{*}{$\begin{pmatrix}0\\1\\-1\end{pmatrix}$ }}
& \multirow{2}{*}{$\begin{pmatrix}1\\\frac{i}{\sqrt{2}}\\\frac{i}{\sqrt{2}}\end{pmatrix}$}
& \multirow{2}{*}{$\begin{pmatrix}0\\0\\0\end{pmatrix}$} \\
\cline{1-1}
$\begin{array}{l}\{ T^3S,T^3ST,\\ T^3ST^2,\\ STS^3\}\end{array}$
&
& \multicolumn{3}{c|}{}
& \multicolumn{3}{c|}{}
&
& \\
\hline
$\begin{array}{l}\{TS, TST,\\ TST^2,\\ TST^3\}\end{array}$
& \multirow{2}{*}[-3ex]{$1$}
& \multicolumn{3}{c|}{\multirow{2}{*}[-2ex]{$\begin{pmatrix}1\\\frac{i}{\sqrt{2}}\\\frac{i}{\sqrt{2}}\end{pmatrix}$ }}
& \multicolumn{3}{c|}{\multirow{2}{*}[-2ex]{$\begin{pmatrix}0\\1\\-1\end{pmatrix}$}}
& \multirow{2}{*}[-2ex]{$\begin{pmatrix}1\\-\frac{i}{\sqrt{2}}\\-\frac{i}{\sqrt{2}}\end{pmatrix}$}
& \multirow{2}{*}[-2ex]{$\begin{pmatrix}0\\0\\0\end{pmatrix}$} \\
\cline{1-1}
$\begin{array}{l}\{TS^3, TS^3T,\\TS^3T^2,\\ TS^3T^3\}\end{array}$
&
& \multicolumn{3}{c|}{}
& \multicolumn{3}{c|}{}
&
& \\
\hline
$\begin{array}{l}\{T^2S,T^2ST,\\T^2ST^2\\,T^2ST^3\}\end{array}$
&  \multirow{2}{*}[-4ex]{$2$}
& \multicolumn{3}{c|}{ \multirow{2}{*}[-2ex]{$\begin{pmatrix}1\\-\frac{1}{\sqrt{2}}\\\frac{1}{\sqrt{2}}\end{pmatrix}$ }}
& \multicolumn{3}{c|}{ \multirow{2}{*}[-2ex]{$\begin{pmatrix}0\\1\\1\end{pmatrix}$ }}
&  \multirow{2}{*}[-2ex]{$\begin{pmatrix}1\\\frac{1}{\sqrt{2}}\\-\frac{1}{\sqrt{2}}\end{pmatrix}$}
&  \multirow{2}{*}[-2ex]{$\begin{pmatrix}0\\0\\0\end{pmatrix}$} \\
\cline{1-1}
$\begin{array}{l}\{T^2S^3,\\T^2S^3T,\\T^2S^3T^2,\\T^2S^3T^3\}\end{array}$
&
& \multicolumn{3}{c|}{}
& \multicolumn{3}{c|}{}
&
& \\
\hline
$\begin{array}{l}\{ST^2S,\\ ST^2ST,\\(ST^2)^2,\\TST^2S^3\}\end{array}$
&  \multirow{2}{*}[-4ex]{$-\frac{1}{2}$}
& \multicolumn{3}{c|}{ \multirow{2}{*}[-3ex]{$\begin{pmatrix}0\\1\\0\end{pmatrix}$ }}
& \multicolumn{3}{c|}{ \multirow{2}{*}[-3ex]{$\begin{pmatrix}1\\0\\0\end{pmatrix}$ }}
&  \multirow{2}{*}[-3ex]{$\begin{pmatrix}0\\0\\1\end{pmatrix}$}
&  \multirow{2}{*}[-3ex]{$\begin{pmatrix}0\\0\\0\end{pmatrix}$} \\
\cline{1-1}
$\begin{array}{l}\{ST^2S^3,\\ST^2S^3T,\\ ST^2S^3T^2,\\ TST^2S\}\end{array}$
&
& \multicolumn{3}{c|}{}
& \multicolumn{3}{c|}{}
&
& \\
\hline
\end{longtable}
\end{center}
}

\section{\label{app:N5}Finite modular group $\Gamma'_5\cong A_5^\prime$ and modular forms of level $N=5$ }

The finite modular group $\Gamma'_5\cong A'_5$ is the double cover of the icosahedral group $A_5$, and its multiplication rules are given by~\cite{Yao:2020zml,Ding:2023htn}
\begin{equation}
S^4=T^5=(ST)^3=1,~~~~ S^2T=TS^2\,.
\end{equation}
The quotient group $A'_5/Z^{S^2}_2$ is isomorphic to $A_5$. Besides the irreducible representations $\bm{1}$, $\bm{3}$, $\bm{3'}$, $\bm{4}$, $\bm{5}$ in common with $A_5$ group,  it has four additional irreducible representations $\widehat{\bm{2}}$, $\widehat{\bm{2}}'$, $\widehat{\bm{4}}$ and $\widehat{\bm{6}}$. In these representations, the generators $S$ and $T$ are represented by
\begin{eqnarray}
\nonumber\bm{1}: & S=1, & T=1, \\
\nonumber \widehat{\bm{2}}:& S=i\sqrt{\frac{1}{\sqrt{5}\phi}}\begin{pmatrix}
 \phi  & 1 \\
 1 & -\phi  \\
\end{pmatrix}, ~& T=\begin{pmatrix}
 \omega_5^2 & 0 \\
 0 & \omega_5^3 \\
\end{pmatrix}\,, \\
\nonumber \widehat{\bm{2}}' : & S=i\sqrt{\frac{1}{\sqrt{5}\phi}}\begin{pmatrix}
 1 ~& \phi  \\
 \phi  ~& -1 \\
\end{pmatrix}, ~& T=\begin{pmatrix}
 \omega_5 ~& 0 \\
 0 ~& \omega_5^4 \\
\end{pmatrix} \,,\\
\nonumber \bm{3}: & S=\frac{1}{\sqrt{5}}\begin{pmatrix}
 1 & -\sqrt{2} & -\sqrt{2} \\
 -\sqrt{2} & -\phi  & \frac{1}{\phi } \\
 -\sqrt{2} & \frac{1}{\phi } & -\phi  \\
\end{pmatrix}, ~&~ T=\begin{pmatrix}
 1 & 0 & 0 \\
 0 & \omega _5 & 0 \\
 0 & 0 & \omega _5^4 \\
\end{pmatrix}\,, \\
\nonumber \bm{3'}: & S=\frac{1}{\sqrt{5}}\begin{pmatrix}
 -1 &\sqrt{2} &\sqrt{2} \\
\sqrt{2} & -\frac{1}{\phi } & \phi  \\
\sqrt{2} & \phi  & -\frac{1}{\phi } \\
\end{pmatrix}, ~&~ T=\begin{pmatrix}
 1 & 0 & 0 \\
 0 & \omega _5^2 & 0 \\
 0 & 0 & \omega _5^3 \\
\end{pmatrix}\,, \\
\nonumber \bm{4}: & S=\frac{1}{\sqrt{5}}\begin{pmatrix}
 1 & \frac{1}{\phi } & \phi  & -1 \\
 \frac{1}{\phi } & -1 & 1 & \phi  \\
 \phi  & 1 & -1 & \frac{1}{\phi } \\
 -1 & \phi  & \frac{1}{\phi } & 1 \\
\end{pmatrix}, ~&~ T=\begin{pmatrix}
 \omega _5 & 0 & 0 & 0 \\
 0 & \omega _5^2 & 0 & 0 \\
 0 & 0 & \omega _5^3 & 0 \\
 0 & 0 & 0 & \omega _5^4 \\
\end{pmatrix}\,,\\
\nonumber \widehat{\bm{4}}: & S=i\sqrt{\frac{1}{5\sqrt{5}\phi}}\begin{pmatrix}
 -\phi ^2 &\sqrt{3} \phi  & -\sqrt{3} & -\frac{1}{\phi } \\
\sqrt{3} \phi  & \frac{1}{\phi } & -\phi ^2 & -\sqrt{3} \\
 -\sqrt{3} & -\phi ^2 & -\frac{1}{\phi } & -\sqrt{3} \phi  \\
 -\frac{1}{\phi } & -\sqrt{3} & -\sqrt{3} \phi  & \phi ^2 \\
\end{pmatrix}, ~&~ T=\begin{pmatrix}
 \omega _5 & 0 & 0 & 0 \\
 0 & \omega _5^2 & 0 & 0 \\
 0 & 0 & \omega _5^3 & 0 \\
 0 & 0 & 0 & \omega _5^4 \\
\end{pmatrix}\,, \\
\nonumber \bm{5}: & S=\frac{1}{5}\begin{pmatrix}
 -1 &\sqrt{6} &\sqrt{6} &\sqrt{6} &\sqrt{6} \\
\sqrt{6} & \frac{1}{\phi ^2} & -2 \phi  & \frac{2}{\phi } & \phi ^2 \\
\sqrt{6} & -2 \phi  & \phi ^2 & \frac{1}{\phi ^2} & \frac{2}{\phi } \\
\sqrt{6} & \frac{2}{\phi } & \frac{1}{\phi ^2} & \phi ^2 & -2 \phi  \\
\sqrt{6} & \phi ^2 & \frac{2}{\phi } & -2 \phi  & \frac{1}{\phi ^2} \\
\end{pmatrix}, ~&~ T=\begin{pmatrix}
 1 & 0 & 0 & 0 & 0 \\
 0 & \omega _5 & 0 & 0 & 0 \\
 0 & 0 & \omega _5^2 & 0 & 0 \\
 0 & 0 & 0 & \omega _5^3 & 0 \\
 0 & 0 & 0 & 0 & \omega _5^4 \\
\end{pmatrix}\,, \\
\label{eq:Irr-matrix-N5}\widehat{\bm{6}}: & S=i\sqrt{\frac{1}{5\sqrt{5}\phi}}\begin{pmatrix}
 -1 & \phi  & \frac{1}{\phi } &\sqrt{2} \phi  &\sqrt{2} & \phi ^2 \\
 \phi  & 1 & \phi ^2 &\sqrt{2} & -\sqrt{2} \phi  & -\frac{1}{\phi } \\
 \frac{1}{\phi } & \phi ^2 & 1 & -\sqrt{2} &\sqrt{2} \phi  & -\phi  \\
\sqrt{2} \phi  &\sqrt{2} & -\sqrt{2} & -\phi  & -1 &\sqrt{2} \phi  \\
\sqrt{2} & -\sqrt{2} \phi  &\sqrt{2} \phi  & -1 & \phi  &\sqrt{2} \\
 \phi ^2 & -\frac{1}{\phi } & -\phi  &\sqrt{2} \phi  &\sqrt{2} & -1 \\
\end{pmatrix}, ~&~ T=\begin{pmatrix}
 1 & 0 & 0 & 0 & 0 & 0 \\
 0 & 1 & 0 & 0 & 0 & 0 \\
 0 & 0 & \omega _5 & 0 & 0 & 0 \\
 0 & 0 & 0 & \omega _5^2 & 0 & 0 \\
 0 & 0 & 0 & 0 & \omega _5^3 & 0 \\
 0 & 0 & 0 & 0 & 0 & \omega _5^4 \\
\end{pmatrix}\,,
\end{eqnarray}
where $\omega_5=e^{2\pi i/5}$ and $\phi=(1+\sqrt{5})/2$ is the golden ratio. Note $S^2=\mathbb{1}$ in the unhatted representations $\bm{1}$, $\bm{3}$, $\bm{3'}$, $\bm{4}$, $\bm{5}$ and $S^2=-\mathbb{1}$ in the hatted representations $\widehat{\bm{2}}$, $\widehat{\bm{2}}'$, $\widehat{\bm{4}}$, $\widehat{\bm{6}}$.

The weight one modular forms of level $N=5$ can be organized into a sextet of $A'_5$, as shown in Eq.~\eqref{eq-sixletsF1F2}. The higher weight modular form multiplets can be obtained from the tensor products of $Y^{(1)}_{\widehat{\bm{6}}}(\tau)$~\cite{Yao:2020zml,Ding:2023htn}. In order to be self-contained, we present their lengthy expressions in the following. The weight 2 modular forms of level $N=5$ can be arranged into three multiplets in the $A'_5$ representations $\bm{3}$, $\bm{3'}$ and $\bm{5}$ as follows,
\begin{eqnarray}
\nonumber Y_{\bm{3}}^{(2)}&=&\left(Y_{\bm{\widehat{6}}}^{(1)}Y_{\bm{\widehat{6}}}^{(1)}\right)_{\bm{3}}=\left(\begin{array}{c}
 -2 \left(Y_1 Y_2+Y_4 Y_5-Y_3 Y_6\right) \\
\sqrt{2} \left(Y_5^2-2 Y_2 Y_3\right) \\
 -\sqrt{2} \left(Y_4^2+2 Y_1 Y_6\right) \\
\end{array}\right)\,,\\
\nonumber Y_{\bm{3'}}^{(2)}&=&\left(Y_{\bm{\widehat{6}}}^{(1)}Y_{\bm{\widehat{6}}}^{(1)}\right)_{\bm{3'}}=\left(\begin{array}{c}
Y_1 Y_2+Y_3 Y_6 \\
Y_5 Y_6-Y_2 Y_4 \\
 -Y_1 Y_5-Y_3 Y_4 \\
\end{array}\right)\,,\\
Y_{\bm{5}}^{(2)}&=&\left(Y_{\bm{\widehat{6}}}^{(1)}Y_{\bm{\widehat{6}}}^{(1)}\right)_{\bm{5}}=\left(\begin{array}{c}
 -\sqrt{6} \left(Y_1^2+Y_2^2\right) \\
 2 \left(Y_5^2+Y_1 Y_3+Y_2 Y_3+\sqrt{2} Y_4 Y_6\right) \\
 2 \left(Y_3^2+\sqrt{2} \left(Y_2 Y_4+Y_5 Y_6\right)\right) \\
 2 \left(Y_6^2+\sqrt{2} Y_3 Y_4-\sqrt{2} Y_1 Y_5\right) \\
 2 \left(Y_4^2-\sqrt{2} Y_3 Y_5+\left(Y_2-Y_1\right) Y_6\right)
\end{array}\right)\,.
\end{eqnarray}
At weight $k=3$, there are a quartet and two independent sextets of modular forms which are given by
\begin{eqnarray}
\nonumber  Y_{\bm{\widehat{4}}}^{(3)}&=&\left(Y_{\bm{\widehat{6}}}^{(1)}Y_{\bm{3'}}^{(2)}\right)_{\bm{\widehat{4}}}=\left(\begin{array}{c}
 -\sqrt{6} Y_3 Y_{\bm{3}',1}^{(2)}-\sqrt{3} Y_6 Y_{\bm{3}',2}^{(2)}+\sqrt{6} Y_5 Y_{\bm{3}',3}^{(2)} \\
 -2 Y_4 Y_{\bm{3}',1}^{(2)}+Y_1 Y_{\bm{3}',2}^{(2)}-3 Y_2 Y_{\bm{3}',2}^{(2)}+Y_6 Y_{\bm{3}',3}^{(2)} \\
 -2 Y_5 Y_{\bm{3}',1}^{(2)}-Y_3 Y_{\bm{3}',2}^{(2)}+\left(3 Y_1+Y_2\right) Y_{\bm{3}',3}^{(2)} \\
 -\sqrt{6} Y_6 Y_{\bm{3}',1}^{(2)}-\sqrt{6} Y_4 Y_{\bm{3}',2}^{(2)}+\sqrt{3} Y_3 Y_{\bm{3}',3}^{(2)}
\end{array}\right)\,, \\
\nonumber Y_{\bm{\widehat{6}}I}^{(3)}&=&\left(Y_{\bm{\widehat{6}}}^{(1)}Y_{\bm{3}}^{(2)}\right)_{\bm{\widehat{6}}}=\left(\begin{array}{c}
 -Y_1 Y_{\bm{3},1}^{(2)}-\sqrt{2} Y_3 Y_{\bm{3},3}^{(2)} \\
 Y_2 Y_{\bm{3},1}^{(2)}+\sqrt{2} Y_6 Y_{\bm{3},2}^{(2)} \\
 Y_3 Y_{\bm{3},1}^{(2)}-\sqrt{2} Y_1 Y_{\bm{3},2}^{(2)} \\
\sqrt{2} Y_5 Y_{\bm{3},3}^{(2)}-Y_4 Y_{\bm{3},1}^{(2)} \\
 Y_5 Y_{\bm{3},1}^{(2)}+\sqrt{2} Y_4 Y_{\bm{3},2}^{(2)} \\
\sqrt{2} Y_2 Y_{\bm{3},3}^{(2)}-Y_6 Y_{\bm{3},1}^{(2)}
\end{array}\right) \,, \\
Y_{\bm{\widehat{6}}II}^{(3)}&=&\left(Y_{\bm{\widehat{6}}}^{(1)}Y_{\bm{5}}^{(2)}\right)_{\bm{\widehat{6}}}=\left(\begin{array}{c}
\sqrt{2} Y_1 Y_{\bm{5},1}^{(2)}+\sqrt{3} \left(Y_6 Y_{\bm{5},2}^{(2)}-\sqrt{2} Y_5 Y_{\bm{5},3}^{(2)}-\sqrt{2} Y_4 Y_{\bm{5},4}^{(2)}+Y_3 Y_{\bm{5},5}^{(2)}\right) \\
\sqrt{2} Y_2 Y_{\bm{5},1}^{(2)}+\sqrt{3} \left(Y_6 Y_{\bm{5},2}^{(2)}-\sqrt{2} Y_5 Y_{\bm{5},3}^{(2)}+\sqrt{2} Y_4 Y_{\bm{5},4}^{(2)}-Y_3 Y_{\bm{5},5}^{(2)}\right) \\
\sqrt{3} \left(Y_1 Y_{\bm{5},2}^{(2)}-Y_2 Y_{\bm{5},2}^{(2)}+\sqrt{2} Y_4 Y_{\bm{5},5}^{(2)}\right)-2\sqrt{2} Y_3 Y_{\bm{5},1}^{(2)} \\
\sqrt{2} Y_4 Y_{\bm{5},1}^{(2)}+\sqrt{6} \left(Y_3 Y_{\bm{5},2}^{(2)}+\left(Y_2-Y_1\right) Y_{\bm{5},3}^{(2)}\right) \\
\sqrt{2} Y_5 Y_{\bm{5},1}^{(2)}-\sqrt{6} \left(Y_1 Y_{\bm{5},4}^{(2)}+Y_2 Y_{\bm{5},4}^{(2)}-Y_6 Y_{\bm{5},5}^{(2)}\right) \\
\sqrt{3} \left(\sqrt{2} Y_5 Y_{\bm{5},2}^{(2)}+\left(Y_1+Y_2\right) Y_{\bm{5},5}^{(2)}\right)-2\sqrt{2} Y_6 Y_{\bm{5},1}^{(2)}
\end{array}\right)\,,
\end{eqnarray}
Analogously the weight 4 modular multiplets can be obtained from the tensor product of $Y^{(1)}_{\bm{6}}$ with the weight 3 modular multiplets $Y_{\bm{\widehat{4}}}^{(3)}$, $Y_{\bm{\widehat{6}}I}^{(3)}$ and $Y_{\bm{\widehat{6}}II}^{(3)}$, and we have
\begin{eqnarray}
\nonumber Y_{\bm{1}}^{(4)}&=&\left(Y_{\bm{\widehat{6}}}^{(1)}Y_{\bm{\widehat{6}}I}^{(3)}\right)_{\bm{1}}=Y_2 Y_{\bm{6}I,1}^{(3)}-Y_1 Y_{\bm{6}I,2}^{(3)}+Y_6 Y_{\bm{6}I,3}^{(3)}-Y_3 Y_{\bm{6}I,6}^{(3)}+Y_5 Y_{\bm{6}I,4}^{(3)}-Y_4 Y_{\bm{6}I,5}^{(3)}\,, \\
\nonumber Y_{\bm{3}}^{(4)}&=&\left(Y_{\bm{\widehat{6}}}^{(1)}Y_{\bm{\widehat{6}}II}^{(3)}\right)_{\bm{3}}=\left(\begin{array}{c}
-Y_1Y_{\bm{6}II,2}^{(3)}-Y_2Y_{\bm{6}II,1}^{(3)}+Y_3Y_{\bm{6}II,6}^{(3)}-Y_4Y_{\bm{6}II,5}^{(3)}-Y_5Y_{\bm{6}II,4}^{(3)}+Y_6Y_{\bm{6}II,3}^{(3)}\\
-\sqrt{2}Y_2Y_{\bm{6}II,3}^{(3)}-\sqrt{2}Y_3Y_{\bm{6}II,2}^{(3)}+\sqrt{2}Y_5Y_{\bm{6}II,5}^{(3)}\\
-\sqrt{2}Y_1Y_{\bm{6}II,6}^{(3)}-\sqrt{2}Y_4Y_{\bm{6}II,4}^{(3)}-\sqrt{2}Y_6Y_{\bm{6}II,1}^{(3)}
\end{array}\right)\,,\\
\nonumber Y_{\bm{3'}}^{(4)}&=&\left(Y_{\bm{\widehat{6}}}^{(1)}Y_{\bm{\widehat{6}}I}^{(3)}\right)_{\bm{3'}}=\left(\begin{array}{c}
 -Y_1 Y_{\bm{6}I,1}^{(3)}+Y_2 Y_{\bm{6}I,2}^{(3)}-Y_6 Y_{\bm{6}I,3}^{(3)}-Y_5 Y_{\bm{6}I,4}^{(3)}-Y_4 Y_{\bm{6}I,5}^{(3)}-Y_3 Y_{\bm{6}I,6}^{(3)} \\
 Y_4 \left(Y_{\bm{6}I,2}^{(3)}-Y_{\bm{6}I,1}^{(3)}\right)-\sqrt{2} Y_3 Y_{\bm{6}I,3}^{(3)}+\left(Y_2-Y_1\right) Y_{\bm{6}I,4}^{(3)} \\
 Y_5 \left(Y_{\bm{6}I,1}^{(3)}+Y_{\bm{6}I,2}^{(3)}\right)+(Y_1+Y_2) Y_{\bm{6}I,5}^{(3)}+\sqrt{2} Y_6 Y_{\bm{6}I,6}^{(3)}
\end{array}\right) \,, \\
\nonumber Y_{\bm{4}}^{(4)}&=&\left(Y_{\bm{\widehat{6}}}^{(1)}Y_{\bm{\widehat{6}}I}^{(3)}\right)_{\bm{4}}=\left(\begin{array}{c}
 -\sqrt{2} Y_3 Y_{\bm{6}I,2}^{(3)}+\sqrt{2} Y_2 Y_{\bm{6}I,3}^{(3)}-Y_6 Y_{\bm{6}I,4}^{(3)}+Y_4 Y_{\bm{6}I,6}^{(3)} \\
 Y_4 \left(Y_{\bm{6}I,1}^{(3)}-Y_{\bm{6}I,2}^{(3)}\right)-Y_1 Y_{\bm{6}I,4}^{(3)}+Y_2 Y_{\bm{6}I,4}^{(3)}+Y_6 Y_{\bm{6}I,5}^{(3)}-Y_5 Y_{\bm{6}I,6}^{(3)} \\
 Y_5 \left(Y_{\bm{6}I,1}^{(3)}+Y_{\bm{6}I,2}^{(3)}\right)-Y_4 Y_{\bm{6}I,3}^{(3)}+Y_3 Y_{\bm{6}I,4}^{(3)}-Y_1 Y_{\bm{6}I,5}^{(3)}-Y_2 Y_{\bm{6}I,5}^{(3)} \\
\sqrt{2} Y_6 Y_{\bm{6}I,1}^{(3)}-Y_5 Y_{\bm{6}I,3}^{(3)}+Y_3 Y_{\bm{6}I,5}^{(3)}-\sqrt{2} Y_1 Y_{\bm{6}I,6}^{(3)}
\end{array}\right)  \,, \\
\nonumber Y_{\bm{5}I}^{(4)}&=&\left(Y_{\bm{\widehat{6}}}^{(1)}Y_{\bm{\widehat{6}}I}^{(3)}\right)_{\bm{5}}=\left(\begin{array}{c}
 -\sqrt{6} \left(Y_1 Y_{\bm{6}I,1}^{(3)}+Y_2 Y_{\bm{6}I,2}^{(3)}\right) \\
 Y_3 \left(Y_{\bm{6}I,1}^{(3)}+Y_{\bm{6}I,2}^{(3)}\right)+Y_1 Y_{\bm{6}I,3}^{(3)}+Y_2 Y_{\bm{6}I,3}^{(3)}+\sqrt{2} Y_6 Y_{\bm{6}I,4}^{(3)}+2 Y_5 Y_{\bm{6}I,5}^{(3)}+\sqrt{2} Y_4 Y_{\bm{6}I,6}^{(3)} \\
\sqrt{2} Y_4 Y_{\bm{6}I,2}^{(3)}+2 Y_3 Y_{\bm{6}I,3}^{(3)}+\sqrt{2} \left(Y_2 Y_{\bm{6}I,4}^{(3)}+Y_6 Y_{\bm{6}I,5}^{(3)}+Y_5 Y_{\bm{6}I,6}^{(3)}\right) \\
 -\sqrt{2} Y_5 Y_{\bm{6}I,1}^{(3)}+\sqrt{2} Y_4 Y_{\bm{6}I,3}^{(3)}+\sqrt{2} Y_3 Y_{\bm{6}I,4}^{(3)}-\sqrt{2} Y_1 Y_{\bm{6}I,5}^{(3)}+2 Y_6 Y_{\bm{6}I,6}^{(3)} \\
 Y_6 \left(Y_{\bm{6}I,2}^{(3)}-Y_{\bm{6}I,1}^{(3)}\right)-\sqrt{2} Y_5 Y_{\bm{6}I,3}^{(3)}+2 Y_4 Y_{\bm{6}I,4}^{(3)}-\sqrt{2} Y_3 Y_{\bm{6}I,5}^{(3)}-Y_1 Y_{\bm{6}I,6}^{(3)}+Y_2 Y_{\bm{6}I,6}^{(3)}
\end{array}\right) \,, \\
Y_{\bm{5}II}^{(4)}&=&\left(Y_{\bm{\widehat{6}}}^{(1)}Y_{\bm{\widehat{6}}I}^{(3)}\right)_{\bm{5}}=\left(\begin{array}{c}
 -Y_2 Y_{\bm{6}I,1}^{(3)}+Y_1 Y_{\bm{6}I,2}^{(3)}-Y_6 Y_{\bm{6}I,3}^{(3)}+2 Y_5 Y_{\bm{6}I,4}^{(3)}-2 Y_4 Y_{\bm{6}I,5}^{(3)}+Y_3 Y_{\bm{6}I,6}^{(3)} \\
\sqrt{6} \left(Y_1 Y_{\bm{6}I,3}^{(3)}-Y_3 Y_{\bm{6}I,1}^{(3)}\right) \\
\sqrt{3} \left(Y_4 Y_{\bm{6}I,2}^{(3)}-Y_2 Y_{\bm{6}I,4}^{(3)}+Y_6 Y_{\bm{6}I,5}^{(3)}-Y_5 Y_{\bm{6}I,6}^{(3)}\right) \\
\sqrt{3} \left(-Y_5 Y_{\bm{6}I,1}^{(3)}-Y_4 Y_{\bm{6}I,3}^{(3)}+Y_3 Y_{\bm{6}I,4}^{(3)}+Y_1 Y_{\bm{6}I,5}^{(3)}\right) \\
\sqrt{6} \left(Y_2 Y_{\bm{6}I,6}^{(3)}-Y_6 Y_{\bm{6}I,2}^{(3)}\right) \\
\end{array}\right)\,.
\end{eqnarray}
The weight 5 modular multiplets can be decomposed into two $A'_5$ doublets $Y_{\bm{\widehat{2}}}^{(5)}(\tau)$, $Y_{\bm{\widehat{2}'}}^{(5)}(\tau)$, $Y_{\bm{\widehat{4}}}^{(5)}(\tau)$, $Y_{\bm{\widehat{6}}II}^{(5)}(\tau)$, $Y_{\bm{\widehat{6}}II}^{(5)}(\tau)$ and $Y_{\bm{\widehat{6}}III}^{(5)}(\tau)$ as follows,
\begin{eqnarray}
\nonumber Y_{\bm{\widehat{2}}}^{(5)}&=&\left(Y_{\bm{\widehat{6}}}^{(1)}Y_{\bm{5}II}^{(4)}\right)_{\widehat{\bm{2}}}=\left(\begin{array}{c}
\sqrt{3} Y_4 Y_{\bm{5}II,1}^{(4)}-2 Y_3 Y_{\bm{5}II,2}^{(4)}+Y_1 Y_{\bm{5}II,3}^{(4)}+2 Y_2 Y_{\bm{5}II,3}^{(4)}+Y_6 Y_{\bm{5}II,4}^{(4)}-\sqrt{2} Y_5 Y_{\bm{5}II,5}^{(4)} \\
 -\sqrt{3} Y_5 Y_{\bm{5}II,1}^{(4)}-\sqrt{2} Y_4 Y_{\bm{5}II,2}^{(4)}+Y_3 Y_{\bm{5}II,3}^{(4)}+2 Y_1 Y_{\bm{5}II,4}^{(4)}-Y_2 Y_{\bm{5}II,4}^{(4)}+2 Y_6 Y_{\bm{5}II,5}^{(4)}
\end{array}\right) \,,\\
\nonumber Y_{\bm{\widehat{2}'}}^{(5)}&=&\left(Y_{\bm{\widehat{6}}}^{(1)}Y_{\bm{5}II}^{(4)}\right)_{\bm{\widehat{2}'}}=\left(\begin{array}{c}
\sqrt{6} Y_3 Y_{\bm{5}II,1}^{(4)}+3 Y_1 Y_{\bm{5}II,2}^{(4)}+Y_2 Y_{\bm{5}II,2}^{(4)}-2 Y_6 Y_{\bm{5}II,3}^{(4)}+2\sqrt{2} Y_5 Y_{\bm{5}II,4}^{(4)}+\sqrt{2} Y_4 Y_{\bm{5}II,5}^{(4)} \\
\sqrt{6} Y_6 Y_{\bm{5}II,1}^{(4)}+\sqrt{2} Y_5 Y_{\bm{5}II,2}^{(4)}-2\sqrt{2} Y_4 Y_{\bm{5}II,3}^{(4)}+2 Y_3 Y_{\bm{5}II,4}^{(4)}-Y_1 Y_{\bm{5}II,5}^{(4)}+3 Y_2 Y_{\bm{5}II,5}^{(4)}
\end{array}\right)\,, \\
\nonumber Y_{\bm{\widehat{4}}}^{(5)}&=&\left(Y_{\bm{\widehat{6}}}^{(1)}Y_{\bm{5}II}^{(4)}\right)_{\bm{\widehat{4}}}=\left(\begin{array}{c}
\sqrt{6} Y_3 Y_{\bm{5}II,1}^{(4)}-2 Y_2 Y_{\bm{5}II,2}^{(4)}+Y_6 Y_{\bm{5}II,3}^{(4)}-\sqrt{2} Y_5 Y_{\bm{5}II,4}^{(4)}+\sqrt{2} Y_4 Y_{\bm{5}II,5}^{(4)} \\
\sqrt{3} \left(Y_1 Y_{\bm{5}II,3}^{(4)}+Y_2 Y_{\bm{5}II,3}^{(4)}-Y_6 Y_{\bm{5}II,4}^{(4)}+\sqrt{2} Y_5 Y_{\bm{5}II,5}^{(4)}\right) \\
\sqrt{3} \left(\sqrt{2} Y_4 Y_{\bm{5}II,2}^{(4)}-Y_3 Y_{\bm{5}II,3}^{(4)}+\left(Y_1-Y_2\right) Y_{\bm{5}II,4}^{(4)}\right) \\
 -\sqrt{6} Y_6 Y_{\bm{5}II,1}^{(4)}-\sqrt{2} Y_5 Y_{\bm{5}II,2}^{(4)}-\sqrt{2} Y_4 Y_{\bm{5}II,3}^{(4)}+Y_3 Y_{\bm{5}II,4}^{(4)}-2 Y_1 Y_{\bm{5}II,5}^{(4)}
\end{array}\right)\,, \\
\nonumber Y_{\bm{\widehat{6}}I}^{(5)}&=&\left(Y_{\bm{\widehat{6}}}^{(1)}Y_{\bm{1}}^{(4)}\right)_{\bm{\widehat{6}}}=\left(\begin{array}{c}
 Y_1 Y_{\bm{1},1}^{(4)} \\
 Y_2 Y_{\bm{1},1}^{(4)} \\
 Y_3 Y_{\bm{1},1}^{(4)} \\
 Y_4 Y_{\bm{1},1}^{(4)} \\
 Y_5 Y_{\bm{1},1}^{(4)} \\
 Y_6 Y_{\bm{1},1}^{(4)}
\end{array}\right)\,, \\
\nonumber Y_{\bm{\widehat{6}}II}^{(5)}&=&\left(Y_{\bm{\widehat{6}}}^{(1)}Y_{\bm{3'}}^{(4)}\right)_{\bm{\widehat{6}}}=\left(\begin{array}{c}
 Y_1 Y_{\bm{3}',1}^{(4)}-Y_4 Y_{\bm{3}',3}^{(4)} \\
 Y_5 Y_{\bm{3}',2}^{(4)}-Y_2 Y_{\bm{3}',1}^{(4)} \\
 Y_3 Y_{\bm{3}',1}^{(4)}+Y_5 Y_{\bm{3}',3}^{(4)} \\
 Y_6 Y_{\bm{3}',3}^{(4)}-Y_1 Y_{\bm{3}',2}^{(4)} \\
 Y_3 Y_{\bm{3}',2}^{(4)}+Y_2 Y_{\bm{3}',3}^{(4)} \\
 Y_4 Y_{\bm{3}',2}^{(4)}-Y_6 Y_{\bm{3}',1}^{(4)}
\end{array}\right) \,, \\
Y_{\bm{\widehat{6}}III}^{(5)}&=&\left(Y_{\bm{\widehat{6}}}^{(1)}Y_{\bm{5}II}^{(4)}\right)_{\bm{\widehat{6}}}=\left(\begin{array}{c}
\sqrt{3} Y_1 Y_{\bm{5}II,1}^{(4)}-2 Y_5 Y_{\bm{5}II,3}^{(4)}-Y_4 Y_{\bm{5}II,4}^{(4)}+\sqrt{2} Y_3 Y_{\bm{5}II,5}^{(4)} \\
\sqrt{3} Y_2 Y_{\bm{5}II,1}^{(4)}+\sqrt{2} Y_6 Y_{\bm{5}II,2}^{(4)}-Y_5 Y_{\bm{5}II,3}^{(4)}+2 Y_4 Y_{\bm{5}II,4}^{(4)} \\
 -\sqrt{3} Y_3 Y_{\bm{5}II,1}^{(4)}+\sqrt{2} Y_1 Y_{\bm{5}II,2}^{(4)}-Y_5 Y_{\bm{5}II,4}^{(4)}+2 Y_4 Y_{\bm{5}II,5}^{(4)} \\
 2 Y_3 Y_{\bm{5}II,2}^{(4)}-Y_1 Y_{\bm{5}II,3}^{(4)}+2 Y_2 Y_{\bm{5}II,3}^{(4)}+Y_6 Y_{\bm{5}II,4}^{(4)} \\
 -Y_3 Y_{\bm{5}II,3}^{(4)}-2 Y_1 Y_{\bm{5}II,4}^{(4)}-Y_2 Y_{\bm{5}II,4}^{(4)}+2 Y_6 Y_{\bm{5}II,5}^{(4)} \\
 -\sqrt{3} Y_6 Y_{\bm{5}II,1}^{(4)}+2 Y_5 Y_{\bm{5}II,2}^{(4)}+Y_4 Y_{\bm{5}II,3}^{(4)}+\sqrt{2} Y_2 Y_{\bm{5}II,5}^{(4)}
\end{array}\right)\,.
\end{eqnarray}
Finally we report the linearly independent weight 6 modular multiplets of level 5,
\begin{eqnarray}
\nonumber Y_{\bm{1}}^{(6)}&=&\left(Y_{\bm{\widehat{6}}}^{(1)}Y_{\bm{\widehat{6}}II}^{(5)}\right)_{\bm{1}}=
 Y_2 Y_{\bm{6}II,1}^{(5)}-Y_1 Y_{\bm{6}II,2}^{(5)}+Y_6 Y_{\bm{6}II,3}^{(5)}+Y_5 Y_{\bm{6}II,4}^{(5)}-Y_4 Y_{\bm{6}II,5}^{(5)}-Y_3 Y_{\bm{6}II,6}^{(5)}\,, \\
\nonumber Y_{\bm{3}I}^{(6)}&=&\left(Y_{\bm{\widehat{6}}}^{(1)}Y_{\bm{\widehat{6}}I}^{(5)}\right)_{\bm{3}}=\left(\begin{array}{c}
 -Y_2 Y_{\bm{6}I,1}^{(5)}-Y_1 Y_{\bm{6}I,2}^{(5)}+Y_6 Y_{\bm{6}I,3}^{(5)}-Y_5 Y_{\bm{6}I,4}^{(5)}-Y_4 Y_{\bm{6}I,5}^{(5)}+Y_3 Y_{\bm{6}I,6}^{(5)} \\
 -\sqrt{2} \left(Y_3 Y_{\bm{6}I,2}^{(5)}+Y_2 Y_{\bm{6}I,3}^{(5)}-Y_5 Y_{\bm{6}I,5}^{(5)}\right) \\
 -\sqrt{2} \left(Y_6 Y_{\bm{6}I,1}^{(5)}+Y_4 Y_{\bm{6}I,4}^{(5)}+Y_1 Y_{\bm{6}I,6}^{(5)}\right)
\end{array}\right)\,,\\
\nonumber Y_{\bm{3}II}^{(6)}&=&\left(Y_{\bm{\widehat{6}}}^{(1)}Y_{\bm{\widehat{6}}III}^{(5)}\right)_{\bm{3}}=\left(\begin{array}{c}
 -Y_2 Y_{\bm{6}III,1}^{(5)}-Y_1 Y_{\bm{6}III,2}^{(5)}+Y_6 Y_{\bm{6}III,3}^{(5)}-Y_5 Y_{\bm{6}III,4}^{(5)}-Y_4 Y_{\bm{6}III,5}^{(5)}+Y_3 Y_{\bm{6}III,6}^{(5)} \\
 -\sqrt{2} \left(Y_3 Y_{\bm{6}III,2}^{(5)}+Y_2 Y_{\bm{6}III,3}^{(5)}-Y_5 Y_{\bm{6}III,5}^{(5)}\right) \\
 -\sqrt{2} \left(Y_6 Y_{\bm{6}III,1}^{(5)}+Y_4 Y_{\bm{6}III,4}^{(5)}+Y_1 Y_{\bm{6}III,6}^{(5)}\right)
\end{array}\right)\,,\\
\nonumber Y_{\bm{3'}I}^{(6)}&=&\left(Y_{\bm{\widehat{6}}}^{(1)}Y_{\bm{\widehat{6}}I}^{(5)}\right)_{\bm{3'}}=\left(\begin{array}{c}
 Y_2 Y_{\bm{6}I,1}^{(5)}+Y_1 Y_{\bm{6}I,2}^{(5)}+Y_6 Y_{\bm{6}I,3}^{(5)}+Y_3 Y_{\bm{6}I,6}^{(5)} \\
 -Y_4 Y_{\bm{6}I,2}^{(5)}-Y_2 Y_{\bm{6}I,4}^{(5)}+Y_6 Y_{\bm{6}I,5}^{(5)}+Y_5 Y_{\bm{6}I,6}^{(5)} \\
 -Y_5 Y_{\bm{6}I,1}^{(5)}-Y_4 Y_{\bm{6}I,3}^{(5)}-Y_3 Y_{\bm{6}I,4}^{(5)}-Y_1 Y_{\bm{6}I,5}^{(5)}
\end{array}\right)\,,\\
\nonumber Y_{\bm{3'}II}^{(6)}&=&\left(Y_{\bm{\widehat{6}}}^{(1)}Y_{\bm{\widehat{6}}II}^{(5)}\right)_{\bm{3'}}=\left(\begin{array}{c}
Y_1 Y_{\bm{6}II,2}^{(5)}+Y_2 Y_{\bm{6}II,1}^{(5)}+Y_3 Y_{\bm{6}II,6}^{(5)}+Y_6 Y_{\bm{6}II,3}^{(5)} \\
-Y_2 Y_{\bm{6}II,4}^{(5)}-Y_4 Y_{\bm{6}II,2}^{(5)}+Y_5 Y_{\bm{6}II,6}^{(5)}+Y_6 Y_{\bm{6}II,5}^{(5)} \\
-Y_1 Y_{\bm{6}II,5}^{(5)}-Y_3 Y_{\bm{6}II,4}^{(5)}-Y_4 Y_{\bm{6}II,3}^{(5)} -Y_5 Y_{\bm{6}II,1}^{(5)}
\end{array}\right)\,, \\
\nonumber Y_{\bm{4}I}^{(6)}&=&\left(Y_{\bm{\widehat{6}}}^{(1)}Y_{\bm{\widehat{2}'}}^{(5)}\right)_{\bm{4}}=\left(\begin{array}{c}
 Y_4 Y_{\bm{2}',2}^{(5)}-\sqrt{2} Y_1 Y_{\bm{2}',1}^{(5)} \\
\sqrt{2} Y_3 Y_{\bm{2}',1}^{(5)}+Y_5 Y_{\bm{2}',2}^{(5)} \\
 Y_4 Y_{\bm{2}',1}^{(5)}+\sqrt{2} Y_6 Y_{\bm{2}',2}^{(5)} \\
 -Y_5 Y_{\bm{2}',1}^{(5)}-\sqrt{2} Y_2 Y_{\bm{2}',2}^{(5)}
\end{array}\right)\,, \\
\nonumber Y_{\bm{4}II}^{(6)}&=&\left(Y_{\bm{\widehat{6}}}^{(1)}Y_{\bm{\widehat{4}}}^{(5)}\right)_{\bm{4}}=\left(\begin{array}{c}
 Y_1 Y_{\bm{\widehat{4}},1}^{(5)}-\sqrt{3} Y_6 Y_{\bm{\widehat{4}},2}^{(5)}-\sqrt{2} Y_4 Y_{\bm{\widehat{4}},4}^{(5)} \\
 -Y_3 Y_{\bm{\widehat{4}},1}^{(5)}-\sqrt{3} Y_2 Y_{\bm{\widehat{4}},2}^{(5)}-\sqrt{2} Y_5 Y_{\bm{\widehat{4}},4}^{(5)} \\
\sqrt{2} Y_4 Y_{\bm{\widehat{4}},1}^{(5)}-\sqrt{3} Y_1 Y_{\bm{\widehat{4}},3}^{(5)}+Y_6 Y_{\bm{\widehat{4}},4}^{(5)} \\
-\sqrt{2} Y_5 Y_{\bm{\widehat{4}},1}^{(5)}-\sqrt{3} Y_3 Y_{\bm{\widehat{4}},3}^{(5)}-Y_2 Y_{\bm{\widehat{4}},4}^{(5)}
\end{array}\right)\,, \\
\nonumber Y_{\bm{5}I}^{(6)}&=&\left(Y_{\bm{\widehat{6}}}^{(1)}Y_{\bm{\widehat{4}}}^{(5)}\right)_{\bm{5}}=\left(\begin{array}{c}
\sqrt{6} \left(Y_5 Y_{\bm{\widehat{4}},2}^{(5)}+Y_4 Y_{\bm{\widehat{4}},3}^{(5)}\right) \\
\sqrt{6} Y_2 Y_{\bm{\widehat{4}},1}^{(5)}+\sqrt{2} Y_6 Y_{\bm{\widehat{4}},2}^{(5)}+Y_5 Y_{\bm{\widehat{4}},3}^{(5)}-\sqrt{3} Y_4 Y_{\bm{\widehat{4}},4}^{(5)} \\
\sqrt{2} \left(-\sqrt{3} Y_3 Y_{\bm{\widehat{4}},1}^{(5)}+Y_1 Y_{\bm{\widehat{4}},2}^{(5)}+Y_2 Y_{\bm{\widehat{4}},2}^{(5)}+Y_6 Y_{\bm{\widehat{4}},3}^{(5)}\right) \\
\sqrt{2} \left(-Y_3 Y_{\bm{\widehat{4}},2}^{(5)}+Y_1 Y_{\bm{\widehat{4}},3}^{(5)}-Y_2 Y_{\bm{\widehat{4}},3}^{(5)}+\sqrt{3} Y_6 Y_{\bm{\widehat{4}},4}^{(5)}\right) \\
 -\sqrt{3} Y_5 Y_{\bm{\widehat{4}},1}^{(5)}-Y_4 Y_{\bm{\widehat{4}},2}^{(5)}+\sqrt{2} Y_3 Y_{\bm{\widehat{4}},3}^{(5)}+\sqrt{6} Y_1 Y_{\bm{\widehat{4}},4}^{(5)}
\end{array}\right)\,, \\
Y_{\bm{5}II}^{(6)}&=&\left(Y_{\bm{\widehat{6}}}^{(1)}Y_{\bm{\widehat{6}}II}^{(5)}\right)_{\bm{5}}=\left(\begin{array}{c}
 -Y_2 Y_{\bm{6}II,1}^{(5)}+Y_1 Y_{\bm{6}II,2}^{(5)}-Y_6 Y_{\bm{6}II,3}^{(5)}+2 Y_5 Y_{\bm{6}II,4}^{(5)}-2 Y_4 Y_{\bm{6}II,5}^{(5)}+Y_3 Y_{\bm{6}II,6}^{(5)} \\
\sqrt{6} \left(Y_1 Y_{\bm{6}II,3}^{(5)}-Y_3 Y_{\bm{6}II,1}^{(5)}\right) \\
\sqrt{3} \left(Y_4 Y_{\bm{6}II,2}^{(5)}-Y_2 Y_{\bm{6}II,4}^{(5)}+Y_6 Y_{\bm{6}II,5}^{(5)}-Y_5 Y_{\bm{6}II,6}^{(5)}\right) \\
\sqrt{3} \left(-Y_5 Y_{\bm{6}II,1}^{(5)}-Y_4 Y_{\bm{6}II,3}^{(5)}+Y_3 Y_{\bm{6}II,4}^{(5)}+Y_1 Y_{\bm{6}II,5}^{(5)}\right) \\
\sqrt{6} \left(Y_2 Y_{\bm{6}II,6}^{(5)}-Y_6 Y_{\bm{6}II,2}^{(5)}\right)
\end{array}\right)\,.
\end{eqnarray}
The above modular multiplets and their representations under $A'_5$ are listed in table~\ref{tab:MFs-N5}. Notice that the modular form singlet is absent at weight $k=2$.

\begin{table}[hptb!]
\centering
\begin{tabular}{|c|c|}
\hline \hline
Modular weight $k$ & Modular form  multiplets $Y^{(k)}_{\bm{r}}$ \\ \hline

$k=1$ & $Y^{(1)}_{\widehat{\bm{6}}}$\\  \hline

$k=2$ & $Y^{(2)}_{\bm{3}}$,\; $Y^{(2)}_{\bm{3'}}$,\; $Y^{(2)}_{\bm{5}}$\\ \hline

$k=3$ & $Y^{(3)}_{\widehat{\bm{4}}}$,\; $Y^{(3)}_{\widehat{\bm{6}}I}$,\; $Y^{(3)}_{\widehat{\bm{6}}II}$\\ \hline

$k=4$ & $Y^{(4)}_{\bm{1}}$,\; $Y^{(4)}_{\bm{3}}$,\; $Y^{(4)}_{\bm{3'}}$, ~$Y^{(4)}_{\bm{4}}$, \; $Y^{(4)}_{\bm{5}I}$, \; $Y^{(4)}_{\bm{5}II}$\\ \hline

$k=5$ & $Y^{(5)}_{\widehat{\bm{2}}}$,\; $Y^{(5)}_{\widehat{\bm{2}}'}$, \; $Y^{(5)}_{\widehat{\bm{4}}}$, \; $Y^{(5)}_{\widehat{\bm{6}}I}$,\; $Y^{(5)}_{\widehat{\bm{6}}II}$,~$Y^{(5)}_{\widehat{\bm{6}}III}$\\ \hline

$k=6$ & $Y^{(6)}_{\bm{1}}$,\;$Y^{(6)}_{\bm{3}I}$,\; $Y^{(6)}_{\bm{3}II}$,\;$Y^{(6)}_{\bm{3'}I}$,\;$Y^{(6)}_{\bm{3'}II}$,\;$Y^{(6)}_{\bm{4}I}$,\; $Y^{(6)}_{\bm{4}II}$,\; $Y^{(6)}_{\bm{5}I}$,\;$Y^{(6)}_{\bm{5}II}$\\ \hline \hline
\end{tabular}
\caption{\label{tab:MFs-N5} Modular form multiplets of level $N=5$ up to weight 6. }
\end{table}

In the modular Littlest seesaw models based on $A_5$ or $A'_5$, the Yukawa couplings are triplet modular forms $Y^{(2)}_{\bm{3}}(\tau)$, $Y^{(2)}_{\bm{3'}}(\tau)$, $Y^{(4)}_{\bm{3}}(\tau)$ or $Y^{(4)}_{\bm{3'}}(\tau)$, From the alignment of modular triplets at the fixed points $\tau_S=i$, $\tau_{ST}$, $\tau_{ST}$ and $\tau_T$ listed in table~\ref{tab:MF-FPs-N5} together with the representation matrices in Eq.~\eqref{eq:Irr-matrix-N5}, using the general formula of Eq.~\eqref{eq:Ytauf}, we can straightforwardly determine the alignments of the weight 2 and weight 4 triplet modular forms at any modular fixed point in the upper half plane. The values of the modular forms $Y^{(2)}_{\bm{3}}(\gamma\tau_S)$ and $Y^{(2)}_{\bm{3'}}(\gamma\tau_S)$ are given in table~\ref{tab:MF-33p-A5pw24}. It is too lengthy to present the remaining triplet modular forms here.

\begin{center}
\setlength\LTcapwidth{\textwidth}
\begin{longtable}{|c|c|c|}
\caption{\label{tab:MF-33p-A5pw24}
The alignments of weight 2 and level $N=5$ modular triplets at the modular fixed points $\gamma\tau_S$. }
\endfirsthead
\multicolumn{3}{c}{{\bfseries \tablename\ \thetable{} -- continued from previous page}}
\endhead

\endlastfoot
\hline
\hline
$\gamma$ & $Y^{(2)}_{\bm{3}}(\gamma\tau_S
)$ & $Y^{(2)}_{\bm{3}^{\prime}}(\gamma\tau_S)$ \\
\hline

$\{1,S,S^2,S^3\}$
& $\begin{pmatrix}1\\\frac{11(1+\sqrt{5})+\sqrt{1090-58\sqrt{5}}}{22\sqrt{2}}\\\frac{11(1+\sqrt{5})-\sqrt{1090-58\sqrt{5}}}{22\sqrt{2}}\end{pmatrix} $
& $\begin{pmatrix}1\\\frac{1-\sqrt{5}-2\sqrt{85-38\sqrt{5}}}{2\sqrt{2}}\\\frac{1-\sqrt{5}+2\sqrt{85-38\sqrt{5}}}{2\sqrt{2}}\end{pmatrix}$\\
\hline
$\{ST,S^3TS^3,S^3T,S^3TS\}$
& $\begin{pmatrix}1\\\frac{11(1-\sqrt{5})-\sqrt{1090+58\sqrt{5}}}{22\sqrt{2}}\,e^{\frac{4\pi i}{5}}\\\frac{11(1-\sqrt{5})+\sqrt{1090+58\sqrt{5}}}{22\sqrt{2}}\,e^{\frac{6\pi i}{5}}\end{pmatrix}$
& $\begin{pmatrix}1\\\frac{1+\sqrt{5}-2\sqrt{85+38\sqrt{5}}}{2\sqrt{2}}\,e^{\frac{8\pi i}{5}}\\\frac{1+\sqrt{5}+2\sqrt{85+38\sqrt{5}}}{2\sqrt{2}}\,e^{\frac{2\pi i}{5}}\end{pmatrix}$\\
\hline
$\{T,TS,TS^2,TS^3\}$
& $
\begin{pmatrix}
1\\
\frac{11(1+\sqrt{5})+\sqrt{1090-58\sqrt{5}}}{22\sqrt{2}}\,e^{\frac{2\pi i}{5}}\\
\frac{11(1+\sqrt{5})-\sqrt{1090-58\sqrt{5}}}{22\sqrt{2}}\,e^{\frac{8\pi i}{5}}
\end{pmatrix} $
& $
\begin{pmatrix}
1\\
\frac{1-\sqrt{5}-2\sqrt{85-38\sqrt{5}}}{2\sqrt{2}}\,e^{\frac{4\pi i}{5}}\\
\frac{1-\sqrt{5}+2\sqrt{85-38\sqrt{5}}}{2\sqrt{2}}\,e^{\frac{6\pi i}{5}}
\end{pmatrix}$\\
\hline
$\{S^3T^4,TS^3T,ST^4,TST\}$
& $
\begin{pmatrix}
1\\
\frac{11(1-\sqrt{5})-\sqrt{1090+58\sqrt{5}}}{22\sqrt{2}}\,e^{\frac{6\pi i}{5}}\\
\frac{11(1-\sqrt{5})+\sqrt{1090+58\sqrt{5}}}{22\sqrt{2}}\,e^{\frac{4\pi i}{5}}
\end{pmatrix} $
& $
\begin{pmatrix}
1\\
\frac{1+\sqrt{5}-2\sqrt{85+38\sqrt{5}}}{2\sqrt{2}}\,e^{\frac{2\pi i}{5}}\\
\frac{1+\sqrt{5}+2\sqrt{85+38\sqrt{5}}}{2\sqrt{2}}\,e^{\frac{8\pi i}{5}}
\end{pmatrix}$\\
\hline
$\{T^4,T^4S,T^4S^2,T^4S^3\}$
& $
\begin{pmatrix}
1\\
\frac{11(1+\sqrt{5})+\sqrt{1090-58\sqrt{5}}}{22\sqrt{2}}\,e^{\frac{8\pi i}{5}}\\
\frac{11(1+\sqrt{5})-\sqrt{1090-58\sqrt{5}}}{22\sqrt{2}}\,e^{\frac{2\pi i}{5}}
\end{pmatrix} $
& $
\begin{pmatrix}
1\\
\frac{1-\sqrt{5}-2\sqrt{85-38\sqrt{5}}}{2\sqrt{2}}\,e^{\frac{6\pi i}{5}}\\
\frac{1-\sqrt{5}+2\sqrt{85-38\sqrt{5}}}{2\sqrt{2}}\,e^{\frac{4\pi i}{5}}
\end{pmatrix}$\\
\hline
\makecell[c]{$\{T^3ST,T^2S^3T^4,$\\$T^3S^3T,T^2ST^4\}$}
& $
\begin{pmatrix}
1\\
\frac{11(1-\sqrt{5})-\sqrt{1090+58\sqrt{5}}}{22\sqrt{2}}\\
\frac{11(1-\sqrt{5})+\sqrt{1090+58\sqrt{5}}}{22\sqrt{2}}
\end{pmatrix} $
& $
\begin{pmatrix}
1\\
\frac{1+\sqrt{5}-2\sqrt{85+38\sqrt{5}}}{2\sqrt{2}}\\
\frac{1+\sqrt{5}+2\sqrt{85+38\sqrt{5}}}{2\sqrt{2}}
\end{pmatrix}$\\
\hline
\makecell[c]{$\{T^2ST^2,T^2ST^2S,$\\$T^2S^3T^2,T^2ST^2S^3\}$}
& $
\begin{pmatrix}
1\\
\frac{(3+11\,i)(1+\sqrt{5}+\sqrt{10-2\sqrt{5}}\,i)}{52\sqrt{2}}\\
\frac{(3+11\,i)(1+\sqrt{5}-\sqrt{10-2\sqrt{5}}\,i)}{52\sqrt{2}}
\end{pmatrix} $
& $
\begin{pmatrix}
1\\
\frac{(3+i)(-1+\sqrt{5}+\sqrt{10+2\sqrt{5}}\,i)}{8\sqrt{2}}\\
\frac{(3+i)(-1+\sqrt{5}-\sqrt{10+2\sqrt{5}}\,i)}{8\sqrt{2}}
\end{pmatrix}$\\
\hline
\makecell[c]{$\{T^3ST^3,T^3ST^3S$\\$T^3S^3T^3,T^3ST^3S^3\}$}
& $
\begin{pmatrix}
1\\
\frac{(3-11\,i)(1+\sqrt{5}-\sqrt{10-2\sqrt{5}}\,i)}{52\sqrt{2}}\\
\frac{(3-11\,i)(1+\sqrt{5}+\sqrt{10-2\sqrt{5}}\,i)}{52\sqrt{2}}
\end{pmatrix} $
& $
\begin{pmatrix}
1\\
\frac{(3-i)(-1+\sqrt{5}-\sqrt{10+2\sqrt{5}}\,i)}{8\sqrt{2}}\\
\frac{(3-i)(-1+\sqrt{5}+\sqrt{10+2\sqrt{5}}\,i)}{8\sqrt{2}}
\end{pmatrix}$\\
\hline
\makecell[c]{$\{T^4ST^2,T^4ST^2S$\\$T^4S^3T^2,T^4ST^2S^3\}$}
& $
\begin{pmatrix}
1\\
-\frac{3+11\,i}{13\sqrt{2}}\\
-\frac{3+11i}{13\sqrt{2}}
\end{pmatrix} $
& $
\begin{pmatrix}
1\\
\frac{3+i}{2\sqrt{2}}\\
\frac{3+i}{2\sqrt{2}}
\end{pmatrix}$\\
\hline
\makecell[c]{$\{TST^3,TST^3S$\\$TS^3T^3,TST^3S^3\}$}
& $
\begin{pmatrix}
1\\
-\frac{3-11\,i}{13\sqrt{2}}\\
-\frac{3-11i}{13\sqrt{2}}
\end{pmatrix} $
& $
\begin{pmatrix}
1\\
\frac{3-i}{2\sqrt{2}}\\
\frac{3-i}{2\sqrt{2}}
\end{pmatrix}$\\
\hline
\makecell[c]{$\{T^2ST^3S,T^2ST^3S^2$\\$T^2ST^3S^3,T^2ST^3\}$}
& $
\begin{pmatrix}
1\\
\frac{(-3+11\,i)(-1+\sqrt{5}+\sqrt{10+2\sqrt{5}}\,i)}{52\sqrt{2}}\\
\frac{(-3+11\,i)(-1+\sqrt{5}-\sqrt{10+2\sqrt{5}}\,i)}{52\sqrt{2}}
\end{pmatrix} $
& $
\begin{pmatrix}
1\\
\frac{(-3+i)(1+\sqrt{5}-\sqrt{10-2\sqrt{5}}\,i)}{8\sqrt{2}}\\
\frac{(-3+i)(1+\sqrt{5}+\sqrt{10-2\sqrt{5}}\,i)}{8\sqrt{2}}
\end{pmatrix}$\\
\hline
\makecell[c]{$\{T^3ST^2S,T^3ST^2S^2$\\$T^3ST^2S^3,T^3ST^2\}$}
& $
\begin{pmatrix}
1\\
\frac{(3+11\,i)(1-\sqrt{5}+\sqrt{10+2\sqrt{5}}\,i)}{52\sqrt{2}}\\
\frac{(3+11\,i)(1-\sqrt{5}-\sqrt{10+2\sqrt{5}}\,i)}{52\sqrt{2}}
\end{pmatrix} $
& $
\begin{pmatrix}
1\\
\frac{(3+i)(-1-\sqrt{5}-\sqrt{10-2\sqrt{5}}\,i)}{8\sqrt{2}}\\
\frac{(3+i)(-1-\sqrt{5}+\sqrt{10-2\sqrt{5}}\,i)}{8\sqrt{2}}
\end{pmatrix}$\\
\hline
\makecell[c]{$\{ST^2ST^3,ST^2ST^3S$\\$ST^2S^3T^3,ST^2ST^3S^3\}$}
& $
\begin{pmatrix}
1\\
\frac{11(1+\sqrt{5})-\sqrt{1090-58\sqrt{5}}}{22\sqrt{2}}\,e^{\frac{4\pi i}{5}}\\
\frac{11(1+\sqrt{5})+\sqrt{1090-58\sqrt{5}}}{22\sqrt{2}}\,e^{\frac{6\pi i}{5}}
\end{pmatrix} $
& $
\begin{pmatrix}
1\\
\frac{-1+\sqrt{5}-2\sqrt{85-38\sqrt{5}}}{2\sqrt{2}}\,e^{\frac{3\pi i}{5}}\\
\frac{-1+\sqrt{5}+2\sqrt{85-38\sqrt{5}}}{2\sqrt{2}}\,e^{\frac{7\pi i}{5}}
\end{pmatrix}$\\
\hline
\makecell[c]{$\{ST^3ST^2,ST^3ST^2S$\\$ST^3S^3T^2,ST^3ST^2S^3\}$}
& $
\begin{pmatrix}
1\\
\frac{11(1+\sqrt{5})-\sqrt{1090-58\sqrt{5}}}{22\sqrt{2}}\,e^{\frac{6\pi i}{5}}\\
\frac{11(1+\sqrt{5})+\sqrt{1090-58\sqrt{5}}}{22\sqrt{2}}\,e^{\frac{4\pi i}{5}}
\end{pmatrix} $
& $
\begin{pmatrix}
1\\
\frac{-1+\sqrt{5}-2\sqrt{85-38\sqrt{5}}}{2\sqrt{2}}\,e^{\frac{7\pi i}{5}}\\
\frac{-1+\sqrt{5}+2\sqrt{85-38\sqrt{5}}}{2\sqrt{2}}\,e^{\frac{3\pi i}{5}}
\end{pmatrix}$\\
\hline
\makecell[c]{$\{S^3T^3ST,S^3T^2S^3T^4$\\$S^3T^3S^3T,S^3T^2ST^4\}$}
& $
\begin{pmatrix}
1\\
\frac{11(1-\sqrt{5})+\sqrt{1090+58\sqrt{5}}}{22\sqrt{2}}\\
\frac{11(1-\sqrt{5})-\sqrt{1090+58\sqrt{5}}}{22\sqrt{2}}
\end{pmatrix} $
& $
\begin{pmatrix}
1\\
\frac{1+\sqrt{5}+2\sqrt{85+38\sqrt{5}}}{2\sqrt{2}}\\
\frac{1+\sqrt{5}-2\sqrt{85+38\sqrt{5}}}{2\sqrt{2}}
\end{pmatrix}$\\
\hline
\makecell[c]{$\{ST^2ST^2,ST^2ST^2S,$\\$ST^2S^3T^2,ST^2ST^2S^3\}$}
& $
\begin{pmatrix}
1\\
\frac{11(1-\sqrt{5})+\sqrt{1090+58\sqrt{5}}}{22\sqrt{2}}\,e^{\frac{8\pi i}{5}}\\
\frac{11(1-\sqrt{5})-\sqrt{1090+58\sqrt{5}}}{22\sqrt{2}}\,e^{\frac{2\pi i}{5}}
\end{pmatrix} $
& $
\begin{pmatrix}
1\\
\frac{1+\sqrt{5}+2\sqrt{85+38\sqrt{5}}}{2\sqrt{2}}\,e^{\frac{6\pi i}{5}}\\
\frac{1+\sqrt{5}-2\sqrt{85+38\sqrt{5}}}{2\sqrt{2}}\,e^{\frac{4\pi i}{5}}
\end{pmatrix}$\\
\hline
\makecell[c]{$\{TST^3ST,ST^3ST^3$\\$TST^3S^3T,ST^3S^3T^3\}$}
& $
\begin{pmatrix}
1\\
\frac{11(1-\sqrt{5})+\sqrt{1090+58\sqrt{5}}}{22\sqrt{2}}\,e^{\frac{2\pi i}{5}}\\
\frac{11(1-\sqrt{5})-\sqrt{1090+58\sqrt{5}}}{22\sqrt{2}}\,e^{\frac{8\pi i}{5}}
\end{pmatrix} $
& $
\begin{pmatrix}
1\\
\frac{1+\sqrt{5}+2\sqrt{85+38\sqrt{5}}}{2\sqrt{2}}\,e^{\frac{4\pi i}{5}}\\
\frac{1+\sqrt{5}-2\sqrt{85+38\sqrt{5}}}{2\sqrt{2}}\,e^{\frac{6\pi i}{5}}
\end{pmatrix}$\\
\hline
\makecell[c]{$\{T^4ST,T^3S^3T^4$\\$T^4STS^2,T^3ST^4\}$}
& $
\begin{pmatrix}
1\\
\frac{11(1-\sqrt{5})-\sqrt{1090+58\sqrt{5}}}{22\sqrt{2}}\,e^{\frac{2\pi i}{5}}\\
\frac{11(1-\sqrt{5})+\sqrt{1090+58\sqrt{5}}}{22\sqrt{2}}\,e^{\frac{8\pi i}{5}}
\end{pmatrix} $
& $
\begin{pmatrix}
1\\
\frac{1+\sqrt{5}-2\sqrt{85+38\sqrt{5}}}{2\sqrt{2}}\,e^{\frac{4\pi i}{5}}\\
\frac{1+\sqrt{5}+2\sqrt{85+38\sqrt{5}}}{2\sqrt{2}}\,e^{\frac{6\pi i}{5}}
\end{pmatrix}$\\
\hline
\makecell[c]{$\{TST^4,T^2ST$\\$TST^4S^2,T^2S^3T\}$}
& $
\begin{pmatrix}
1\\
\frac{11(1-\sqrt{5})-\sqrt{1090+58\sqrt{5}}}{22\sqrt{2}}\,e^{\frac{8\pi i}{5}}\\
\frac{11(1-\sqrt{5})+\sqrt{1090+58\sqrt{5}}}{22\sqrt{2}}\,e^{\frac{2\pi i}{5}}
\end{pmatrix} $
& $
\begin{pmatrix}
1\\
\frac{1+\sqrt{5}-2\sqrt{85+38\sqrt{5}}}{2\sqrt{2}}\,e^{\frac{6\pi i}{5}}\\
\frac{1+\sqrt{5}+2\sqrt{85+38\sqrt{5}}}{2\sqrt{2}}\,e^{\frac{4\pi i}{5}}
\end{pmatrix}$\\
\hline
\makecell[c]{$\{ST^2ST,T^4ST^3$\\$ST^2STS^2,T^4S^3T^3\}$}
& $
\begin{pmatrix}
1\\
\frac{(3-11\,i)(1+\sqrt{5}+\sqrt{10-2\sqrt{5}}\,i)}{52\sqrt{2}}\\
\frac{(3-11\,i)(1+\sqrt{5}-\sqrt{10-2\sqrt{5}}\,i)}{52\sqrt{2}}
\end{pmatrix} $
& $
\begin{pmatrix}
1\\
\frac{(3-i)(-1+\sqrt{5}+\sqrt{10+2\sqrt{5}}\,i)}{8\sqrt{2}}\\
\frac{(3-i)(-1+\sqrt{5}-\sqrt{10+2\sqrt{5}}\,i)}{8\sqrt{2}}
\end{pmatrix}$\\
\hline
\makecell[c]{$\{TST^2,TST^2S$\\$TS^3T^2,TST^2S^3\}$}
& $
\begin{pmatrix}
1\\
\frac{(3+11\,i)(1+\sqrt{5}-\sqrt{10-2\sqrt{5}}\,i)}{52\sqrt{2}}\\
\frac{(3+11\,i)(1+\sqrt{5}+\sqrt{10-2\sqrt{5}}\,i)}{52\sqrt{2}}
\end{pmatrix} $
& $
\begin{pmatrix}
1\\
\frac{(3+i)(-1+\sqrt{5}-\sqrt{10+2\sqrt{5}}\,i)}{8\sqrt{2}}\\
\frac{(3+i)(-1+\sqrt{5}+\sqrt{10+2\sqrt{5}}\,i)}{8\sqrt{2}}
\end{pmatrix}$\\
\hline
\makecell[c]{$\{T^2ST^3ST^2,T^2ST^3ST^2S$\\$T^2ST^3ST^2S^2,T^2ST^3ST^2S^3\}$}
& $
\begin{pmatrix}
1\\
\frac{11(1+\sqrt{5})-\sqrt{1090-58\sqrt{5}}}{22\sqrt{2}}\\
\frac{11(1+\sqrt{5})+\sqrt{1090-58\sqrt{5}}}{22\sqrt{2}}\
\end{pmatrix} $
& $
\begin{pmatrix}
1\\
\frac{1-\sqrt{5}+2\sqrt{85-38\sqrt{5}}}{2\sqrt{2}}\\
\frac{1-\sqrt{5}-2\sqrt{85-38\sqrt{5}}}{2\sqrt{2}}\
\end{pmatrix}$\\
\hline
\makecell[c]{$\{T^3ST^3ST,T^3ST^2S^3T^4$\\$T^3ST^3STS^2,T^3ST^2ST^4\}$}
& $
\begin{pmatrix}
1\\
\frac{11(1-\sqrt{5})+\sqrt{1090+58\sqrt{5}}}{22\sqrt{2}}\,e^{\frac{6\pi i}{5}}\\
\frac{11(1-\sqrt{5})-\sqrt{1090+58\sqrt{5}}}{22\sqrt{2}}\,e^{\frac{4\pi i}{5}}
\end{pmatrix} $
& $
\begin{pmatrix}
1\\
\frac{1+\sqrt{5}+2\sqrt{85+38\sqrt{5}}}{2\sqrt{2}}\,e^{\frac{2\pi i}{5}}\\
\frac{1+\sqrt{5}-2\sqrt{85+38\sqrt{5}}}{2\sqrt{2}}\,e^{\frac{8\pi i}{5}}
\end{pmatrix}$\\
\hline
\makecell[c]{$\{TST^3ST^3,T^2ST^3S^3T$\\$TST^3ST^3S^2,T^2ST^3ST\}$}
& $
\begin{pmatrix}
1\\
\frac{11(1-\sqrt{5})+\sqrt{1090+58\sqrt{5}}}{22\sqrt{2}}\,e^{\frac{4\pi i}{5}}\\
\frac{11(1-\sqrt{5})-\sqrt{1090+58\sqrt{5}}}{22\sqrt{2}}\,e^{\frac{6\pi i}{5}}
\end{pmatrix} $
& $
\begin{pmatrix}
1\\
\frac{1+\sqrt{5}+2\sqrt{85+38\sqrt{5}}}{2\sqrt{2}}\,e^{\frac{8\pi i}{5}}\\
\frac{1+\sqrt{5}-2\sqrt{85+38\sqrt{5}}}{2\sqrt{2}}\,e^{\frac{2\pi i}{5}}
\end{pmatrix}$\\
\hline
\makecell[c]{$\{ST^2ST^3ST,T^4ST^2S^3T^3$\\$ST^2ST^3STS^2,T^4ST^2ST^3\}$}
& $
\begin{pmatrix}
1\\
\frac{11(1+\sqrt{5})-\sqrt{1090-58\sqrt{5}}}{22\sqrt{2}}\,e^{\frac{2\pi i}{5}}\\
\frac{11(1+\sqrt{5})+\sqrt{1090-58\sqrt{5}}}{22\sqrt{2}}\,e^{\frac{8\pi i}{5}}
\end{pmatrix} $
& $
\begin{pmatrix}
1\\
\frac{-1+\sqrt{5}-2\sqrt{85-38\sqrt{5}}}{2\sqrt{2}}\,e^{\frac{9\pi i}{5}}\\
\frac{-1+\sqrt{5}+2\sqrt{85-38\sqrt{5}}}{2\sqrt{2}}\,e^{\frac{\pi i}{5}}
\end{pmatrix}$\\
\hline
\makecell[c]{$\{TST^3ST^2S,TST^3S^3T^2$\\$TST^3ST^2S^3,TST^3ST^2\}$}
& $
\begin{pmatrix}
1\\
\frac{11(1+\sqrt{5})-\sqrt{1090-58\sqrt{5}}}{22\sqrt{2}}\,e^{\frac{8\pi i}{5}}\\
\frac{11(1+\sqrt{5})+\sqrt{1090-58\sqrt{5}}}{22\sqrt{2}}\,e^{\frac{2\pi i}{5}}
\end{pmatrix} $
& $
\begin{pmatrix}
1\\
\frac{-1+\sqrt{5}-2\sqrt{85-38\sqrt{5}}}{2\sqrt{2}}\,e^{\frac{\pi i}{5}}\\
\frac{-1+\sqrt{5}+2\sqrt{85-38\sqrt{5}}}{2\sqrt{2}}\,e^{\frac{9\pi i}{5}}
\end{pmatrix}$\\
\hline
\makecell[c]{$\{T^3,T^3S$\\$T^3S^2,T^3S^3\}$}
& $
\begin{pmatrix}
1\\
\frac{11(1+\sqrt{5})+\sqrt{1090-58\sqrt{5}}}{22\sqrt{2}}\,e^{\frac{6\pi i}{5}}\\
\frac{11(1+\sqrt{5})-\sqrt{1090-58\sqrt{5}}}{22\sqrt{2}}\,e^{\frac{4\pi i}{5}}
\end{pmatrix} $
& $
\begin{pmatrix}
1\\
\frac{-1+\sqrt{5}+2\sqrt{85-38\sqrt{5}}}{2\sqrt{2}}\,e^{\frac{7\pi i}{5}}\\
\frac{-1+\sqrt{5}-2\sqrt{85-38\sqrt{5}}}{2\sqrt{2}}\,e^{\frac{3\pi i}{5}}
\end{pmatrix}$\\
\hline
\makecell[c]{$\{ST^3,S^3T^3S^3$\\$S^3T^3,S^3T^3S\}$}
& $
\begin{pmatrix}
1\\
\frac{(3-11\,i)(1-\sqrt{5}+\sqrt{10+2\sqrt{5}}\,i)}{52\sqrt{2}}\\
\frac{(3-11\,i)(1-\sqrt{5}-\sqrt{10+2\sqrt{5}}\,i)}{52\sqrt{2}}
\end{pmatrix} $
& $
\begin{pmatrix}
1\\
\frac{(3-i)(-1-\sqrt{5}-\sqrt{10-2\sqrt{5}}\,i)}{8\sqrt{2}}\\
\frac{(3-i)(-1-\sqrt{5}+\sqrt{10-2\sqrt{5}}\,i)}{8\sqrt{2}}
\end{pmatrix}$\\
\hline
\makecell[c]{$\{S^3T^2,S^3T^2S$\\$ST^2,S^3T^2S^3\}$}
& $
\begin{pmatrix}
1\\
\frac{(3+11\,i)(1-\sqrt{5}-\sqrt{10+2\sqrt{5}}\,i)}{52\sqrt{2}}\\
\frac{(3+11\,i)(1-\sqrt{5}+\sqrt{10+2\sqrt{5}}\,i)}{52\sqrt{2}}
\end{pmatrix} $
& $
\begin{pmatrix}
1\\
\frac{(3+i)(-1-\sqrt{5}+\sqrt{10-2\sqrt{5}}\,i)}{8\sqrt{2}}\\
\frac{(3+i)(-1-\sqrt{5}-\sqrt{10-2\sqrt{5}}\,i)}{8\sqrt{2}}
\end{pmatrix}$\\
\hline
\makecell[c]{$\{T^2,T^2S$\\$T^2S^2,T^2S^3\}$}
& $
\begin{pmatrix}
1\\
\frac{11(1+\sqrt{5})+\sqrt{1090-58\sqrt{5}}}{22\sqrt{2}}\,e^{\frac{4\pi i}{5}}\\
\frac{11(1+\sqrt{5})-\sqrt{1090-58\sqrt{5}}}{22\sqrt{2}}\,e^{\frac{6\pi i}{5}}
\end{pmatrix} $
& $
\begin{pmatrix}
1\\
\frac{-1+\sqrt{5}+2\sqrt{85-38\sqrt{5}}}{2\sqrt{2}}\,e^{\frac{3\pi i}{5}}\\
\frac{-1+\sqrt{5}-2\sqrt{85-38\sqrt{5}}}{2\sqrt{2}}\,e^{\frac{7\pi i}{5}}
\end{pmatrix}$\\
\hline
\end{longtable}
\end{center}

\section{\label{app:basis-transformation} Basis transformation}

In this section, we present the procedure and formulas for performing basis transformations in the context of modular flavor symmetry. Most of the following results hold independently of modular symmetry. Let $(\alpha_1, \alpha_2, \alpha_3)^T$ be a triplet in the three-dimensional  representation $\bm{3}$ of the finite modular group $\Gamma'_N$, it transforms under the modular symmetry as follows,
\begin{equation}
\begin{pmatrix}
\alpha_1 \\
\alpha_2 \\
\alpha_3
\end{pmatrix}\xrightarrow{\gamma}\rho_{\bm{3}}(\gamma)\begin{pmatrix}
\alpha_1 \\
\alpha_2 \\
\alpha_3
\end{pmatrix}\,,~~\gamma\in\Gamma\,,
\end{equation}
where the automorphy factor is dropped for notation simplicity. When performing a unitary basis transformation $V$ for the triplet $(\alpha_1, \alpha_2, \alpha_3)^T$, we have
\begin{equation}
\begin{pmatrix}
\alpha'_1 \\
\alpha'_2 \\
\alpha'_3
\end{pmatrix}=V\begin{pmatrix}
\alpha_1 \\
\alpha_2 \\
\alpha_3
\end{pmatrix}\,,~~\text{equivalently}~~\begin{pmatrix}
\alpha_1 \\
\alpha_2 \\
\alpha_3
\end{pmatrix}=V^{\dagger}\begin{pmatrix}
\alpha'_1 \\
\alpha'_2 \\
\alpha'_3
\end{pmatrix}\,,
\end{equation}
where the basis transformation $V$ is a $3\times3$ unitary matrix. Then the modular transformation of $(\alpha'_1, \alpha'_2, \alpha'_3)^T$ is given by
\begin{equation}
\begin{pmatrix}
\alpha'_1 \\
\alpha'_2 \\
\alpha'_3
\end{pmatrix}\xrightarrow{\gamma}\rho'_{\bm{3}}(\gamma)\begin{pmatrix}
\alpha'_1 \\
\alpha'_2 \\
\alpha'_3
\end{pmatrix}\,,~~~\rho'_{\bm{3}}(\gamma)=V\rho_{\bm{3}}(\gamma)V^{\dagger}\,,~~\gamma\in\Gamma\,.
\end{equation}
Consequently $\rho_{\bm{3}}(\gamma)$ and $\rho'_{\bm{3}}(\gamma)$ are equivalent representations of $\Gamma'_N$. In the working basis of Appendices~\ref{app:N4-group-MF} and~\ref{app:N5} for the concerned $S'_4$ and $A'_5$ modular groups,
two triplets $\alpha=(\alpha_1, \alpha_2, \alpha_3)^T\sim\bm{3}$ and $\beta=(\beta_1, \beta_2, \beta_3)^T\sim\bm{3}$ contract into an invariant singlet in the following way,
\begin{eqnarray}
(\alpha\otimes\beta)_{\bm{1}}=\alpha_1\beta_1+\alpha_2\beta_3+\alpha_3\beta_2=\begin{pmatrix}
\alpha_1, \alpha_2, \alpha_3\end{pmatrix}P_{132}\begin{pmatrix}
\beta_1 \\
\beta_2 \\
\beta_3
\end{pmatrix}=\begin{pmatrix}
\alpha'_1, \alpha'_2, \alpha'_3\end{pmatrix}V^{*}P_{132}V^{\dagger}\begin{pmatrix}
\beta'_1 \\
\beta'_2 \\
\beta'_3
\end{pmatrix}\,.
\end{eqnarray}
Consequently the Clebsch–Gordan contraction matrix is transformed into $V^{*}P_{132}V^{\dagger}$ after the basis transformation. Regarding the alignment of modular forms at the fixed point $\tau_0$ satisfying $\gamma_0\tau_0=\tau_0$, we have
\begin{eqnarray}
\rho_{\bm{3}}(\gamma_0)Y^{(k)}_{\bm{3}}(\tau_0)=(c_0\tau_0+d_0)^{-k}Y^{(k)}_{\bm{3}}(\tau_0)\,,~~~
\rho'_{\bm{3}}(\gamma_0)Y^{(k)'}_{\bm{3}}(\tau_0)=(c_0\tau_0+d_0)^{-k}Y^{(k)'}_{\bm{3}}(\tau_0)\,,
\end{eqnarray}
Using the basis transformation $\rho'_{\bm{3}}(\gamma)=V\rho_{\bm{3}}(\gamma)V^{\dagger}$, we obtain
\begin{equation}
Y^{(k)'}_{\bm{3}}(\tau_0)=VY^{(k)}_{\bm{3}}(\tau_0)\,.
\end{equation}
Hence the atmospheric and solar neutrino alignments $Y'_{\text{atm}}\equiv Y_{\bm{3}}(\tau_{\text{atm}})$ and $Y'_{\text{sol}}\equiv Y_{\bm{3}}(\tau_{\text{sol}})$ in the new basis take the following form,
\begin{equation}
Y'_{\text{atm}}=VY_{\text{atm}}\,,~~~Y'_{\text{sol}}=VY_{\text{sol}}\,.
\end{equation}
As a result, the light neutrino mass matrix $M_{\nu}$ in Eq.~\eqref{eq:Mnu-ma-ms} undergoes the following basis transformation:
\begin{eqnarray}
M_{\nu}=V^{*}M_{\nu}V^{\dagger}\,.
\end{eqnarray}
Consequently, the neutrino diagonalization matrix $U_{\nu}$ transforms according to
\begin{eqnarray}
U'_{\nu}=VU_{\nu}\,,
\end{eqnarray}
where $U_{\nu}$ and $U'_{\nu}$ diagonalize $m_{\nu}$ and $m'_{\nu}$ respectively such
that $U^{T}_{\nu}m_{\nu}U_{\nu}=U'^{T}_{\nu}m'_{\nu}U'_{\nu}=\text{diag}(m_1, m_2, m_3)$.

In a similar manner, one finds that under a basis transformation the charged-lepton mass matrix satisfies
\begin{eqnarray}
M'^{\dagger}_\ell M'_\ell= VM^{\dagger}_\ell(\tau_\ell)M_\ell(\tau_\ell)V^{\dagger} \,,
\end{eqnarray}
which implies
\begin{equation}
U'_{\ell}=VU_{\ell}\,.
\end{equation}
Consequently, the lepton mixing matrix is invariant under a change of basis, namely
\begin{equation}
U'_{PMNS}=U'^{\dagger}_{\ell}U'_{\nu}=\left(VU_{\ell}\right)^{\dagger}VU_{\nu}=U^{\dagger}_{\ell}U_{\nu}=U_{PMNS}\,.
\end{equation}

\end{appendix}

\providecommand{\href}[2]{#2}\begingroup\raggedright\endgroup

\end{document}